\PassOptionsToClass{twocolumn}{revtex4-1}

\documentclass[%
  floatfix,%
  aps,%
  pra,%
  superscriptaddress,%
  secnumarabic,%
  a4paper,%
]{revtex4-1}

\usepackage[T1]{fontenc}
\usepackage[latin1]{inputenc}
\usepackage[ngerman,english]{babel}

\usepackage{textcomp}
\usepackage{amsmath}
\usepackage{amsfonts}
\usepackage{amssymb}

\usepackage{bbm}

\usepackage{braket}

\usepackage{xcolor}
\usepackage{graphicx}
\usepackage[textfontcmd=small,vecinclude=overwrite]{combinedgraphics}

\usepackage{ifpdf}
\ifpdf
  \newcommand*{\hyperrefcolor}{violet}
  \usepackage[%
    pdfsubject={Preprint},%
    pdftitle={Many-Body Physics with Trapped Ions},%
    pdfauthor={Christian Schneider, Diego Porras, and Tobias Schaetz},%
    bookmarksopen=true,   
    plainpages=false,     
    colorlinks,           
    linkcolor=\hyperrefcolor,%
    anchorcolor=\hyperrefcolor,%
    citecolor=\hyperrefcolor,%
    urlcolor=\hyperrefcolor,%
    menucolor=\hyperrefcolor,%
    filecolor=\hyperrefcolor%
  ]{hyperref}
\fi

\usepackage{userdef_macros}

\numberwithin{equation}{section}

\secnumarabic

\def\mybibstyle{unsrt_inlinelink}

\def\mypicturescale{1}
\def\mypicturescalepre{0.5}
\def\myfigurescale{0.677}  
\def\myfigurescalepre{0.775}

\newcommand{\theintroductorynote}{%
  \pagenumbering{roman}
  \thispagestyle{empty}
  \section*{Introductory Note}
  This version of the manuscript contains the differences between the
  previous revision and the current one. Any change at each page is preceeded
  by a unique mark, \eg \changecntbox{1}. Changes between the revisions
  are emphasized like this: {\removedcolor\removedfont{text removed
  from the previous revision}} and {\addedcolor\addedfont{newly added text
  in the current revision}}. A list with comments explaining each change
  is appended at the end of the manuscript (see section \revisionchangesname).
  \clearpage~
  \thispagestyle{empty}
  \clearpage
}

\newcommand*{\site}[2][i]{\addtosuperscript{#2}{(#1)}}
\newcommand*{\mode}[2][m]{\addtosubscript{#2}{#1}}

\newcommand*{\rwa}[1]{\addtosuperscript{#1}{\text{(RWA)}}}
\newcommand*{\ldr}[1]{\addtosuperscript{#1}{\text{(LDR)}}}

\newcommand*{\pauli}[1]{\op{\sigma}_{#1}}
\newcommand*{\kauli}{\op{\kappa}}

\providecommand*{\mathbbm}[1]{#1}

\makeatletter
\newcommand*{\ultrashortmacroson}{%
  \let\@usm@H@orig=\H
  \let\@usm@a@orig=\a
  \let\@usm@P@orig=\P
  \let\@usm@L@orig=\L

  \newcommand*{\I}{\op{\mathbbm{1}}}
  \renewcommand*{\H}{\op{\mathcal{H}}}
  \renewcommand*{\a}{\op{a}}
  \newcommand*{\ad}{\op{a}^\dagger}
  \newcommand*{\U}{\op{U}}
  \newcommand*{\D}{\op{D}}
  \renewcommand*{\P}{\op{P}}
  \newcommand*{\R}{\op{R}}
  \renewcommand*{\L}{\mathcal{L}}
}

\newcommand*{\ultrashortmacrosoff}{%
  \let\I=\undefined
  \let\H=\@usm@H@orig
  \let\a=\@usm@a@orig
  \let\ad=\undefined
  \let\U=\undefined
  \let\D=\undefined
  \let\P=\@usm@P@orig
  \let\R=\undefined
  \let\L=\@usm@L@orig
}
\makeatother

\newcommand*{\OmegaI}[1][I]{\Omega_\text{#1}}
\newcommand*{\omegaI}[1][I]{\omega_\text{#1}}
\newcommand*{\kI}[1][I]{k_\text{#1}}
\newcommand*{\phiI}[1][I]{\phi_\text{#1}}
\newcommand*{\HI}[1][I]{\H_\text{#1}}
\newcommand*{\UI}[1][I]{\U_\text{#1}}

\makeatletter
\@ifclassloaded{revtex4}{%
  \iftwocolumn{\let@environment{widetext}{widetext@grid}}%
    {\newenvironment{widetext}{}{}}%
}{}
\makeatother

\ultrashortmacroson 

\newcommand{\myfigurekink}{
  \begin{figure}[!htb]
    \centering
    \includegraphics[width=\mypicturescale\hsize,keepaspectratio]%
      {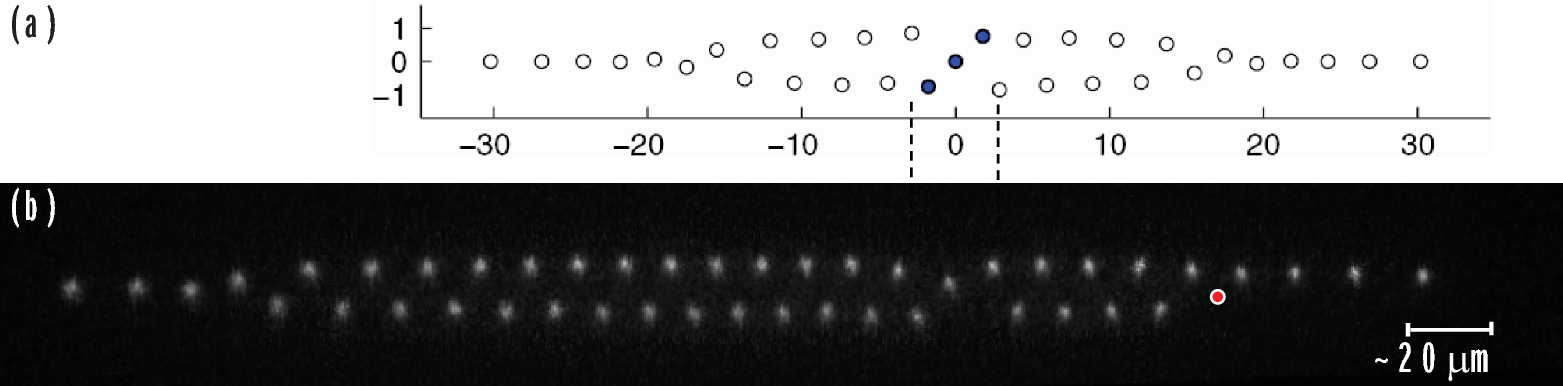}
    \caption{Topological defects in two-dimensional Coulomb crystals
      (\cmp{} \figref{coulomb:crystals}c for a comparable crystal without
      defects).
      Changing the experimental parameters non-adiabatically during a
      structural phase transition from a linear chain of ions to a zigzag
      structure, the order within the crystal breaks up in domains, framed by
      topologically protected defects that are suited to simulate solitons.
      (a) Numerical simulations for $33$ ions predicting a localized
      topological defect at the position of the marked (blue) ions.
      (courtesy of Benni Reznik and Haggai Landa)
      (b) CCD image of $45$ laser cooled $\atom{Mg}[+]$ ions
      providing clear evidence of the topological defect
      indicated by the zigzag--zagzig transition.
      The crystal contains a non-fluorescing molecular ion ($\atom{MgH}[+]$) at
      the red mark.
      (courtesy of G\"{u}nther Leschhorn and Steffen Kahra of the group at MPQ)}
    \figlabel{kink}
  \end{figure}
}

\newcommand{\myfigureopticalrftrap}{
  \begin{figure}[!htb]
    \centering
    \includegraphics[width=\mypicturescale\hsize,keepaspectratio]%
      {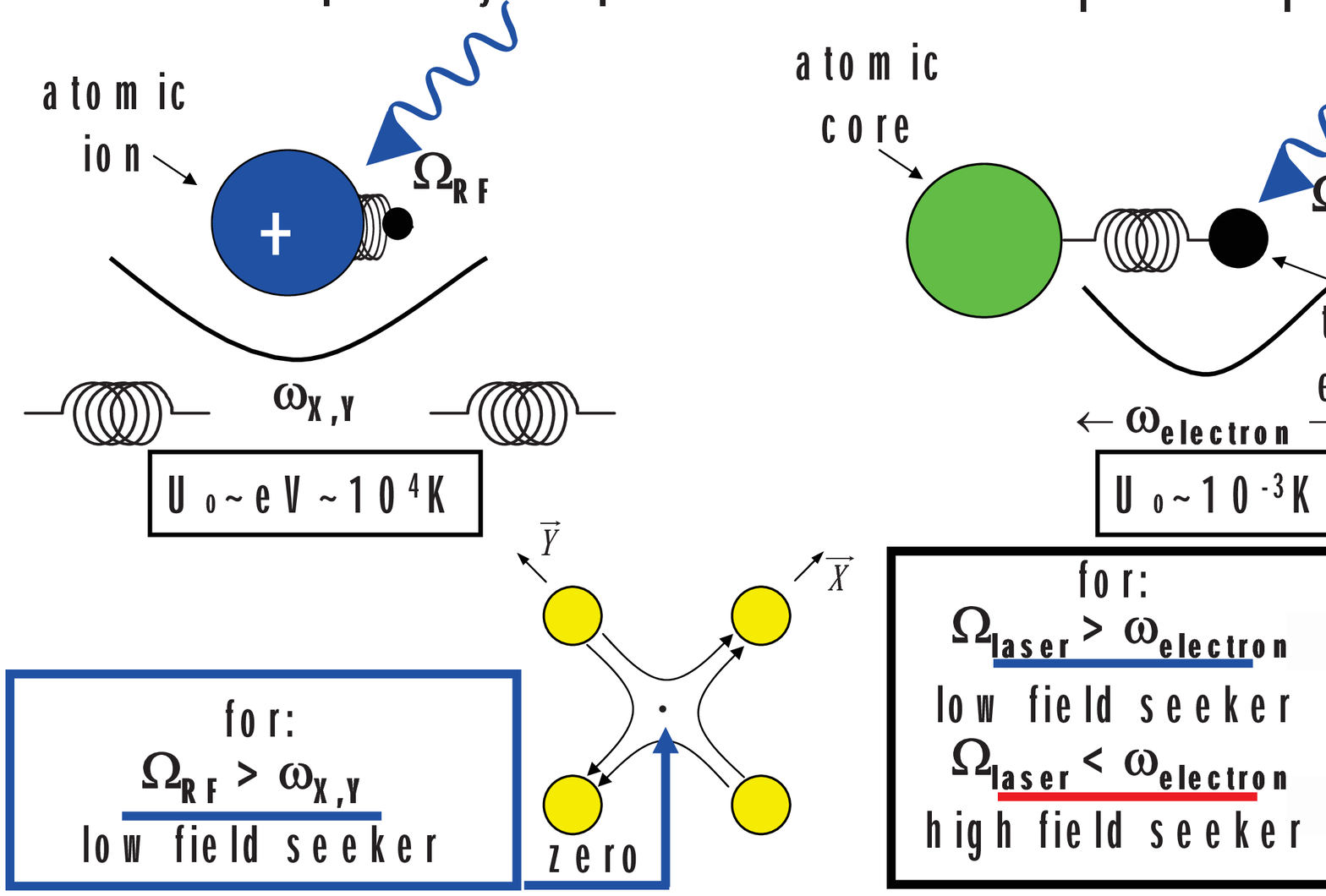}
    \caption{Two concepts for trapping charged particles.
      Both concepts require electro-magnetic fields (blue sinusoidal arrow).
      (a) In RF traps, a RF field at frequency $\Omega_\text{RF}$
      applied to quadrupole electrodes (yellow circles)
      interacts with a charged atom directly.
      The time averaged confining pseudopotential allows the ion to
      oscillate at frequencies $\omega_{X/Y}$
      approximately an order of magnitude smaller than $\Omega_\text{RF}$.
      Since $\Omega_\text{RF} > \omega_{X/Y}$, the trapping field can be
      understood as blue-detuned with respect to the ``resonance'' frequency
      $\omega_{X/Y}$, consequently the ion will seek for the field minimum in
      the centre of the quadrupole (field lines indicated by black arrows).
      Typical depths of the pseudopotential are of the order of
      $\kB \times \unit[10^4]{K}$.
      (b) In optical traps, the optical field is typically applied via laser
      beams that provide an intensity dependent AC Stark shift of the
      electronic levels of the atom or ion.
      The frequency $\Omega_\text{laser}$ of the laser can be detuned blue
      (red) with respect to the relevant electronic resonance frequency
      $\omega_\text{electron}$ and therefore enforces the atom/ion to seek
      for low (high) fields.
      Typical depths of the pseudopotential are of the order of $\kB \times
      \unit[10^{-3}]{K}$.}
    \figlabel{optical:rf:trapping}
  \end{figure}
}

\newcommand{\myfigurerftrap}{
  \begin{figure}[!htb]
    \centering
    \includegraphics[width=\mypicturescale\hsize,keepaspectratio]%
      {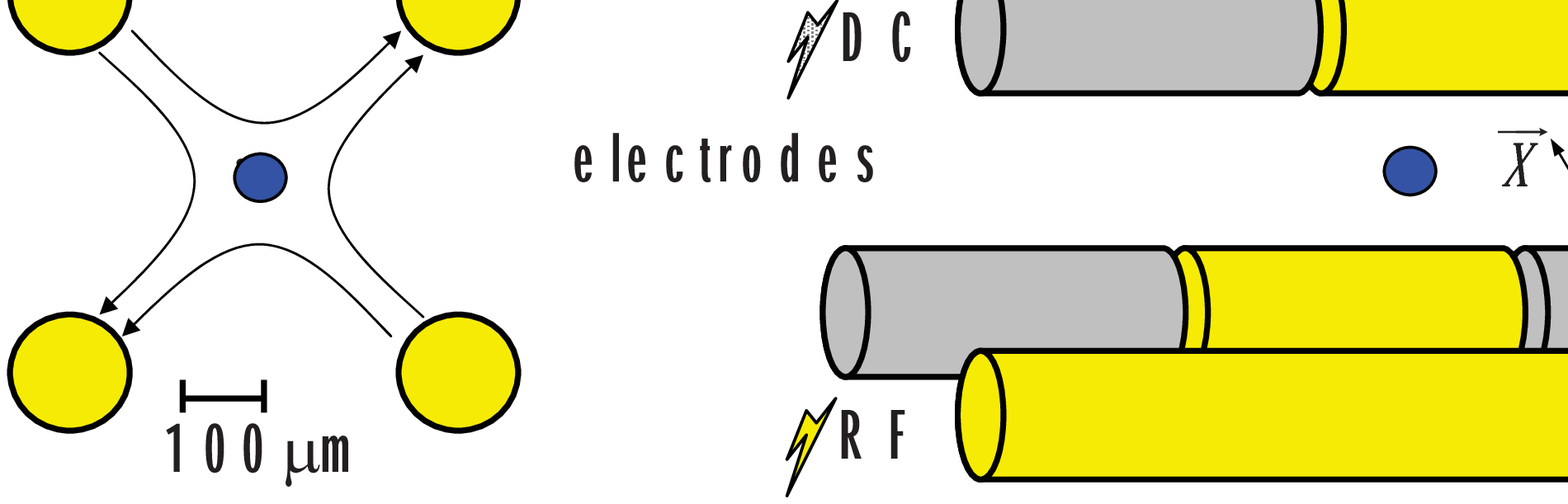}
    \caption{Schematic of the three-dimensional electrode geometry of a
      linear RF trap.
      (a) Cross section through the central quadrupole electrodes (yellow)
      providing the radial confinement for the ion (blue disk).
      (b) Side view, where segments (grey) are used to apply DC voltages
      providing a static potential well along the $Z$-axis.
      Combined with the radial ($X$, $Y$) pseudopotential due to the RF field,
      a three-dimensional confinement is achieved.
      The ion is stored in ultra-high vacuum and is well protected against
      disturbances from the environment.
      However, the fairly open geometry allows to access the external (motional)
      and internal (electronic) degrees of freedom, for example, with focused
      laser beams.}
    \figlabel{rf:trap}
  \end{figure}
}

\newcommand{\myfigurecoulombcrystals}{
  \begin{figure}[!htb]
    \centering
    \includegraphics[width=\mypicturescale\hsize,keepaspectratio]%
      {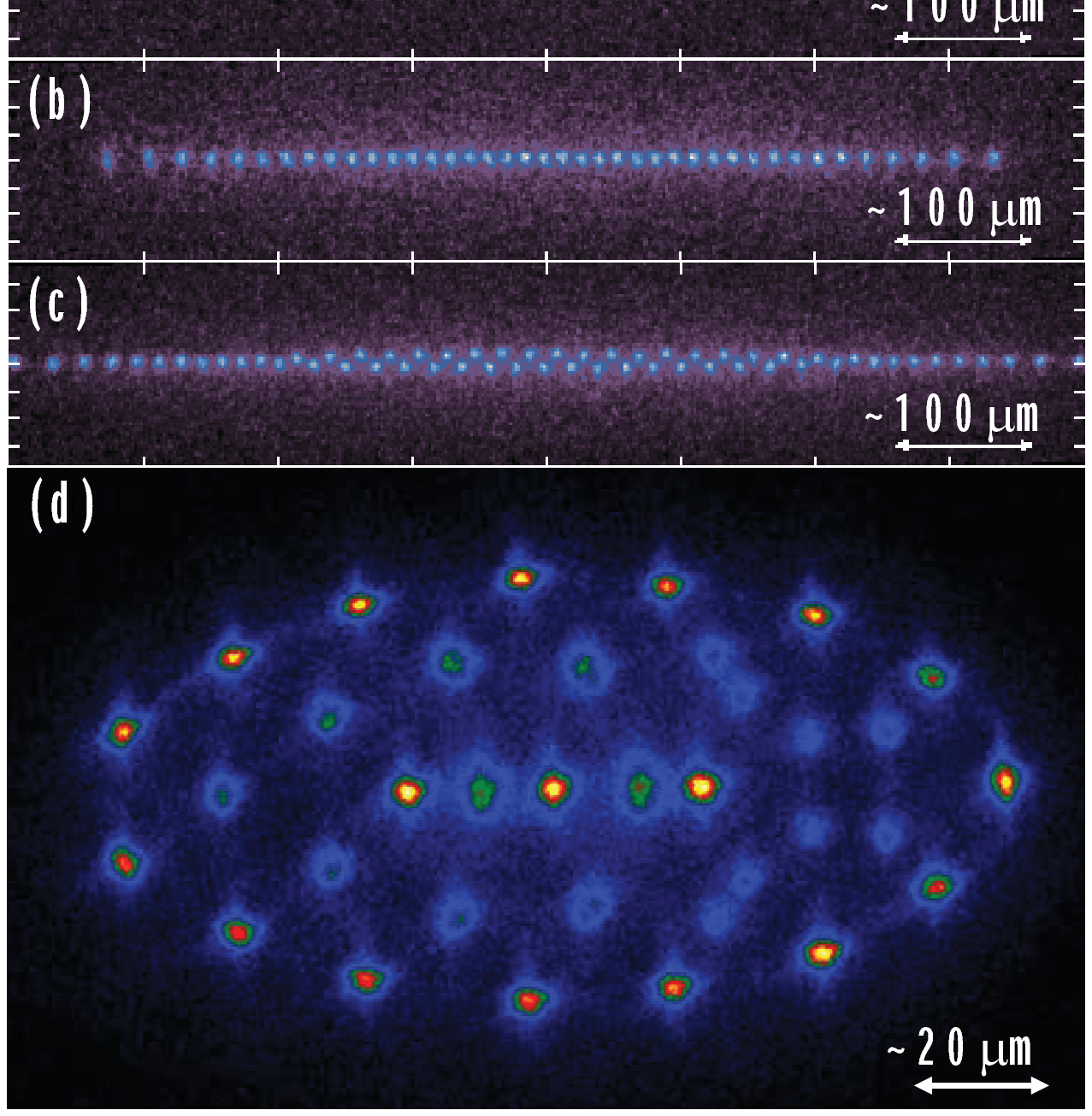}
    \caption{Fluorescence images of laser-cooled ions in a common
      confining potential of a linear RF trap (see \figref{rf:trap}),
      forming differently structured Coulomb crystals.
      (a) A single ion ($\atom{Mg}[+]$).
      (b) A linear string of $40$ ions at $\omega_{X/Y} \gg \omega_{Z}$.
      The axis of the chain coincides with the trap $Z$-axis, which is
      identically orientated in the rest of the images.
      (c) A linear string embedding a two-dimensional zigzag structure of $60$
      ions for $\omega_{X/Y} > \omega_{Z}$.
      (d) A three-dimensional structure of more than $40$ ions at
      $\omega_{X/Y} \gtrsim \omega_{Z}$.
      The enhanced signal-to-noise ratio in (d) is achieved by extended
      exposure.
      Structural phase transitions can be induced between one-, two- and
      three-dimensional crystals, for example by reducing the ratio of
      radial to axial trapping frequencies.}
    \figlabel{coulomb:crystals}
  \end{figure}
}

\newcommand{\myfigureqm}{
  \begin{figure}[!htb]
    \centering
    \includegraphics[width=\mypicturescale\hsize,keepaspectratio]%
      {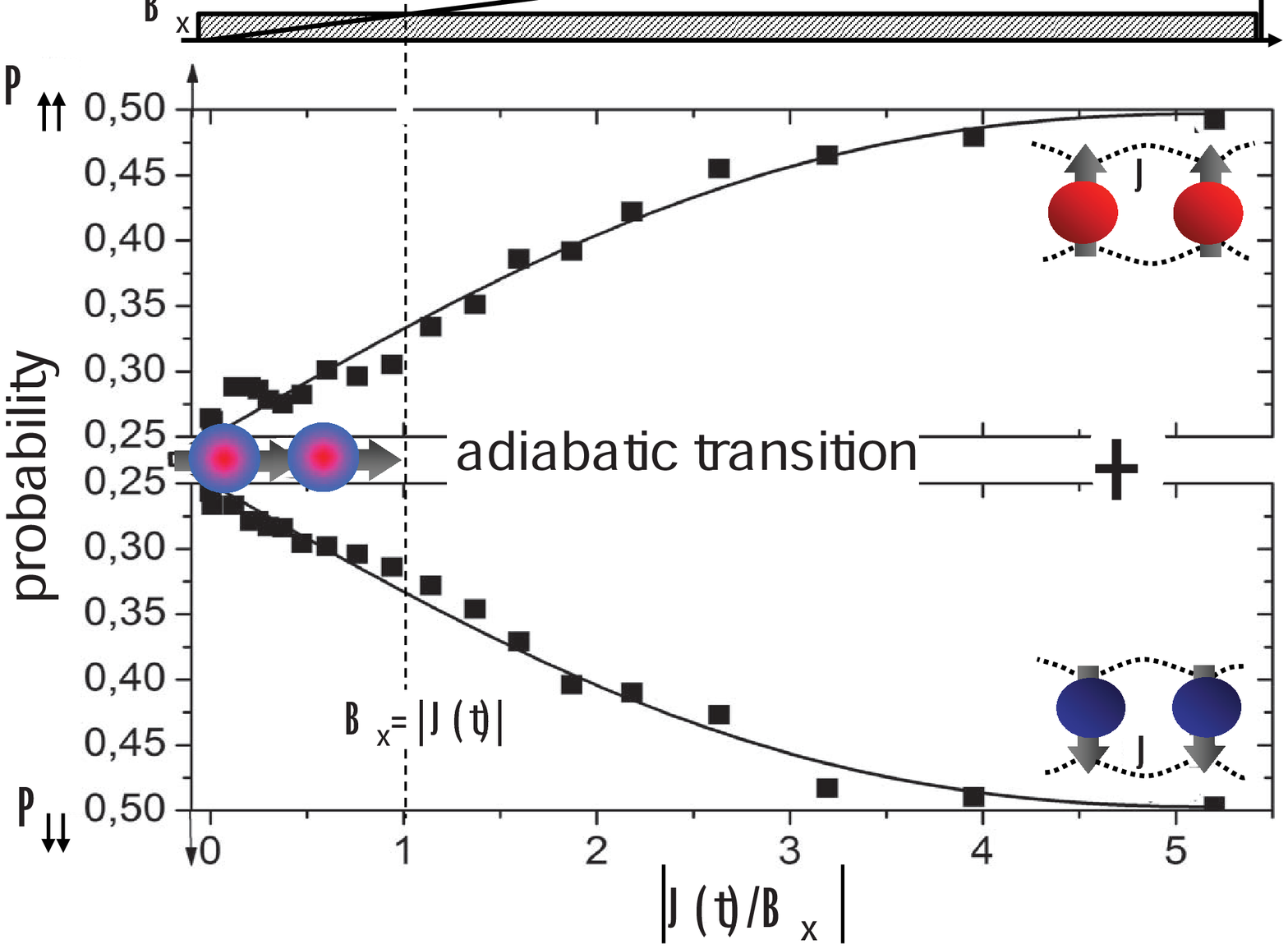}
    \caption{Probability to find two spins in either of the states
      $\ket{\down\down}$ or $\ket{\up\up}$ after an adiabatic QS of the quantum
      Ising Hamiltonian in dependence of $|J / B_x|$, starting within
      paramagnetic order.
      The experimental protocol (top) consists of the interactions
      applied simultaneously including an adiabatic increase of $|J|$ to
      transfer the system from the former ground state $\ket{\rgt\rgt}$ to
      the new one (bottom).
      We achieve a maximal probability of $P_{\up\up} = P_{\down\down} =
      \unit[(49 \pm 1)]{\%}$ to observe one of the states $\ket{\down\down}$
      and $\ket{\up\up}$ corresponding to a ferromagnetic order and define
      the quantum magnetization to be equal to $P_{\up\up}+P_{\down\down} =
      \unit[(98 \pm 2)]{\%}$.
      We derive the fidelity for the entangled state $1 / \sqrt{2}
      (\ket{\down\down} + \ket{\up\up})$ to approximately $F = 0.88$ by a
      parity measurement
      (\cmp{} \refcite{friedenauer:qmagnet} and \secref{phase:gate}).}
    \figlabel{qm}
  \end{figure}
}

\newcommand{\myfiguretwodproposal}{
  \begin{figure}[!htb]
    \centering
    \includegraphics[width=\mypicturescale\hsize,keepaspectratio]%
      {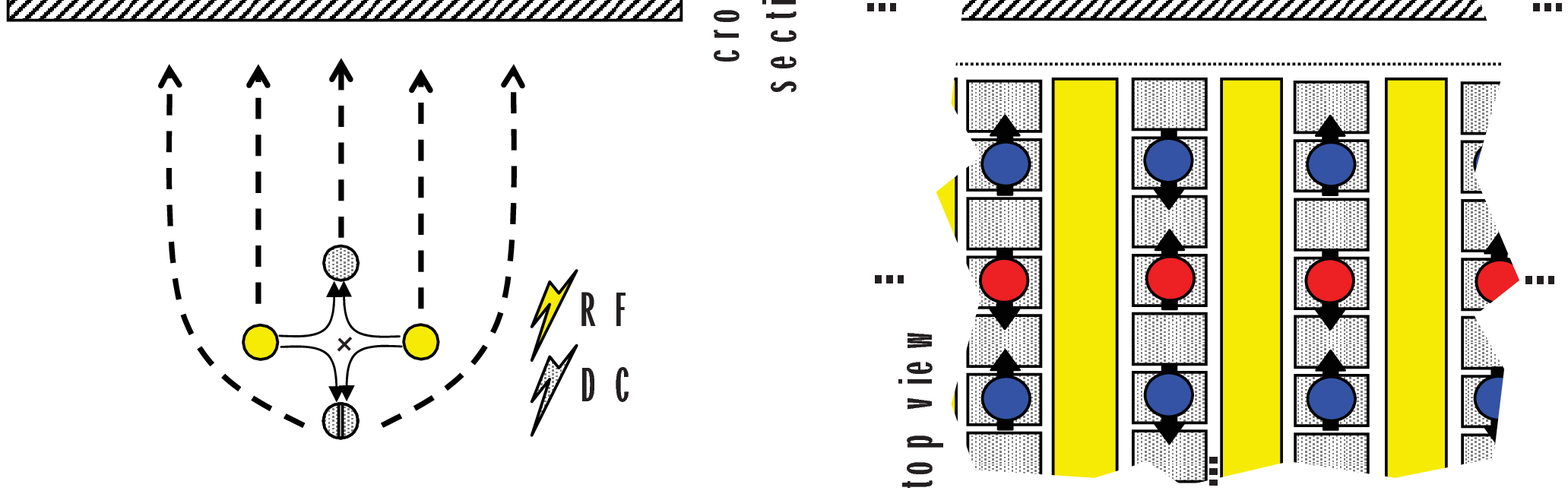}
    \caption{Schematic to illustrate the projection of the electrodes
      of the RF (yellow) and DC (shaded) electrodes on a surface as a way to
      scale towards two-dimensional arrays of ions.
      The black crosses indicate the positions of the minima of the
      pseudopotentials.
      (a) Cross section of the electrodes of a conventional linear RF trap with
      three-dimensional geometry and the electrode structures projected onto a
      surface.
      The dashed arrows point at the new location of the electrodes, the white
      areas represent isolating gaps.
      (b) Cross section (upper part) and top view (lower part) of the stripe
      electrodes.
      It has been proposed to concatenate several of linear RF surface
      electrode traps depicted in (a) as a basic unit to span a two-dimensional
      array of ion \cite{schaetz:qsim} (red and blue disks representing ions in
      opposite spin states).
      For sufficiently small mutual ion distances and decoherence rates of the
      ions, this approach is intended to scale analogue QS.}
    \figlabel{two:d:proposal}
  \end{figure}
}

\newcommand{\myfiguretriangleheight}{
  \begin{figure}[!htb]
    \centering
    \includegraphics[width=\mypicturescale\hsize,keepaspectratio]%
      {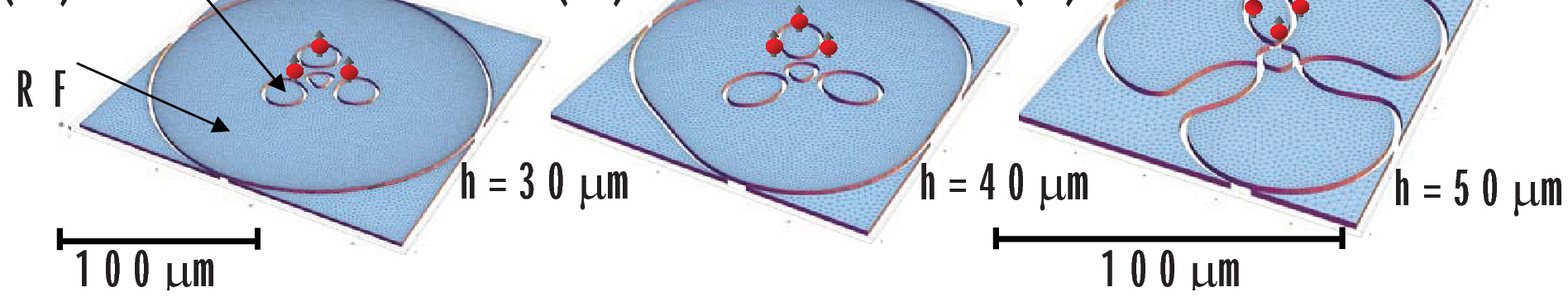}
    \caption{Illustration of the optimization results for the electrode
      structure for a basic triangular lattice in dependence of the height of
      the ions above the traps at constant mutual ion distance.
      The white gap isolates between RF and DC patches.
      The three red disks symbolize three ions at a constant distance of
      $d = \unit[40]{\micro m}$, hovering above the surface in a height of
      (a) $h = \unit[30]{\micro m}$, (b) $h = \unit[40]{\micro m}$, and
      (c) $h = \unit[50]{\micro m}$.
      (courtesy of Roman Schmied)}
    \figlabel{triangle:height}
  \end{figure}
}

\newcommand{\myfiguretriangletilt}{
  \begin{figure}[!htb]
    \centering
    \includegraphics[width=\mypicturescale\hsize,keepaspectratio]%
      {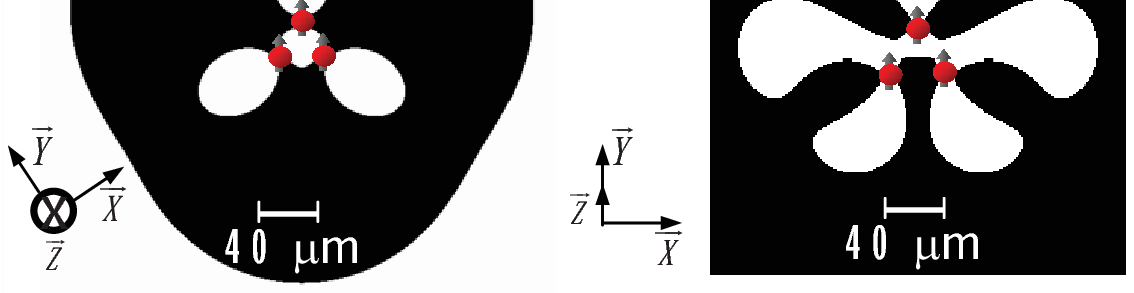}
    \caption{Electrode structures of basic triangular lattices
      with different orientation and tilt of the principal axes.
      Red disks symbolize ions trapped in the potential minima for parameters
      comparable to those in \figref{triangle:height}b.
      (a) One principal axis points in vertical direction with the $X$ and $Y$
      axis lying in the horizontal plane of the electrodes.
      The $X$-principal axis of each trap points towards the centre of the
      triangle.
      (b) The respective principal axes of all traps point in the same
      direction and additionally the $Z$-axis is tilted with respect to the
      surface by more than $\unit[10]{\degree}$, which results in a different
      symmetry of the electrodes.
      The tilt of the $Z$-axis is essential to reach all spatial degrees of
      freedom with laser beams restricted to a plane parallel with respect to
      the electrodes.
      (courtesy of Roman Schmied)}
    \figlabel{triangle:tilt}
  \end{figure}
}

\newcommand{\myfiguretrianglescaled}{
  \begin{figure}[!htb]
    \centering
    \includegraphics[width=\mypicturescale\hsize,keepaspectratio]%
      {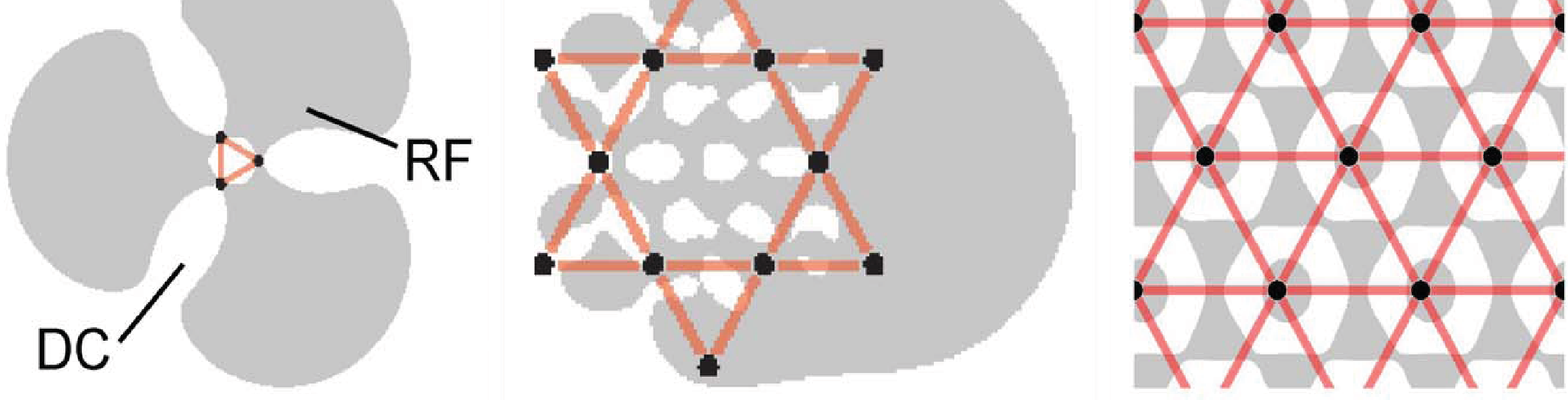}
    \caption{Electrode structures for RF surface electrode traps scaled for
      analogue QSs.
      Black dots symbolize the RF minima, red lines serve as guide to the
      eye to emphasize the lattice structures for (a) three, (b) twelve, and
      (c) an infinite number of ions/spins, the latter both considering the
      parallel orientation and the tilt of the principal axis (see
      \figref{triangle:tilt}). (courtesy of Roman Schmied)}
    \figlabel{triangle:scaled}
  \end{figure}
}

\newcommand{\myfigureperspectives}{
  \begin{figure}[!htb]
    \centering
    \includegraphics[width=\mypicturescale\hsize,keepaspectratio]%
      {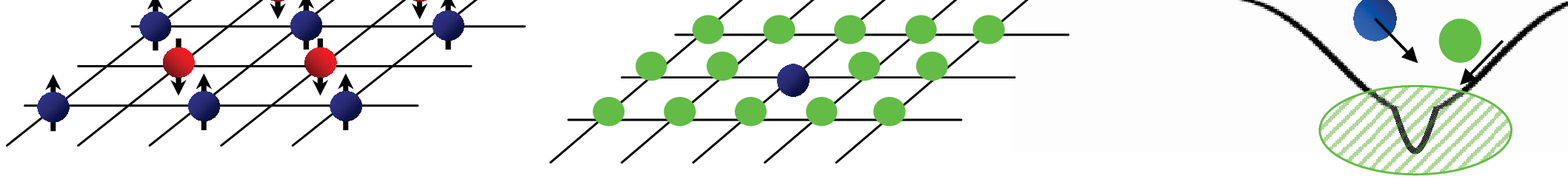}
    \caption{Illustration of new options for analogue QS based on ions (red
      and blue) and atoms (green) in optical potentials (black lines as guide
      to the eye).
      (a) Ions populate an optical lattice on well separated sites.
      The Coulomb force still provides a large strength of dipolar
      (long-ranging) interaction allowing for analogue QS on many-body effects,
      similar to the proposed approach in arrays of RF surface electrode traps
      (see \secref{surface:traps}).
      (b) An ion and atoms populate a common optical lattice and,
      for example, share the charge via tunnelling electrons.
      (c) An ion could be cooled sympathetically by cold atoms (for example, a
      BEC indicated as green ellipse) \cite{zipkes:ion:bec,schmid:ion:bec}.
      Since the micromotion of the ion and the related differential motion
      between atoms and ion becomes negligible in the common optical trap
      \cite{cormick:optical:trapping}, deep equilibrium temperatures are
      predicted to be achievable, down to a regime, where ultra-cold chemistry
      might dominate the collisions.}
    \figlabel{perspectives}
  \end{figure}
}

\newcommand{\myfigureferroantiferro}{
  \begin{figure}[!htb]
    \centering
    \includecombinedgraphics{figure10}
    \caption{Ion chains superimposed by standing waves providing
      state-dependent forces in (a) axial direction and (b) radial direction.
      All ions are placed at the same phases of the standing waves.
      (a) If all spins are in the same state, the ions will all
      be shifted in the same direction without changing their mutual Coulomb
      energy.
      However, if every second spin is in the opposite spin state, distances
      between neighbouring ions will be alternately increased and decreased
      and the mutual Coulomb energy is increased due to its $1 / d$ dependence.
      Here $d$ denotes the distance between neighbouring spins.
      The ferromagnetic order is energetically preferred, such that $J < 0$ for
      this interaction.
      (b) A chain of ions with same spin states will again only be displaced and
      the mutual Coulomb energy will not change.
      For alternating spin states the distances between neighbouring ions will
      increase, such that the mutual Coulomb energy will be decreased.
      (Note that this should not to be confused with the structural zigzag
      phase-transition (see \figref{coulomb:crystals}), where the the radial
      displacements are typically orders of magnitude larger and
      spin-independent.)
      The anti-ferromagnetic order is energetically preferred, such that $J > 0$.}
    \figlabel{sw:ferro:antiferro}
  \end{figure}
}

\newcommand{\myfigureeffbj}{
  \begin{figure}[!htb]
    \centering
    \includecombinedgraphics{figure5}
    \caption{Implementations of different interaction types for
      hyperfine/Zeeman qubits.
      a) An operation of type (a) can be implemented,
      for example, by two-photon stimulated-Raman transitions driven by a pair
      of laser beams (shown without motional dependence) or directly by a
      microwave field.
      These types of interactions can be used for single-qubit gates in QC and
      to simulate the effective magnetic field in the simulation of quantum
      spin Hamiltonians.
      b) State-dependent forces (see type (c) in the text) can be created by
      two beams detuned by approximately the frequency of a motional mode.
      This interaction is used in the geometric phase gate for the displacement
      pulse \cite{leibfried:gate, schmitz:arch} or in the simulation of the
      quantum Ising Hamiltonian to create the effective spin--spin interaction
      \cite{porras:eff-spin,friedenauer:qmagnet}.
    }
    \figlabel{eff:b:j}
  \end{figure}
}

\newcommand{\myfigurelevelscheme}{
  \begin{figure}[!htb]
    \centering
    \includecombinedgraphics{figure4}%
    \caption{Excerpt of the level scheme of $\atom[25]{Mg}[+]$ as an example
      of a hyperfine qubit (not to scale).
      $\atom[25]{Mg}[+]$ has a nuclear spin of $I = 5 / 2$ and thus a
      hyperfine-split ground state ($\level{S}[1/2], F = 3$ and
      $\level{S}[1/2], F = 2$).
      By applying a static magnetic field of few Gauss, the degeneracy of the
      Zeeman sublevels is lifted.
      The Doppler cooling laser (labeled ``BD'') is $\sigma^+$ polarized and
      detuned red by $\Gamma / 2 \approx 2 \pi \times \unit[20]{MHz}$ from the
      cycling transition $\level{S}[1/2], F = 3, M_F = 3 \leftrightarrow
      \level{P}[3/2], F = 4, M_F = 4$.
      Here, $\Gamma$ denotes the linewidth of the $\level{P}$ levels.
      The level $\ket{\down} \defeq \ket{\level{S}[1/2], F = 3, M_F = 3}$ and
      the level $\ket{\up} \defeq \ket{\level{S}[1/2], F = 2, M_F = 2}$ are
      chosen as qubit states or (simulated) spin states, respectively.
      The ion is optically pumped into $\ket{\down}$ during cooling.
      The electronic state is read out by a variant of ``BD'', which is resonant
      on the cycling transition.
      Hence, an ion in the state $\ket{\down}$ will fluoresce, while an ion
      in state $\ket{\up}$ is off-resonant by almost $50 \Gamma$ and will remain
      dark.
      The motional states of one of the motional modes are indicated as
      ``ladders'' on top of the electronic states.
      Two laser beams (labelled ``Raman'') detuned by $\Delta$ from the
      $\level{P}[3/2]$ level can be used to drive two-photon
      stimulated-Raman transitions between $\ket{\down}$ and $\ket{\up}$.
      A flop on the first red sideband is exemplary indicated by the arrows
      from $\ket{\down}\ket{2} \to \ket{\up}\ket{1}$.
    }
    \figlabel{level:scheme:mg25}
  \end{figure}
}

\newcommand{\myfiguredidiaxrad}{
  \begin{figure}[!htb]
    \centering
    \includecombinedgraphics{figure6}%
    \caption{Comparison between parameters of geometric phase gates
      \cite{leibfried:gate} with two ions using the axial motional modes and
      radial motional modes.
      (a) The parameters correspond to the gate from \refcite{schmitz:arch}.
      The axial centre-of-mass (COM) and stretch (STR) mode have a large
      frequency difference ($2 \pi \times \unit[1.6]{MHz}$).
      The detuning of the Raman beams from the STR mode amounts to
      $\mode[\text{STR}]{\delta} = - 2 \pi \times \unit[266]{MHz}$.
      That is why the main contribution to the differential geometric phase
      between $\ket{\down\down}$/$\ket{\up\up}$ and
      $\ket{\down\up}$/$\ket{\up\down}$ is due to a (single) loop in the phase
      space of the STR mode.
      However, as already suggested in \refcite{leibfried:gate}, the detuning
      from the COM mode is chosen to be an integer multiple of the detuning
      from the STR mode ($\mode[\text{COM}]{\delta} = - 5 \times
      \mode[\text{STR}]{\delta}$).
      Hence, there is no entanglement left between the electronic and motional
      modes at the gate duration $T_\text{g} =
      |2 \pi / \mode[\text{STR}]{\delta}| = \unit[3.75]{\micro s}$.
      (Note that the spin-echo sequence is not included in $T_\text{g}$.)
      (b) The parameters correspond to a phase gate on two of the radial
      motional modes.
      The radial centre-of-mass (COM) and rocking (ROC) mode have a
      comparatively small frequency difference of only
      $2 \pi \times \unit[130]{kHz}$.
      The detunings from both modes are chosen to have same absolute values
      resulting in (approximately) equal contributions to the acquired geometric
      phase from both modes.
      The gate duration according to the original implementation would amount
      to $T_\text{g} = |2 \pi / \mode[\text{STR}]{\delta}| =
      \unit[15.4]{\micro s}$.
      As the displacement pulse is repeated in the second gap of the spin-echo
      sequence (\cmp{} \figref{didi:pulsescheme}) to cancel dynamic phases
      (\cmp{} \refcite{home:gate} and see text), the duration increases by an
      additional factor of two.
    }
    \figlabel{didi:ax:vs:rad}
  \end{figure}
}

\newcommand{\myfiguredidipulse}{
  \begin{figure}[!htb]
    \centering
    \includecombinedgraphics{figure7}%
    \caption{Pulse scheme of the geometric phase gate.
      It consists of a spin-echo sequence ($\R(\pi / 2, \pi / 2)$,
      $\R(\pi, \pi / 2)$, $\R(\pi / 2, \pi / 2)$ pulses) with a displacement
      pulse (labeled ``$\op{D}_1$'') in the first gap of the spin-echo sequence
      for the original implementation of the phase gate \cite{leibfried:gate}.
      For the gate on the radial modes of motion a second displacement pulse
      (labelled ``$\op{D}_2$'') is introduced to cancel dynamic phases from
      $\op{D}_1$ due to different absolute values of the forces on
      $\ket{\up}$ and $\ket{\down}$ (\cmp{} \refcite{home:gate}).
      For all gates the duration of each displacement pulses is chosen to be
      $T_D = |2 \pi / \mode[\text{STR/ROC}]{\delta}|$ such that
      each displacement pulse leads to a closed loop in each phase space.
      Hence, the total gate duration amounts to $T_\text{g} = T_D$ for the
      original implementation and $T_\text{g} = 2 T_D$ for the gate on the
      radial modes of motion.
      (Note that the spin-echo sequence is not included in $T_\text{g}$.)
      The dashed $\R(\pi / 2, \pi / 2 + \phi)$ analysis pulse is added for the
      measurement of the gate fidelity (see \figref{didi:parity}).
    }
    \figlabel{didi:pulsescheme}
  \end{figure}
}

\newcommand{\myfiguredidiparity}{
  \begin{figure}[!htb]
    \centering
    \includecombinedgraphics[vecscale=\myfigurescale]%
      {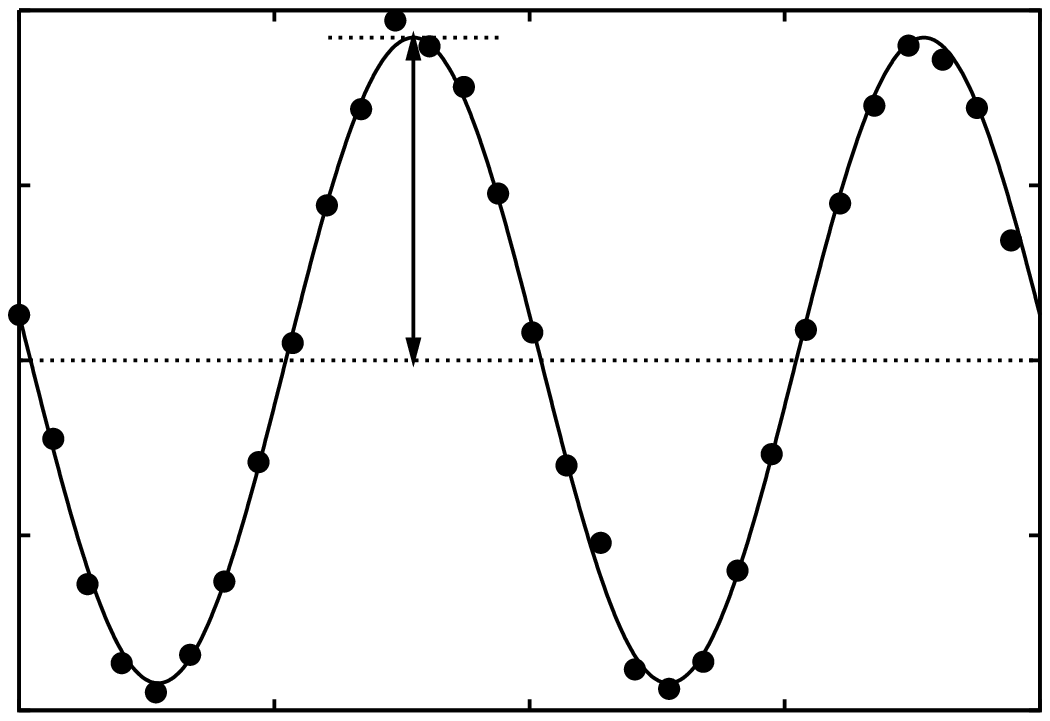}%
    \caption{Parity measurement after the geometric phase gate on two radial
      modes of motion with two ions.
      The parity is defined as $P \defeq P_{\down \down} + P_{\up \up}
      - (P_{\up \down} + P_{\down \up})$, where $P_{s, s'}$ denotes the
      population of the electronic state $\ket{s, s'}$ with
      $s, s' \in \set{\down, \up}$.
      It is measured as a function of the phase $\phi$ of the analysis pulse
      $\R(\pi / 2, \pi / 2 + \phi)$ (\cmp{} \figref{didi:pulsescheme}).
      Each data point represents the mean of $2500$ measurements.
      The contrast $C = \unit[92.2]{\%}$ is determined from the fitted curve.
      Considering the populations $P_{\down \down} + P_{\up \up} >
      \unit[98]{\%}$ for the entangled state we obtain a Bell state fidelity
      $F > \unit[95]{\%}$.
    }
    \figlabel{didi:parity}
  \end{figure}
}

\newcommand{\myfiguredidievol}{
  \begin{figure}[!htb]
    \centering
    \includecombinedgraphics[vecscale=\myfigurescale]%
      {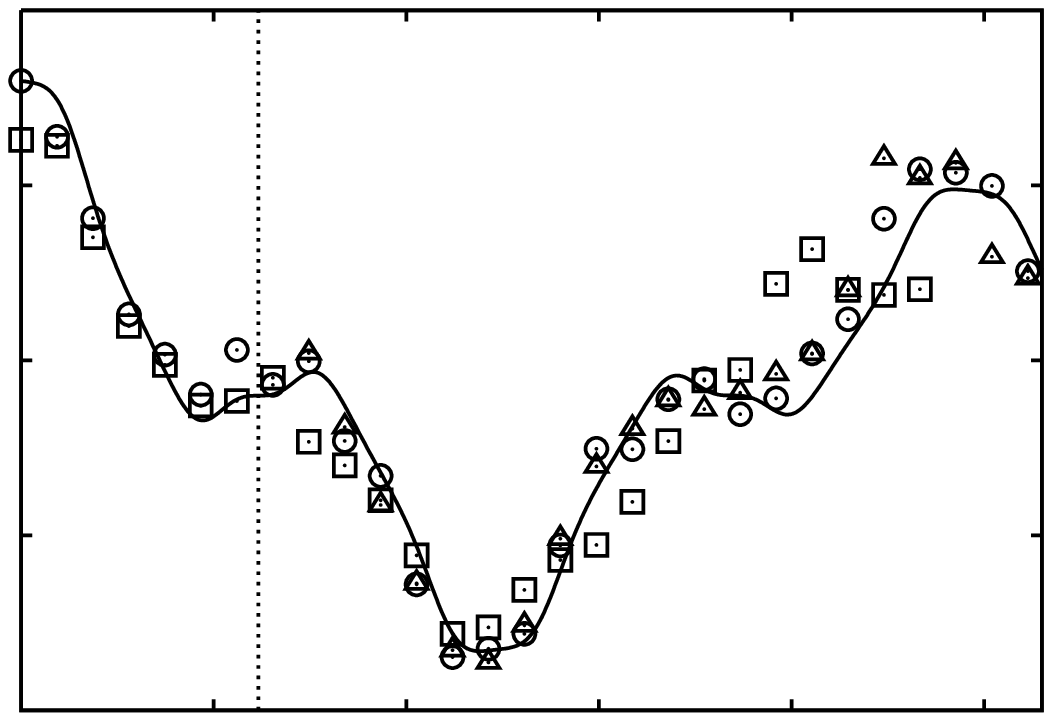}%
    \caption{Total fluorescence from the two ions as a function of the
      total displacement duration $2 T_D$ (\cmp{} \figref{didi:pulsescheme}).
      The detected fluorescence signal from both ions amounts to approximately
      $\unit[7]{counts} / \unit[20]{\micro s}$ for state $\ket{\down\down}$ and
      close to zero for $\ket{\up\up}$.
      The duration between the displacement pulses is chosen to be
      $T_\text{w} = |2 \pi / \mode[\text{COM/ROC}]{\delta}|$ in the experiment
      (\cmp{} \figref{didi:pulsescheme}).
      At $T_\text{g} \approx \unit[30.8]{\micro s}$ the state
      $\ket{\psi} \approx \left(\ket{\up\up}
      + \ii \ket{\down\down}\right) \ket{\mode[\text{COM}]{n} = 0,
      \mode[\text{ROC}]{n} = 0}$ is prepared.
      Each data point represents the average of $400$ measurements (squares
      and triangles) and $200$ measurements (circles), respectively.
      The statistical errors are on the order of the size of the symbols.
      The curve is based on a fit of the time evolution of
      \eqref{disp:geo:phase} with an additional empirical exponential decay to
      mimic decoherence effects.
      The only fit parameters are the fluorescence for $\ket{\down\down}$
      amounting
      to $\unit[7.2]{counts} / \unit[20]{\micro s}$ and the decay constant
      $\tau \approx \unit[290]{\micro s}$.
      The gate serves as experimental reference for the isolated interaction
      strength and is not optimized to provide the highest gate fidelity.
    }
    \figlabel{didi:evol}
  \end{figure}
}

\ultrashortmacrosoff 

\begin{document}

\iftwocolumn{}{\gdef\myfigurescale{\myfigurescalepre}}
\iftwocolumn{}{\gdef\mypicturescale{\mypicturescalepre}}

\pagenumbering{arabic}

\title{Many-Body Physics with Trapped Ions}

\author{Christian Schneider}
\affiliation{%
  Max-Planck-Institut f\"{u}r Quantenoptik,
  Hans-Kopfermann-Stra\ss{}e 1,
  85748 Garching,
  Germany%
}
\author{Diego Porras}
\affiliation{%
  Departamento de F\'{i}sica Te\'{o}rica I,
  Universidad Complutense,
  28040 Madrid,
  Spain%
}
\author{Tobias Schaetz}
\email{tobias.schaetz@mpq.mpg.de}
\affiliation{%
  Max-Planck-Institut f\"{u}r Quantenoptik,
  Hans-Kopfermann-Stra\ss{}e 1,
  85748 Garching,
  Germany%
}

\begin{abstract}

Direct experimental access to some of the most intriguing
quantum phenomena is not granted due to the lack of precise control of the
relevant parameters in their naturally intricate environment.
Their simulation on conventional computers is impossible, since quantum
behaviour arising with superposition states or entanglement is not efficiently
translatable into the classical language.
However, one could gain deeper insight into complex quantum dynamics by
experimentally simulating the quantum behaviour of interest in another quantum
system, where the relevant parameters and interactions can be controlled and
robust effects detected sufficiently well.
We report on the progress in experimentally simulating quantum
many-body physics with trapped ions.

\end{abstract}

\maketitle

\tableofcontents

\clearpage


\section{Introduction}
\seclabel{intro}

Simulations and deeper understanding of the dynamics of some tens of
interacting spins are already intractable with the most powerful
classical computers.
For instance, the generic state of $50$ spin-$1/2$ particles is defined by
$2^{50}$ numbers and to describe its evolution a $2^{50} \times  2^{50}$
matrix has to be exponentiated \cite{lloyd:qsim}.
Recently, one of the ten most powerful supercomputers, JUGENE in
J\"{u}lich, was exploring this regime.
The current record was set by simulating a system of 42 quantum bits (qubits),
equivalent to $42$ spin-$1/2$ particles
\cite{schinarakis:jugene,deraedt:quantum:computer}.

In any case, it will not help to increase its impressive classical
calculation capabilities to simulate only slightly larger quantum
systems.
Each doubling of the computational power will just allow to add one spin/qubit
to the system (approximately after two years, according to Moore's law
\cite{moore:law}).
Furthermore, simply pursuing this path of exponential growth in the computer's
classical capabilities would require an exponential shrinking of its electronic
components\footnote{%
The electronic components are arranged in two dimensions and, since recently,
the third dimension is exploited.
However, sufficient cooling has to be provided.}.
The structure size currently amounts to approximately $\unit[30]{nm}$, a
distance spanned by roughly $100$ atoms.
If gifted engineers further miniaturize the sizes of their structures,
``currents'' of a few electrons will ``flow'' on ``wires'' spanned by a few
atoms only.
As a consequence, quantum effects will have to be considered for future
classical computers, leading to serious consequences.
Electrons charging a capacitor, for example, currently realize a storage of
logical information: a charged capacitor represents a ``one'', a
discharged capacitor a ``zero''.
What, if the few electrons, classically well caught within the potential of
the capacitor, follow their natural quantum mechanical paths and simply escape
through the walls by tunnelling?

However, allowing for quantum effects in a controlled way might also
be exploited as a feature.
Richard Feynman originally proposed \cite{feynman:qsim} to use a well
controlled quantum system to efficiently track problems that are very hard to
address on classical computers and named the device a ``quantum computer''
(QC).
Nowadays his proposal can be seen closer to the description of a
quantum simulator (QS) \footnote{Depending on the context, the abbreviations
``QC'' and ``QS'' may also stand for quantum computation and quantum simulation,
respectively.}.
In any case, his idea has been theoretically investigated and
further developed to the concept of a universal QC.
Fulfilling a well defined set of prerequisites, known as deVincenzo's
criteria \cite{divincenzo:twobitgates,divincenzo:criteria}, should make
possible to run any classical and quantum algorithm by a stroboscopic sequence
of operations.
These have to act on single qubits, for example, changing their state, and on
pairs of qubits performing changes on one qubit, conditional on the state of
its mate.

Hundreds of groups worldwide work on many approaches in
different fields of atomic, molecular and solid state systems to
realize their version of the envisioned QC.
For a concise review see, for example, \refcite{ladd:quantum:computers:review}.

However, even assuming an ideal system and perfect operations will
require the control of the order of $10^3$ logical qubits as basis
for translating any algorithm or the quantum dynamics of a complex
system into a sequence of stroboscopic gate operations on a
potential universal QC \cite{nielsen:qinf}.
Residual decoherence will cause computational errors and must be minimized to
allow for high operational fidelities ($\sim \unit[(99.99 \dotso 99.9)]{\%}$)
\cite{knill:quantum:computer:noisy}.
At present, only then, the errors could be overcome by quantum error
correction, at the price of a reasonable but still tremendous overhead of
ancilla qubits, approximately another $100$ per logical qubit.
In total, of the order of $10^5$ qubits are required.
Even though there appear to be no fundamental obstacles for enhancing the
fidelities of the operations and for scaling the size of the systems
\cite{home:methods:set}, there is still challenging technological development
ahead.
The realization of a universal QC is not expected within the next decades.

A shortcut via analogue QS has been taken
into consideration \cite{feynman:qsim} to allow to gain deeper insight into the
dynamics of quantum systems.
``Analogue'' emphasizes that the dynamics of the system are not translated
into an algorithm of gate operations on subsets of qubits.
In contrast, a system of quantum particles is required, where (1)
the initial state and its dynamics can be precisely controlled, (2)
as many relevant parameters as possible manipulated and (3) the
readout of the important characteristics of the final state
performed in an efficient way.
If the system's evolution was governed by a Hamiltonian suspected to account
for the quantum effects of interest, we would be enabled to experimentally
investigate the physics of interest isolated from disturbances,
close to Feynman's original proposal.
The requirements on the number of quantum particles
and fidelities of operations for analogue QS are predicted to be substantially
relaxed compared to QC \cite{buluta:quantum:simulators}.
However, it remains to be investigated which realistic assumption on
different sources of decoherence in the particular system will
lead to a sufficiently small (or even appropriate) impact on the dedicated QS.
It is predicted that QS are less prone to decoherence, \eg{}, in simulating
robust effects like quantum phase transistions (QPT).
Therefore, they do not require any precautions in contrast to QC, which suffer
from the costly overhead due to quantum error correction.
It was even proposed to establish decoherence as an asset \cite{lloyd:qsim}.
In this context, decoherence is not to be seen as a source of errors, like
in the field of universal QC, but as a resource to simulate its
natural counterpart.
For example, decoherence is suspected to be responsible and required
for enhanced efficiencies of (quantum) processes in biological
systems at $T \sim \unit[300]{K}$ \cite{engel:photosynthesis,%
mohseni:photosynthesis}.

To discuss the different requirements for different analogue QS, we
can distinguish between two categories of simulations.
One category deals with problems where QS provide a simulated counterpart that
allows to experimentally address intriguing questions that are not
directly tractable in the laboratory.
Examples are highly relativistic effects like Hawking and Unruh radiation
or the Zitterbewegung of a freely moving particle predicted by Dirac's
equation (see also \refcite{buluta:quantum:simulators}).
The second category of simulations deals with objectives that are
(probably fundamentally) not accessible with classical computation,
for example, the complex quantum dynamics of spins in solid state
systems, as mentioned above.
A promising strategy is to initialize an analogue QS in a state
that can be prepared easily in the system of choice according to
step (1) introduced above.
Evolving the system adiabatically by changing its parameters
according to (2) allows to reach a new state that is hard or
impossible to reach otherwise, for example, via a QPT.
The aim here is not to simulate the effects including all disturbances and
peculiarities, because the analogue QS would than become as complex
as the system to be simulated.
The aim can be to investigate, whether the simplified model still yields
the effects observable in nature and, thereby, to gain a concise deeper
understanding of their relevant ingredients.
However, there remains room for the important discussion, whether the specific
dynamics emulate nature or simulate the implemented model (Hamiltonian) and
whether the results allow to draw further conclusions.

In any case, it has to be emphasized that analogue QS are
intrinsically not universal. That is, different realizations of a QS
will allow to simulate different systems. Even more important,
different approaches for the identical models (Hamiltonians) might
allow to cross-check the validity of the
QSs \cite{leibfried:visions}.

There are several systems proposed to implement analogue QS,
offering different advantages \cite{buluta:quantum:simulators} to address
the physics in many-body systems.
One of them consists of neutral atoms within
optical lattices \cite{bloch:many:body,simon:qs:antiferromagnet}.
Another promising candidate is based on trapped ions
\cite{porras:eff-spin,porras:bosehubbard,wunderlich:condspinres},
originally suggested by I. Cirac and P. Zoller in 1995 \cite{cirac:qcomputer}
in the context of QC.
Trapped ions already compete at the forefront of many fields, were ultimate
accuracy and precision is required, like metrology (see, \eg{},
\refcite{chou:comparison:clocks}).
Trapped ions offer unique operational fidelities, individual
addressability and short- as well as long-range interactions due to
Coulomb forces.

Many models of both categories of QS are promising candidates or already
addressed by trapped ions.
Examples for the first category are emanating from the fields of cosmology
\cite{schuetzhold:universe,horstmann:black:hole,horstmann:black:hole:2},
relativistic dynamics
\cite{lamata:dirac,gerritsma:dirac:eq,gerritsma:klein:paradox},
quantum optics \cite{leibfried:interferometer} including quantum walks as a
potential tool for QSs \cite{travaglione:qrandomwalk,schmitz:qrw,%
zaehringer:walk,mohseni:photosynthesis},
chemistry \cite{cai:chemical:compass}, and
biology \cite{cai:entanglement:molecules,plenio:networks}.
For the second category, quantum spin Hamiltonians \cite{porras:eff-spin},
Bose--Hubbard \cite{porras:bosehubbard} and spin--boson
\cite{porras:spin:boson} models were proposed to describe solid-state
systems and their simulation would allow to observe and investigate a rich
variety of QPTs \cite{sachdev:quantum:phase:trans:a}.
A summary and concise description of theoretical proposals on QS of
both categories based on trapped ions and first experimental results
until the year 2008 can be found in \cite{johanning:review}.

This reports aims to describe the current status of the field of
experimental, analogue QS addressing many-body physics, its
challenges and possible ways to address them.
The first proof-of-principle experiment was achieved \cite{friedenauer:qmagnet}
and extended recently \cite{kim:frustration,islam:qmagnet} on
a few trapped ions in linear radio-frequency (RF) traps.
The main challenge for QS remains to scale up towards $50 \dotso 100$ ions or
even beyond.
A simulated system of this size would already reach far beyond the regime
accessible via classical computation and, even more important, already allow
to address open scientific questions.

The report is organized as follows:
In \secref{tools} we introduce the tools available for QS by briefly
summarizing the types of traps, different ions species, and different technical
implementations of the control of the electronic and motional degrees of
freedom.
In the following \secref{theory} we derive the mathematical description
based on \refmcite{wineland:bible,leibfried:habil,porras:eff-spin}.
We aim at extending the existing formalism to be directly applicable to more
dimensions and individual trapping conditions envisioned in arrays
of ions.
We apply this formalism to a basic building block of QC, a two qubit phase gate
on the radial modes measured in our group, and emphasize similarities and
differences between the application of similar operations for analogue
QS.
This section is supplemented by a detailed appendix.
In \secref{interpretation}, we interpret the interactions in the context of
analogue QS, which should be sufficient for understanding the subsequent
discussion of the experimental implementations without going through the
details of \secref{theory}.
In the following \secref{many:body}, we first depict the proof-of-principle
experiments on a few trapped ions in linear RF traps.
Based on the state of the art capabilities we present in the second
part of this section a summary of proposals to
study many-body physics in a variety of solid state systems.
The two following sections are dedicated to two proposals aiming for
scaling up the systems:
In \secref{surface:traps}, we discuss potential realizations of a
two-dimensional array of RF surface electrode traps.
They are conceptually similar to promising approaches in Penning
traps \cite{wineland:bible,biercuk:hifi:penning,bollinger:priv:comm}.
We will also introduce an alternative approach based on
ions in optical traps in \secref{lattices}, thus, trying to combine
the advantages of trapped ions and optical lattices.
Finally, we conclude in \secref{outlook}.


\section{Tools Required for Experimental Quantum Simulations}
\seclabel{tools}

In this section, we describe the requirements to
implement analogue QS based on trapped ions.
Most of these tools have been developed over the last decades, many for
the purpose of quantum information processing (QIP) with main focus on QC.

\subsection{Scientists}
\seclabel{tools:scientists}

Currently, more than 30 experimental groups worldwide focus on QIP
based on trapped ions.
A rapidly growing percentage of those is extending their objectives into the
field of analogue QS, including (to the best of our knowledge) groups at the
University of California Berkeley, Duke University,
Eidgen\"{o}sische Technische Hochschule (ETH), University of Freiburg,
Gorgia Institute of Technology, Imperial College, University of
Hannover, University of Innsbruck, University of Mainz, University
of Maryland, Massachusetts Institute of Technology (MIT),
Max Planck Institute for Quantum Optics (MPQ), National Institute of Standards
and Technology (NIST), Sandia National Laboratories (SNL), and University of
Siegen.

\subsection{Ion Traps and Coulomb Crystals}
\seclabel{tools:traps}

Isolating and trapping of individual particles as well as the
precise control of their motional (external) degrees of freedom is key for many
high precision measurements.
Several trapping concepts have been developed for and implemented with ions,
like RF traps \cite{paul:lecture}, Penning traps \cite{dehmelt:lecture} and
optical traps \cite{phillips:lecture}.
The physics of these devices, for example, of RF traps and optical dipole
traps, is closely related.
Electro-magnetic multipole fields act on the charge or induce electric dipole
moments.
The resulting forces on the particles lead in time average to a confining
pseudopotential.
The two concepts are compared in \figref{optical:rf:trapping}.

\figurehandler{\myfigureopticalrftrap}

However, there was a delay of more than a decade between trapping
charged atoms in RF fields \cite{paul:massfilter,paul:kaefig} and trapping
neutral particles with optical fields \cite{ashkin:trap:particles}.
One explanation is that RF traps provide potential depths of the order of
several $\unit{eV} \approx \kB \times \unit[10^4]{K}$, while optical traps
typically store particles up to $\kB \times \unit[10^{-3}]{K}$ only.
This discrepancy is mainly due to the comparatively large Coulomb force that
RF fields can exert on charges.
The RF field at typical frequencies $\Omega_\text{RF} / (2 \pi) =
\unit[(10\dotso100)]{MHz}$ directly acts on the massive ion.
The related motional frequencies within the deep pseudopotential amount to a
few $\unit{MHz}$.
Optical fields, in contrast, oscillate more than six orders of magnitude
faster: too fast for the massive atomic core to follow.
In a simplified picture, the optical field has to induce a dipole
moment of the electron and the atomic core first to allow for a
subsequent interaction of the dipole with the optical field.
Therefore, the resulting optical pseudopotential for neutral atoms and
charged ions remains close to identical \cite{cormick:optical:trapping}.

\figurehandler{\myfigurerftrap}

Here we focus first on ions in linear RF traps.
The concept for the radial confinement is depicted in
\figsref{optical:rf:trapping}{rf:trap}.
The RF field applied to two opposing electrodes of the quadrupole can
provide a radially confining pseudopotential.
Similar to a quadrupole mass filter, one can find voltages for given parameters
(electrode geometry and mass/charge ratio of the ion species) that
allow for stable confinement in two dimensions.
Additional DC voltages add a static harmonic potential to complete the
three-dimensional confinement that can be assumed to be harmonic.
Dependent on the application, these DC voltages can be applied to electrodes
realized as rings or needles along the axis or by a segmentation of the
quadrupole electrodes (see \figref{rf:trap}b).
A confined ion will oscillate with frequency $\omega_Z / (2 \pi)$ along the
the trap axis and with frequencies $\omega_{X/Y} / (2 \pi)$ in the radial
directions.
The radial oscillation is superimposed by a fast oscillation at frequency
$\Omega_\text{RF} / (2 \pi)$ (so-called micromotion), which increases with
increasing distance of the ion from the trap centre, such that the RF field
does not vanish anymore.

Typical parameters for conventional setups are a minimal ion--electrode
distance $h \sim \unit[(100 \dotso 1000)]{\micro m}$ allowing for RF voltages
of the order of $\unit[1000]{V}$.

\figurehandler{\myfigurecoulombcrystals}

Different laser cooling schemes can be applied to reduce the total energy of
motion of the ion \cite{leibfried:habil}.
Doppler cooling \cite{wineland:doppler,haensch:cooling,wineland:doppler:exp,%
neuhauser:cooling} of several ions already allows to enter a
regime, where the kinetic energy ($\kB T \sim \unit{mK}$) of the ions becomes
significantly smaller than the energy related to the mutual Coulomb repulsion.
Hence, the ions cannot exchange their position anymore.
A phase transition from the gaseous (liquid) plasma to a crystalline
structure occurs \cite{waki:ordered:structures,birkl:coulomb:crystal}.
On the one hand, the resulting Coulomb crystals (see \figref{coulomb:crystals})
provide many similarities with solid state crystals already partially
explaining why Coulomb crystals appear naturally suited to simulate many-body
physics:
(1) The ions reside on individual lattice sites.
(2) The motion of the ions (external degree of freedom) can be described
easiest in terms of common motional modes with the related quanta
being phonons.
The phonons in Coulomb crystals allow to mediate long-range
interactions between the ions.
In a different context, the phonons can also be interpreted as bosonic
particles, for example, capable of tunnelling between lattice sites simulated by
the ions (see also \secref{many:body:theo}).
On the other hand, there are advantageous differences compared to solid state
crystals:
(3) Coulomb crystals typically build up in ultra-high vacuum
($\unit[(10^{-9} \dotso 10^{-11})]{mbar}$) and are very well
shielded against disturbances from the environment, thus providing long
coherence times.
(4) Coulomb crystals feature lattice constants of a few
micrometres (see \figref{coulomb:crystals}), dependent on the trapping
potential counteracting the mutual Coulomb repulsion.
Compared to a solid, where distances are of the order of \AA{}ngstrom
($\unit[10^{-10}]{m}$) the density of the structure in one dimension is reduced
by five, in three dimensions by fifteen orders of magnitude.
This allows for individual addressing of the ions and for individual
preparation, control and readout of their electronic and motional states.
(5) The Coulomb interaction between the charged ions is not shielded
within the crystal, as in Coulomb crystals the charge of all ions has the same
sign in contrast to ionic crystals in solid state systems.

It has to be pointed out that it is possible to deterministically achieve
phase transitions between different structures of Coulomb crystals for
large numbers of ions \cite{waki:ordered:structures,birkl:coulomb:crystal}.
When the ratio of radial to axial confinement is reduced or the amount of
confined ions is increased, we observe the transition from a linear chain of
ions via a two-dimensional zigzag structure to a three-dimensional structure
(see \figref{coulomb:crystals}b--d).

Despite the unique conditions in Coulomb crystals in linear RF traps
and the high fidelities of operations, current experimental
approaches on QS (and QC) are still limited to a small number of
ions.
The approaches include of the order of ten ions arranged in a linear chain
\cite{islam:qmagnet,monz:14:qubit}.
This is accomplished by choosing the radial confinement much stronger than the
axial one.
The linear chain orientates along the weakest ($Z$) direction, where tiny
oscillations of the cooled ions around the minimum of the pseudopotential
($X$ and $Y$) and thus micromotion still remain negligible.

For purposes of a QC and QS, scaling to a larger number of spins and more
dimensions while keeping sufficient control over all required degrees of
freedom remains the challenge of the field.
Using longer linear chains confined in anharmonic axial potentials
\cite{lin:anharmonic} might provide a way to reach a number of ions in
the system that in principle already exceeds capabilities of a classical
supercomputer.
Another way might be the use of RF ring traps offering periodic boundary
conditions for static Coulomb crystals
\cite{waki:ordered:structures,birkl:coulomb:crystal} and
even (more-dimensional) crystalline beams of ions
\cite{schaetz:pallas,schramm:bunched:beams,schramm:beams:3d}.
A microfabricated ring trap is currently developed and fabricated at
Sandia National Laboratories \cite{moehring:priv:comm}.

The two main limitations for further scaling of the number of ions
in a common potential are, from a practical point of view:
(1) the emergence of $3 N$ normal modes for $N$ ions plus their sum and
difference frequencies that lead to an increasingly crowded phonon
spectrum (already for each spatial dimension separately).
Individual spectral components become difficult to identify and off-resonant
couplings to ``spectator'' transitions \cite{wineland:bible} are hard to avoid.
However, under certain conditions, QS are predicted to allow for coupling to
all modes simultaneously, see \eg{} \refcite{porras:eff-spin}.
(2) QS based on ions in large, more-dimensional Coulomb crystals suffer from
additional challenges, for example, intrinsic micromotion (due to the
displacement from the minimum of the pseudopotential), an inhomogeneous ion
spacing (due to space charge effects) and the coupling between modes of all
three spatial dimensions.

One approach for scalability might be to generate a spin-off of the
QIP community based on their invention of a new concept of a surface
electrode geometry for RF traps \cite{chiaverini:surface-trap,%
seidelin:surface-trap} (see \figref{two:d:proposal}).
Currently, this design is tested with the aim to allow for networks of
interconnected linear traps.
This constitutes a promising possibility to realize the multiplex architecture
of memory and processor traps for universal QC \cite{kielpinski:architecture}.
However, for QS we need a miniaturized array of traps allowing for
more-dimensional interactions, as discussed in \secref{surface:traps}.

It has to be emphasized that there are other concepts for trapping
ions, a prominent one being Penning traps.
Penning traps provide trapping potentials of similar parameters as
RF traps.
A strong, static magnetic field and a DC electric field yield a stable
confinement of large, rotating Coulomb crystals.
Storing many cool ions in a Penning trap, naturally provides, \eg{}, a large
triangular lattice of ions \cite{tan:order:wigner,itano:bragg:crystal,%
mitchell:phase:trans} that is also
predicted to be well suited for QS \cite{porras:2d:crystals}.
Promising results are on their way
\cite{biercuk:hifi:penning,bollinger:priv:comm}.
Another challenging proposal for QS involves trapping ions optionally
simultaneously with atoms in optical lattices \cite{schneider:dipole:trap}
(see \secref{lattices}).

\subsection{Ions}
\seclabel{tools:ions}

\figurehandler{\myfigurelevelscheme}

A huge variety of different atomic ions have already been used for the purpose
of QIP.
Every ion or, more specific, every isotope has different properties, \eg{},
regarding the level scheme or the charge--mass ratio, and thus can meet
different requirements of a QS.
However, they all have a single valence electron leading to an alkali-like
level scheme.
Most prominent are the earth alkali ions $\atom{Be}[+]$,
$\atom{Mg}[+]$, $\atom{Ca}[+]$, $\atom{Sr}[+]$, and $\atom{Ba}[+]$.
A similar electronic structure possess $\atom{Zn}[+]$, $\atom{Cd}[+]$, and
$\atom{Hg}[+]$, followed by $\atom{Yb}[+]$ \cite{monroe:table}.

Typically, two electronic levels with sufficiently long coherence times
are chosen as qubit or spin states $\ket{\down}$ and $\ket{\up}$, respectively.
(In principle, however, the restriction to two states is not required
and the use of up to $60$ states has been proposed for (neutral)
holmium \cite{saffman:holmium}.)
The types of qubits can be divided into two classes:
In optical qubits, the states are encoded in two states with a dipole-forbidden
transition at an optical frequency.
An example is $\atom[40]{Ca}[+]$ with $\ket{\down} \defeq
\ket{\level{S}_{1/2}}$ and $\ket{\up} \defeq \ket{\level{D}_{5/2}}$.
The lifetime of $\ket{\up}$ is on the order of $\unit[1]{s}$, which defines
the upper bound for its coherence time.
In hyperfine/Zeeman qubits, two sublevels from the ground state manifold
are chosen as $\ket{\down}$ and $\ket{\up}$.
An applied magnetic field lifts the degeneracy within the manifolds of
electronic levels to allow for spectrally resolving the dedicated
states.
The states of hyperfine/Zeeman qubits have extremely long lifetimes and
coherence times on the order of minutes have been observed
\cite{bollinger:frequency:standard,fisk:yb:spectroscopy}.
As an example for a hyperfine/Zeeman qubit, an excerpt of the level scheme of
$\atom[25]{Mg}[+]$ is shown in \figref{level:scheme:mg25}.
The transition frequencies in hyperfine/Zeeman qubits are in the microwave
regime.

\subsection{Basic Operations}
\seclabel{tools:operations}

The quantized oscillation of the ions in the harmonically approximated
potential of the trap gives rise to motional states,
which are typically expressed in terms of Fock states $\ket{n}$.
Independent of the choice of qubit we will require three
different types of couplings to electronic states and/or motional states to
assemble the toolbox for QC and QS based on trapped ions (for details see
\secref{theory}).

\emph{(a) coupling of the electronic states only
($\ket{\down}\ket{n} \rightleftarrows \ket{\up}\ket{n}$)}
This operation can be used to implement Rabi flops between the electronic
states and serves as a one-qubit gate of a potential QC.
In the context of QS it can be interpreted as simulated magnetic field
(see also \secref{rabi:sideband}).

\emph{(b) coupling of the electronic and motional
states ($\ket{\down}\ket{n} \rightleftarrows \ket{\up}\ket{n'}$)}
This operation can drive Rabi flops between electronic states and
different motional states, \eg{} on the first red ($n' = n - 1$) or
blue sideband ($n' = n + 1$) (see also \secref{rabi:sideband}).
It can be used to create entanglement between the electronic and
motional states and is an important ingredient for both sideband
cooling and the readout of the motional state (see below).

\emph{(c) state-dependent forces
(\eg{}, $\ket{\down}\ket{n}\rightarrow \ket{\down}\ket{n+1}$)}
These forces lead to state-dependent displacements.
They can be used for conditional interactions between multiple ions,
which are exploited for quantum gates (see \secsref{phase:space}{phase:gate})
or effective spin--spin interactions in the simulation of quantum spin
Hamiltonians (see \secref{spin:spin}).

Operations (a) to (c) can be realized for both classes of qubits in
the optical regime and for hyperfine/Zeeman qubits additionally via
microwave fields \cite{wineland:bible,leibfried:habil}:

\emph{coupling via optical fields}
Optical qubit states can be linked by a single, nearly resonant laser beam with
frequency $\omegaI$ and wave vector $\vec \kI$.
Due to the lifetime of the $\ket{\up}$ state of $1 / \Gamma \sim \unit[1]{s}$,
the linewidth of the laser has to be very narrow ($\sim \unit[1]{Hz}$).
Operations (a) and (b) can be implemented directly
(see \secref{rabi:sideband}).
State-dependent forces (c) can be provided by a bichromatic light field
(see, \eg{}, \refmcite{soerensen:quantum:comp,roos:gate,lee:phase:control} and
also \secref{phase:space}).

\figurehandler{\myfigureeffbj}

In hyperfine/Zeeman qubits, the single laser beam can be substituted by
two beams with frequencies $\omega_1, \omega_2$ and wavevectors $\vec
k_1, \vec k_2$ driving two-photon stimulated-Raman transitions.
The beams are detuned by $\Delta \gg \Gamma$ from a third
level, \eg{}, a $\level{P}$ level (\cmp{} \figref{level:scheme:mg25}) with
a typical lifetime $1 / \Gamma$ of the order of few nanoseconds.
In the mathematical treatment, this third level can be
adiabatically eliminated for large detunings and the interaction
gains the form of an interaction with a single beam of frequency
$\omegaI = |\omega_1 - \omega_2|$ and wavevector
$\vec \kI = \vec k_1 - \vec k_2$.
The requirement of a narrow linewidth holds only for the difference
frequency $\omegaI$, which can be fulfilled comparatively easily:
The two beams can be generated from the same laser using
acousto-optical modulators driven by a stable microwave source, while
the requirements on the frequency stability of the laser are relaxed.
For operation (a), the frequency $\omegaI$ has to (approximately) meet the
transition frequency of the qubit states (see \figref{eff:b:j}a).
For operation (b), $\vec \kI$ in addition must not vanish to achieve a
sufficient momentum transfer to the ions, see \figref{level:scheme:mg25} and
\secref{rabi:sideband}.
Therefore, the two beams are typically orthogonal
($|\vec \kI| \approx \sqrt{2} |\vec k_1|$) or counter-propagating
($|\vec \kI| \approx 2 |\vec k_1|$).
The state-dependent forces (c) can be implemented by nearly resonant
beams ($\omegaI \approx 0$) and beam geometries as for operation (b) (see
\figref{eff:b:j}b, \secsref[to]{phase:space}{spin:spin},
and \secref{interpretation}).

The main technical drawback of using two-photon stimulated-Raman transitions
is decoherence due to spontaneous emission after off-resonantly populating
the third level.
This limitation can be mitigated by increasing the detuning $\Delta$
and the intensities $I_{1/2}$ of the beams, since the interaction strength
scales with $ I_{1/2}/ \Delta$, while the spontaneous emission rate scales
with $I_{1/2} / \Delta^2$.

\emph{coupling via microwave fields}
Alternatively, transitions between the electronic states in
hyperfine/Zeeman qubits can be driven laser-less by microwave fields.
This allows to directly realize operation (a).
However, due to the comparatively long wavelength and the related small
momentum transfer ($|\hbar \vec \kI| \rightarrow 0$), only negligible
coupling to the motional modes can be achieved directly and additional efforts
are required to provide operations (b) and (c)
\cite{wunderlich:long-wave,johanning:addressing}:
By applying a static magnetic field gradient along the axis of an ion chain,
the transition frequency between $\ket{\down}$ and $\ket{\up}$ becomes
site-dependent due to position-dependent Zeeman shifts.
The ions can be individually addressed by applying microwave fields with these
site-dependent frequencies $\omegaI$.
In addition, this causes state-dependent forces as in the Stern--Gerlach
experiment and allows for coupling to the motional modes.
The main challenge here is to provide sufficiently large magnetic field
gradients and to cope with state-dependent transition frequencies, if high
operational fidelities are required (see, \eg{}, \refcite{timoney:gate}).

As an alternative to the static magnetic field gradients, alternating magnetic
fields due to microwave currents in electrodes of surface electrode traps
(see \secref{surface:traps}) have been proposed \cite{ospelkaus:magnetic:gates}
and first promising results have been achieved \cite{brown:gate,ospelkaus:gate}.
Due to the small height of the ion above the
electrode in this type of traps, a sufficiently large AC Zeeman shift
can be generated, which can be treated analogously to the AC Stark shift
created by laser beams in two-photon stimulated-Raman transitions discussed
above.
However, the small height above the electrodes leads to further challenges
(\cmp{} \secref{surface:traps}) and high microwave powers are required.

\subsection{Initialization and Readout}
\seclabel{tools:readout}

\emph{initialization of motional and electronic states}
The initialization into one of the qubit states, \eg{} $\ket{\down}$, can be
achieved with near-unity efficiency by optical pumping \cite{happer:pumping}.
Regarding the motional modes, the initialization includes Doppler cooling
in all three dimensions leading to a thermal state with an average phonon
number $\mean{n}$ of typically a few to ten quanta.
This pre-cooling is required to reach the Lamb-Dicke regime (see
\secref{rabi:sideband}), where subsequent resolved sideband cooling
\cite{wineland:cooling,diedrich:cooling,monroe:cooling}
or cooling utilizing electro-magnetically induced transparency can be applied
\cite{morigi:cooling:eit,roos:cooling:eit}.
These cooling schemes lead close to the motional ground state $\ket{0}$
($n = 0$ with probability of $\unit[98]{\%}$ in \refcite{monroe:cooling}) of
the dedicated modes.

\emph{readout of electronic and motional states}
We distinguish the two electronic states by observing state-dependent laser
fluorescence.
The dipole allowed transition to an excited state starting in the
state $\ket{\down}$ is driven resonantly (see transition labelled ``BD'' in
\figref{level:scheme:mg25}) in a closed cycle completed by spontaneous emission
back to the state $\ket{\down}$ due to selection rules.
For state $\ket{\up}$ the detection laser is off-resonant.
The ion therefore appears ``bright'' for $\ket{\down}$, while it remains
``dark'' for $\ket{\up}$ \cite{wineland:pumping,nagourney:quantum:jumps,%
sauter:quantum:jumps,bergquist:quantum:jumps}.
Typically, a few per mill of the scattered photons are detected by a
photo-multiplier tube or a CCD camera.
The fidelity of this detection scheme has been shown
experimentally to exceed $\unit[99.99]{\%}$ for averaged and even individual
measurements \cite{myerson:readout,burrell:readout}.
However, additional possibilities to enhance the detection efficiency, for
example by methods developed for QC using ancilla qubits
\cite{schaetz:readout,hume:readout,schmidt:spectroscopy}, cannot be applied to
analogue QS, since all ions participate during the simulation.
For the detection of the motional state, it can be mapped to the electronic
state via an operation of type (b) and derived from the result of the spin
state detection described above \cite{meekhof:squeezed}.


\ultrashortmacroson 

\section{Theoretical Excursion}
\seclabel{theory}

The following calculations (\secsref[to]{intro:theo}{rabi:sideband})
are the mathematical description of the toolbox
that is available for both QCs and QSs.
Detailed discussions can be found, \eg{}, in \refcite{wineland:bible} or
\refcite{leibfried:habil}.
We summarize important equations in the following and extend the
mathematical description to be applicable to scaled approaches of QSs, \eg{},
two-dimensional arrays of ions in individual traps.
We will continue with the description of an implementation of the effective
spin--spin interaction for ions appearing in quantum spin Hamiltonians.
In order to investigate it isolated from other interactions, we will first
discuss it from the point of view of quantum gates \cite{leibfried:gate} in
\secsref{phase:space}{phase:gate}.
Finally, in \secref{spin:spin}, we use all tools to derive and
discuss the quantum Ising Hamiltonian based on
\refcite{porras:eff-spin} as an example.
The mathematical descriptions will be required to pursue proposals described in
\secref{many:body:theo}.

\subsection{Theoretical Basics}
\seclabel{intro:theo}

In the following we consider two-level systems only.
The Hamiltonian describing the energy of the electronic states of $N$ such
systems is given by:
\begin{equation}
  \eqlabel{h:e}
  \H_\text{e} = \sum_{i = 1}^N \frac{\hbar \omega_\updown}{2} \site{\pauli{z}}
    + \underbrace{N \hbar \frac{\omega_\up + \omega_\down}{2}}_{= \const
    \text{ (omitted)}},
\end{equation}
where $\omega_{\up/\down}$ denote the energies of the states $\ket{\down}$
and $\ket{\up}$, respectively, $\omega_{\updown} \defeq \omega_\up
- \omega_\down$, and the operator $\site{\pauli{z}}$ the Pauli
operator (\cmp{} \secref{pauli:trans}, \eqref{pauli}) acting on the $i$-th ion.

The ions are considered to be trapped in a common harmonic potential or several
individual harmonic potentials.
The corresponding Hamiltonian in terms of the normal modes of the oscillation
reads:
\begin{equation}
  \eqlabel{h:m}
  \H_\text{m} = \sum_{m = 1}^{3 N} \hbar \mode{\omega} \left(\mode{\ad}
  \mode{\a} + \frac{1}{2}\right).
\end{equation}
Here, $\mode{\a}$ and $\mode{\ad}$ are the annihilation and creation operators
of the $m$-th mode, respectively, and $\mode{\omega}$ the corresponding
frequency.
In the following, the constant terms $\hbar \mode{\omega} / 2$ will also be
omitted and the abbreviation $\H_0 \defeq \H_\text{e} + \H_\text{m}$ will be
used.

An interaction of an ion with the electric field $\vec E$ of a laser
beam is described by $- \op{\vec \mu} \cdot \vec E(\vec r, t)$, where
$\op{\vec \mu}$ denotes the electric dipole operator for the transition
$\ket{\down} \leftrightarrow \ket{\up}$ and $\vec E(\vec r, t)$ the field at the
site of the ion.
The Hamiltonian describing the interaction of the field with $N$ ions becomes:
\begin{equation}
  \eqlabel{h:i:plain}
  \HI= \sum_{i = 1}^N \hbar \site{\OmegaI} \left(
  \ee^{\ii(\site{\vec \kI} \cdot \site{\op{\vec r}}
  - \omegaI t + \site{\phiI})} + \hc\right) \site{\kauli}.
\end{equation}
Here, $\site{\OmegaI} = - \mu \site{E} / 2 \in \nset{R}$ is the interaction
strength at
site $i$, $\site{\vec \kI}$ the wavevector at site $i$, $\site{\op{\vec r}}$
the position of the $i$-th ion, $\omegaI$ the frequency of the field and
$\site{\phiI}$ an additional phase.
In the most general form, the operator $\site{\kauli}$ can be expressed as a
linear combination of Pauli operators $\site{\pauli{x/y/z}}$ and the identity
operator $\site{\I}$ (see \secref{pauli:trans}):
\begin{equation}
  \eqlabel{kauli}
  \site{\kauli} \defeq \alpha_0 \site{\I} + \alpha_1 \site{\pauli{x}}
   + \alpha_2 \site{\pauli{y}} + \alpha_3 \site{\pauli{z}},
\end{equation}
with the prefactors $\alpha_j \in \nset{R}$.

The position operator $\site{\op{\vec r}}$ in \eqref{h:i:plain} is
decomposed into the equilibrium position $\site{\vec x_0}$ and the displacement
$\site{\op{\vec x}} = \site{\op{\vec r}} - \site{\vec x_0}$.
The terms $\site{\vec \kI} \cdot \site{\vec x_0}$ give rise to a constant
phase, which we absorb into $\site{\phiI}
+ \site{\vec \kI} \cdot \site{\vec x_0} \to \site{\phiI}$.

The displacement of the ion from its equilibrium position $\site{\op{\vec x}}$
is expressed in terms of the normal modes of motion
\begin{equation}
  \site{\op{\vec x}} = \sum_{m = 1}^{3 N} \left(
    b_{m, i} \mode{\op{q}} \vec e_X
    + b_{m, i + N} \mode{\op{q}} \vec e_Y
    + b_{m, i + 2 N} \mode{\op{q}} \vec e_Z\right),
\end{equation}
where $b_{m,i}$ are the elements of an (orthogonal) transformation matrix
(\cmp{} \secref{modes:freqs}, \eqref{modes:trans}).
Expressing the operators $\mode{\op{q}}$ of the normal modes in terms of the
creation and annihilation operators yields
\begin{equation}
  \mode{\op{q}} = \mode{q}_0 \left(\mode{\a} + \mode{\ad}\right)
  \ms \text{with} \ms \mode{q}_0 \defeq \sqrt{\frac{\hbar}{2 M \mode{\omega}}},
\end{equation}
where $M$ denotes the mass of one ion.
Hence, the scalar product appearing in the Hamiltonian yields
\begin{equation}
  \site{\vec \kI} \cdot \site{\op{\vec x}} = \sum_{m = 1}^{3 N}
  \site{\mode{\eta}} \left(\mode{\a} + \mode{\ad}\right),
\end{equation}
where the Lamb-Dicke parameter of the $m$-th mode and $i$-th site has been
introduced:
\begin{equation}
\eqlabel{lamb:dicke}
  \begin{aligned}
    \site{\mode{\eta}}
      &\defeq \mode{q}_0 \left(b_{m, i} \site{\vec \kI} \cdot \vec e_X\right. \\
      &\quad \left.+ b_{m, i + N} \site{\vec \kI} \cdot \vec e_Y
        + b_{m, i + 2 N} \site{\vec \kI} \cdot \vec e_Z\right).
  \end{aligned}
\end{equation}

To summarize, the interaction term of the Hamiltonian gains the form
\begin{equation}
  \eqlabel{h:i}
  \HI = \sum_{i = 1}^N \hbar \site{\OmegaI} \left(
  \ee^{\ii \left[\sum_{m = 1}^{3 N}
  \site{\mode{\eta}} \left(\mode{\a} + \mode{\ad}\right)
  - \omegaI t + \site{\phiI}\right]} + \hc\right) \site{\kauli}.
\end{equation}

\subsection{Interaction Picture}
\seclabel{interaction:pic}

The transformation into the interaction picture
\begin{equation}
  \HI' = \U_0^\dagger \HI \U_0
  \ms \text{with} \ms \U_0 \defeq \ee^{-\ii \H_0 t / \hbar}
\end{equation}
can be carried out for each site separately, hence
\begin{equation}
  \site{\HI'} = \site{\U_0}^\dagger \site{\HI} \site{\U_0}
  \ms \text{with} \ms \site{\U_0} \defeq \ee^{-\ii \site{\H_0} t / \hbar},
\end{equation}
where $\site{\HI}$ is the Hamiltonian corresponding to the $i$-th ion
and $\HI = \sum_{i = 1}^N \site{\HI}$.

The operator $\site{\kauli}$ (\eqref{kauli}) related to the electronic states
reads as follows in the interaction picture (\cmp{} \eqref{pauli:trans}):
\begin{align}
  \site{\kauli'}
    &= \ee^{\ii \omega_\updown t \site{\pauli{z}} / 2} \site{\kauli}
      \ee^{- \ii \omega_\updown t \site{\pauli{z}} / 2} \\
    &= \frac{1}{2} \left[\alpha_0 \site{\I}
      + \left(\alpha_1 + \frac{\alpha_2}{\ii} \right)\ee^{\ii \omega_\updown t}
      \site{\pauli{+}}
      + \alpha_3 \site{\pauli{z}}\right] + \hc
\end{align}
Here, we introduced $\site{\pauli{+}} \defeq \site{\pauli{x}} + \ii
\site{\pauli{y}}$ and $\site{\pauli{-}} \defeq \site{\pauli{x}} - \ii
\site{\pauli{y}}$.
The terms in \eqref{h:i} containing the motional operators transform as follows
(\cmp{} \eqref{exp:aad:trans}):
\begin{multline}
  \eqlabel{h:i:trans:a}
  \ee^{\ii \mode{\omega} t \mode{\ad} \mode{\a}}
    \ee^{\ii \site{\mode{\eta}} \left(\mode{\a} + \mode{\ad}\right)}
    \ee^{- \ii \mode{\omega} t \mode{\ad} \mode{\a}} \\
  = \exp\left(\ii \site{\mode{\eta}}\left[\mode{\a} \ee^{-\ii \mode{\omega} t}
    + \mode{\ad} \ee^{\ii \mode{\omega} t}\right]\right).
\end{multline}

Hence, it yields the following expression for the complete Hamiltonian in the
interaction picture:
\begin{widetext}
  \begin{equation}
    \eqlabel{h:i:i}
    \site{\HI'}
    = \hbar \site{\OmegaI} \left\{\exp\left(\ii \left[
    \sum_{m = 1}^{3 N} \site{\mode{\eta}}
    \left(\mode{\a} \ee^{-\ii \mode{\omega} t}
    + \mode{\ad} \ee^{\ii \mode{\omega} t}\right) - \omegaI t + \site{\phiI}
    \right]\right) + \hc\right\} \site{\kauli'}.
  \end{equation}
\end{widetext}

At this point fast rotating terms which average out on short timescales are
neglected (rotating wave approximation, RWA).
For $\site{\OmegaI} \ll \omega_\updown$ we distinguish between two cases:
In the first case, $\omegaI \ll \omega_\updown$, terms containing
$\ee^{\pm \ii \omega_\updown t}$ are neglected (see \figref{eff:b:j}b for an
example of an implementation).
(If $\alpha_1 = \alpha_2 = 0$, nothing will change and the Hamiltonian will
still be exact.)
In the second case, $\left|\omegaI - \omega_\updown\right| \ll
\omega_\updown$, all terms but $\ee^{\pm \ii \left(\omega_\updown
- \omegaI\right) t}$ are neglected (see \figref{eff:b:j}a).
\begin{widetext}
  \begin{align}
    \eqlabel{rwa:z}
    \rwa{\site{\HI'}} &= \hbar \site{\OmegaI} \exp\left(\ii \left[
      \sum_{m = 1}^{3 N} \site{\mode{\eta}}
      \left(\mode{\a} \ee^{-\ii \mode{\omega} t}
      + \mode{\ad} \ee^{\ii \mode{\omega} t}\right) - \omegaI t + \site{\phiI}
      \right]\right)
      \left(\alpha_0 \site{\I} + \alpha_3 \site{\pauli{z}}\right) + \hc
      && \left(\omegaI \ll \omega_\updown\right), \\
    \eqlabel{rwa:xy}
    \rwa{\site{\HI'}} &= \frac{\hbar}{2} \site{\OmegaI} \exp\left(\ii
      \left[\sum_{m = 1}^{3 N} \site{\mode{\eta}}
      \left(\mode{\a} \ee^{-\ii \mode{\omega} t}
      + \mode{\ad} \ee^{\ii \mode{\omega} t}\right) -
      \left(\omegaI - \omega_\updown\right) t + \site{\phiI}\right]\right)
      \left(\alpha_1 + \frac{\alpha_2}{\ii}\right) \site{\pauli{+}} + \hc
      && \left(\left|\omegaI - \omega_\updown\right|
      \ll \omega_\updown\right).
  \end{align}
\end{widetext}

\subsection{$\mathbf{\pauli{x}}$/$\mathbf{\pauli{y}}$ Interaction}
\seclabel{rabi:sideband}

The time evolution corresponding to $\rwa{\site{\HI'}}$ in
\eqref{rwa:xy} is involved.
The time evolution is calculated for a single ion $i$ and a single motional
mode $m$, \eg{}, in \refscite{wineland:bible}{leibfried:habil}.
As some simplifications (Lamb-Dicke regime, see below) are not always
justified for experiments, we will summarize this calculation here.

In this case the Hamiltonian simplifies:
\begin{widetext}
  \begin{align}
    \rwa{\HI'} &= \frac{\hbar}{2} \OmegaI \exp\left(\ii \left[
      \eta \left(\a \ee^{-\ii \omega t} + \ad \ee^{\ii \omega t}\right) -
      \left(\omegaI - \omega_\updown\right) t + \phiI\right]\right)
      \left(\alpha_1 + \frac{\alpha_2}{\ii}\right) \pauli{+} + \hc
  \end{align}
\end{widetext}
Writing the state vector in the basis of electronic states $\ket{s}$ and
motional Fock states $\ket{n}$,
\begin{equation}
  \ket{\psi(t)} = \sum_{s \in \set{\down, \up}} \sum_n
  c_{s, n}(t) \ket{s, n},
\end{equation}
the Schr\"{o}dinger equation yields
\begin{equation}
  \eqlabel{schroedinger:dgl}
  \ii \hbar \dot c_{s', n'}(t) = \sum_{s \in \set{\down, \up}}
    \sum_n \braket{s', n' | \rwa{\HI'} | s, n} c_{s, n}(t).
\end{equation}

Matrix elements of the Hamiltonian vanish for $s' = s$.
We obtain for the non-vanishing matrix elements
\cite{cahill:expansions,wineland:cooling} (\cmp{} \secref{matrix:disp})
\begin{equation}
  \begin{aligned}
    \braket{\up, n' | \rwa{\HI'} | \down, n}
      &= \frac{\hbar}{2} \OmegaI
        \ee^{\ii \left(-(\omegaI - \omega_\updown) t + \phiI \right)}
        \left(\alpha_1 + \frac{\alpha_2}{\ii}\right) \\
      &\quad \times
        \braket{n' | \D\left(\ii \eta \ee^{\ii \omega t} \right) | n}
        \braket{\up | \pauli{+} | \down} \\
      &= \hbar \Omega_{n', n} \left(\alpha_1 + \frac{\alpha_2}{\ii}\right)
        \ii^{|n' - n|} \\
      &\quad \times \ee^{\ii \left(\left[(n' - n) \omega -(\omegaI
        - \omega_\updown)\right] t + \phiI \right)}
  \end{aligned}
\end{equation}
where $\D(\lambda) \defeq \ee^{\lambda \ad - \lambda^* \a}$ denotes the
displacement operator and
\begin{equation}
  \eqlabel{rabi:nprimen}
  \Omega_{n', n} \defeq \OmegaI\ee^{-\eta^2 / 2} \eta^{|n' - n|}
  \sqrt{\frac{n_< !}{n_> !}} L_{n_<}^{(|n' - n|)}\left(\eta^2\right).
\end{equation}
Here, $L_n^{(\alpha)}(x)$ are the associated Laguerre polynomials,
$n_< \defeq \min(n', n)$, and $n_> \defeq \max(n', n)$.
Analogously, we obtain $\braket{\down, n | \rwa{\HI'} |
\up, n'} =  \braket{\up, n' | \rwa{\HI'} | \down, n}^*$.

We define $\delta \defeq (\omegaI - \omega_\updown) - (n' - n) \omega$.
For small detunings $|\delta| \ll \omega$ and interaction strengths
$\left|\Omega_{n', n}\right| \ll \omega$ (resolved sideband regime), we apply a
RWA neglecting terms rotating faster than $\ee^{\pm \ii \delta t}$.
\Eqref{schroedinger:dgl} can then be solved for each subset
$\ket{n', \up}$ and $\ket{n, \down}$ separately:
\begin{align}
  \dot c_{\up, n'}(t)
    &= - \ii \Omega_{n', n} \left(\alpha_1 + \frac{\alpha_2}{\ii}\right)
      \ii^{|n' - n|} \ee^{- \ii (\delta t - \phiI)} c_{\down, n}(t) \\
  \dot c_{\down, n}(t)
    &= - \ii \Omega_{n', n} \left(\alpha_1 + \frac{\alpha_2}{\ii}\right)^*
      (-\ii)^{|n' - n|} \ee^{\ii (\delta t + \phiI)} c_{\up, n'}(t).
\end{align}
The solution of the system of differential equations yield Rabi oscillations
between the states $\ket{\down, n} \leftrightarrow \ket{\up, n'}$
(\cmp{} \secref{rabi:dgl}):
\begin{widetext}
  \begin{equation}
    \eqlabel{rabi:sol}
    \begin{pmatrix}
      c_{\up, n'}(t) \\
      c_{\down, n}(t)
    \end{pmatrix}
    =
    \begin{bmatrix}
      \left(\cos(X_{n', n} t) + \frac{\delta}{2} \frac{\ii}{X_{n', n}}
        \sin(X_{n', n} t)\right) \ee^{- \ii \delta t / 2} &
        \frac{Y_{n', n}}{X_{n', n}} \sin(X_{n', n} t)
        \ee^{- \ii \delta t / 2} \\
      - \frac{Y_{n', n}^*}{X_{n', n}} \sin(X_{n', n} t)
        \ee^{\ii \delta t / 2} &
        \left(\cos(X_{n', n} t) - \frac{\delta}{2} \frac{\ii}{X_{n', n}}
        \sin(X_{n', n} t)\right) \ee^{\ii \delta t / 2}
    \end{bmatrix}
    \begin{pmatrix}
      c_{\up, n'}(0) \\
      c_{\down, n}(0)
    \end{pmatrix}
  \end{equation}
\end{widetext}
with $Y_{n', n} \defeq -\ii \Omega_{n', n} (\alpha_1 +
\alpha_2 / \ii) \ii^{|n' - n|} \ee^{\ii \phiI}$ and
$X_{n', n} \defeq \sqrt{\frac{\delta^2}{4} + \left|Y_{n', n}\right|^2}$.

In the Lamb-Dicke regime, $\eta \expval{\left(\a + \ad\right)^2}^{1/2}
\ll 1$, \eqref{rabi:nprimen} can be expanded to first order in $\eta$:
\begin{align}
  \eqlabel{omega:red:sb}
  \ldr{\Omega_{n - 1, n}} &= \OmegaI \eta \sqrt{n}
    &&\text{(first red sideband)} \\
  \eqlabel{omega:carrier}
  \ldr{\Omega_{n, n}} &= \OmegaI &&\text{(carrier)} \\
  \eqlabel{omega:blue:sb}
  \ldr{\Omega_{n + 1, n}} &= \OmegaI \eta \sqrt{n + 1}
    &&\text{(first blue sideband)}
\end{align}

Successive red sideband transitions $\ket{\down}\ket{n} \to
\ket{\up}\ket{n - 1}$ followed by dissipative repumping to
$\ket{\down}\ket{n - 1}$ with high probability are routinely used for
sideband cooling close to the motional ground state $\ket{n = 0}$
\cite{wineland:cooling,diedrich:cooling,monroe:cooling}.

If the Lamb-Dicke parameter becomes effectively zero, the motional dependence
will vanish (see \eqsref{omega:red:sb}{omega:blue:sb}).
The only remaining transition is the carrier transition \eqref{omega:carrier}
affecting the electronic states only.
This is the case, \eg{}, for two-photon stimulated Raman transitions with
co-propagating beams or for microwave driven transitions in hyperfine qubits,
where $\vec \kI \approx 0$.
In systems with more than one ion the ions will not be motionally coupled.
That is why, \eqref{rabi:sol} also holds for each site separately in such
systems.

\Eqref{rabi:sol} simplifies for resonant carrier transitions ($\delta = 0$)
and a pure $\pauli{x}$ interaction ($\alpha_1 = 1$ and $\alpha_2 = 0$):
\begin{equation}
  \eqlabel{rabi:sol:resonant}
  \begin{pmatrix}
    c_{\up, n}(t) \\
    c_{\down, n}(t)
  \end{pmatrix}
  = \R(\theta, \phi)
  \begin{pmatrix}
    c_{\up, n}(0) \\
    c_{\down, n}(0)
  \end{pmatrix},
\end{equation}
where
\begin{equation}
  \eqlabel{rot:mat}
  \R(\theta, \phi) \defeq \begin{pmatrix}
    \cos(\theta / 2) & - \ii \ee^{\ii \phi} \sin(\theta / 2) \\
    - \ii \ee^{-\ii \phi} \sin(\theta / 2) & \cos(\theta / 2)
  \end{pmatrix},
\end{equation}
$\theta \defeq 2 \Omega_{n, n} t$ and $\phi \defeq \phiI$.
The rotation matrix $\R(\pi / 2, \phi)$ describes a $\pi / 2$-pulse and
$\R(\pi, \phi)$ a $\pi$-pulse with phase $\phi$.

\subsection{Effective $\mathbf{\pauli{z} \otimes \pauli{z}}$ Interaction}
\seclabel{phase:space}

We will now discuss the case of \eqref{rwa:z} with $\alpha_1 = \alpha_2 = 0$.
Hence, we omit the superscript of the Hamiltonian indicating a RWA.
In the Lamb-Dicke regime, $\site{\mode{\eta}} \expval{\left(\mode{\a}
+ \mode{\ad}\right)^2}^{1/2} \ll 1$, the Hamiltonian can be expanded to first
order in the Lamb-Dicke parameters $\site{\mode{\eta}}$.
A subsequent RWA neglecting terms rotating faster than
$\ee^{\pm \ii \mode{\delta} t}$
with $\mode{\delta} \defeq \omegaI - \mode{\omega}$ yields
\begin{widetext}
  \begin{align}
    \ldr{\site{\HI'}}
      &= \hbar \site{\OmegaI}
        \ee^{\ii \left(- \omegaI t + \site{\phiI}\right)}
        \left[1 + \ii \sum_{m = 1}^{3 N} \site{\mode{\eta}}
        \left(\mode{\a} \ee^{-\ii \mode{\omega} t}
        + \mode{\ad} \ee^{\ii \mode{\omega} t}\right)\right]
        \left(\alpha_0 \site{\I} + \alpha_3 \site{\pauli{z}}\right) + \hc \\
    \eqlabel{rwa:ldr}
    \follows \rwa{\ldr{\site{\HI'}}}
      &= \ii \hbar \site{\OmegaI}
        \sum_{m = 1}^{3 N} \site{\mode{\eta}}
        \ee^{\ii \left(-\mode{\delta} t + \site{\phiI}\right)} \mode{\ad}
        \left(\alpha_0 \site{\I} + \alpha_3 \site{\pauli{z}}\right) + \hc
  \end{align}
\end{widetext}
Note that \eqref{rwa:ldr} breaks up into a sum over terms that depend on only
one mode $m$ and one site $i$ each.

With the excursion in \secref{time:evol} the total time evolution operator in
the interaction picture reads
\begin{widetext}
  \begin{equation}
    \eqlabel{disp:geo:phase}
    \begin{aligned}
      \rwa{\ldr{\UI'}}(t, t_0)
        &= \exp\left(\ii \left[\sum_{i = 1}^N \sum_{m = 1}^{3 N}
          \frac{\site{\OmegaI} \site{\mode{\eta}}}{\mode{\delta}}
          \left(\ee^{- \ii \mode{\delta} (t - t_0)}
          - 1\right) \ee^{- \ii \mode{\delta} t_0} \ee^{\ii \site{\phiI}}
          \mode{\ad} \left(\alpha_0 \site{\I} + \alpha_3 \site{\pauli{z}}\right)
          + \hc \right]\right) \\
        &\quad \times \exp\Biggl(
          - \ii \sum_{i = 1}^N \sum_{j = 1}^N \sum_{m = 1}^{3 N}
          \frac{\site{\OmegaI} \site[j]{\OmegaI}
          \site{\mode{\eta}} \site[j]{\mode{\eta}}}{\mode{\delta}^2} 
          \left(\alpha_0 \site{\I} + \alpha_3 \site{\pauli{z}}\right) \otimes
          \left(\alpha_0 \site[j]{\I} + \alpha_3 \site[j]{\pauli{z}}\right) \\
        &\quad \times \left[\mode{\delta} (t - t_0)
          \cos\left(\site{\phiI} - \site[j]{\phiI}\right)
          - \sin\left(\mode{\delta} (t - t_0) - \left(\site{\phiI}
          - \site[j]{\phiI}\right)\right)\right] \Biggr).
    \end{aligned}
  \end{equation}
\end{widetext}
The interaction described by \eqref{disp:geo:phase} can be interpreted as
follows:
The first exponential function has the form of a displacement operator
$\D(\lambda) = \ee^{\lambda \ad - \lambda^* \a}$, which leads to a displacement
of a coherent state by $\lambda$ in phase space.
Due to the $(\ee^{- \ii \mode{\delta} (t - t_0)} - 1)$ proportionality of the
exponent, the trajectory for a coherent state of each mode describes a circle
in phase space (or a straight line in the limit $\mode{\delta} = 0$).
The coherent state returns to its initial position at times $\mode{T}
= 2 \pi l / \mode{\delta}$ with $l \in \nset{N}$, where the exponent vanishes.
The second exponential can be expanded into a $\pauli{z} \otimes \pauli{z}$
interaction, a $\pauli{z}$ interaction, and a global phase.
The $\pauli{z} \otimes \pauli{z}$ terms give rise to a geometric phase, which
increases in time $t$, and the $\pauli{z}$ terms lead to a dynamic phase
\cite{berry:phase,aharonov:phase}.
The area in phase space enclosed by the trajectory is proportional to these
phases.

\subsection{Geometric Phase Gates}
\seclabel{phase:gate}

The collective interaction of multiple ions with the same laser(s) has been
proposed for the implementation of quantum gates
\cite{soerensen:quantum:comp,solano:bell:states,%
milburn:quantum:computing,sorensen:quantum:comp:2,wang:multiqubit:gates}.
These gates are described in the $z$-basis by \eqref{disp:geo:phase} and
have been first implemented in \refscite{leibfried:gate}{home:gate}.
M\o{}lmer-S\o{}rensen gates can be mathematically treated analogously in a
rotated basis and are described in detail in
\refscite{lee:phase:control}{roos:gate}.
Implementations are reported in \refmcite{sacket:entanglementfour,
haljan:clock:states,benhelm:gate,kim:gate:transverse}.

We will exemplary discuss geometric phase gates based on the $\pauli{z}
\otimes \pauli{z}$ terms in \eqref{disp:geo:phase} in the following.
They offer excellent tools to investigate a pure $\pauli{z} \otimes \pauli{z}$
interaction required for the simulation of more involved Hamiltonians like
quantum spin Hamiltonians.

The interaction according to the Hamiltonian can be implemented
\cite{leibfried:gate,friedenauer:qmagnet,schmitz:arch} by stimulated-Raman
transitions driven by two beams with wavevectors $\vec k_1, \vec k_2$ and
difference frequency close to a (several) motional mode(s)
(see \figref{eff:b:j}b).
On average the differential AC Stark shift between the levels $\ket{\down}$ and
$\ket{\up}$ caused by the two beams can be compensated by choosing appropriate
polarizations of the beams.
Still, on short timescales $\sim 2 \pi / \delta_m$ the ions experience a
state-dependent force that leads to the above displacement in the phase spaces
of the corresponding modes.

In the original implementation of the geometric phase gate \cite{leibfried:gate}
two $\atom[9]{Be}[+]$ ions are used.
The state-dependent forces amount to $\vec F_\down = -2 \vec F_\up$.
This means that the operators $(\alpha_0 \site{\I} + \alpha_3 \site{\pauli{z}})$
have diagonal elements $1$ and $-2$, which is fulfilled for
$\alpha_0 = -1 / 2$ and $\alpha_3 = 3 / 2$.
The effective wavevectors $\site[1]{\vec \kI} = \site[2]{\vec \kI} =
\vec k_1 - \vec k_2$ point along the axis of the linear trap and the laser
beams are detuned by $\mode[\text{STR}]{\delta} = 2 \pi \times \unit[26]{kHz}$
from the stretch (STR) mode.
The effect of the centre-of-mass (COM) mode can be neglected
($\mode[\text{COM}]{\delta} \approx 100 \mode[\text{STR}]{\delta}$).
The ions are placed at the same phase of the stimulated-Raman interaction
($\site[1]{\phiI} = \site[2]{\phiI} = 0$).

For $t = T_\text{g} = 2 \pi / \mode[\text{STR}]{\delta}$ the time evolution
operator \eqref{disp:geo:phase} simplifies:
\begin{widetext}
  \begin{equation}
    \eqlabel{time:evol:str}
    \rwa{\ldr{\UI'}} \left(T_\text{g}, 0\right)
    \approx \exp\left(-2 \pi \ii \sum_{i = 1}^2 \sum_{j = 1}^2
    \frac{(-1)^{i - j} \OmegaI^2
    \mode[\text{STR}]{\eta}^2}{\mode[\text{STR}]{\delta}^2}
    \left[\alpha_3^2 \site{\pauli{z}} \otimes \site[j]{\pauli{z}}
    + \alpha_0 \alpha_3 \left(\site{\pauli{z}} + \site[j]{\pauli{z}}
    \right)\right]\right),
  \end{equation}
\end{widetext}
where we have used $\mode[\text{STR}]{\eta} \defeq
\site[1]{\mode[\text{STR}]{\eta}} = -\site[2]{\mode[\text{STR}]{\eta}}$ and
neglected the global phase arising from the $\site{\I} \otimes \site[j]{\I}$
terms.
The sequence of the gate is similar to the one in \figref{didi:pulsescheme}, but
without the second displacement pulse $\D_2$.
Ideally, the initial state $\ket{\psi} = \ket{\down\down}
\ket{\mode[\text{COM}]{n} = 0, \mode[\text{STR}]{n} = 0}$
is rotated to $1/2 (\ket{\down\down} +
\ket{\down\up} + \ket{\up\down} + \ket{\up\up})\ket{\mode[\text{COM}]{n} = 0,
\mode[\text{STR}]{n} = 0}$ by the first $\R(\pi / 2, \pi / 2)$ pulse (the phase
$\phi$ of the first pulse can be chosen arbitrarily).
The only non-vanishing contributions arise from the
$\pauli{z} \otimes \pauli{z}$ terms for the $\ket{\down\up}$ and
$\ket{\up\down}$ states, which gain a geometric phase
\begin{equation}
  \eqlabel{geo:phase:str}
  \mode[\text{STR}]{\Phi}_{\down\up/\up\down} = -2 \pi \times
  4 \frac{\OmegaI^2 \mode[\text{STR}]{\eta}^2}{\mode[\text{STR}]{\delta}^2}
  \alpha_3^2.
\end{equation}
By choosing appropriate beam intensities and thus $\OmegaI$, these phases
equal $\mode[STR]{\Phi}_{\down\up/\up\down} = -\pi / 2$.
The subsequent $\R(\pi / 2, \pi)$ and $\R(\pi / 2, \pi / 2)$ pulses lead to the
final Bell state $\ket{\tilde \psi} = 1 / \sqrt{2} (\ket{\down \down} +
\ii \ket{\up \up})$, which is achieved experimentally with a fidelity of
$F = \unit[97]{\%}$ \cite{leibfried:gate}.

\figurehandler{\myfiguredidiaxrad}

A similar implementation of the geometric phase gate is
reported in \refcite{schmitz:arch} based on two $\atom[25]{Mg}[+]$ ions.
The state-dependent forces amount to $\vec F_\down = -3 / 2 \vec F_\up$
($\alpha_0 = -1 / 4$ and $\alpha_3 = 5 / 4$).
Furthermore, the detuning from the STR mode amounts to
$\mode[\text{STR}]{\delta} = - 2 \pi \times \unit[266]{kHz}$ and simultaneously
the detuning from the COM mode $\mode[\text{COM}]{\delta} = - 2 \pi \times
\unit[1330]{kHz}$ (\cmp{} \figref{didi:ax:vs:rad}a).
Hence, the effect of the COM mode is also exploited for the gate.
As the detuning from the COM mode is chosen an integer multiple of the
detuning from the STR mode ($\mode[\text{COM}]{\delta} = - 5
\mode[\text{STR}]{\delta}$), the first exponential in \eqref{disp:geo:phase}
still becomes unity for the gate duration of $T_\text{g} = |2 \pi /
\mode[\text{STR}]{\delta}|$.
(In other words, all circular trajectories in all phase spaces return to their
initial position for $T_\text{g}$.)
As a result, there is no entanglement left between the electronic and motional
states.

Analogous to \eqref{geo:phase:str}, but considering $\mode[\text{STR}]{\delta}
< 0$ and $\mode[\text{COM}]{\delta} > 0$ for the detunings and
$\site[1]{\mode[\text{COM}]{\eta}} = \site[2]{\mode[\text{COM}]{\eta}}$ for the
Lamb-Dicke parameters of the COM mode, the geometric phases yield:
\begin{align}
  \eqlabel{geo:phase:str:2}
  \mode[\text{STR}]{\Phi}_{\down\up/\up\down}
    &= 2 \pi \times
      4 \frac{\OmegaI^2
      \mode[\text{STR}]{\eta}^2}{\mode[\text{STR}]{\delta}^2} \alpha_3^2 \\
  \eqlabel{geo:phase:com:2}
  \mode[\text{COM}]{\Phi}_{\down\down/\up\up}
    &= -2 \pi
      \left|\frac{\mode[\text{COM}]{\delta}}{\mode[\text{STR}]{\delta}}\right|
      \times 4 \frac{\OmegaI^2
      \mode[\text{COM}]{\eta}^2}{\mode[\text{COM}]{\delta}^2} \alpha_3^2.
\end{align}
By adjusting the beam intensities appropriately the differential phase between
$\ket{\down\down}/\ket{\up\up}$ and $\ket{\down\up}/\ket{\up\down}$
can be adjusted to fulfil $\mode[\text{STR}]{\Phi}_{\down\up/\up\down} -
\mode[\text{COM}]{\Phi}_{\down\down/\up\up} = \pi / 2$.
As $\mode[\text{STR}]{\Phi}_{\down\up/\up\down}$ has the opposite sign compared
to $\mode[\text{COM}]{\Phi}_{\down\down/\up\up}$, the geometric phase gate
makes use of two motional modes simultaneously.

However, some of the dynamic phases from the COM mode do not vanish:
\begin{equation}
  \eqlabel{dyn:phase:com:2}
  \mode[\text{COM}]{\tilde \Phi}_{\down\down/\up\up}
  = \pm 2 \pi
  \left|\frac{\mode[\text{COM}]{\delta}}{\mode[\text{STR}]{\delta}}\right|
  \times 8 \frac{\OmegaI^2
  \mode[\text{COM}]{\eta}^2}{\mode[\text{COM}]{\delta}^2} \alpha_0 \alpha_3.
\end{equation}
These phases have an absolute value of $2 \alpha_0 / \alpha_3$ of
the geometric phase from the COM mode and lead to a small deviation from the
ideal state at the end of the gate.

Compared to the original implementation in \refcite{leibfried:gate}, the
gate is speeded up by approximately a factor of $10$ and the fidelity
$F$ for the Bell state exceeds $\unit[95]{\%}$.
(Note that the duration of the spin-echo sequence is not included in
$T_\text{g}$, because its rotations could be much faster and empty gaps can in
principle be removed.)


The radial motional modes are interesting, because they are similar to the
normal modes in systems of individual traps for each ion (\cmp{}
\secref{surface:traps}),
which are promising candidates for scalable systems in quantum simulations.
To investigate the differences between the axial and radial modes of motion
the geometric
phase gate with $\atom[25]{Mg}[+]$ is performed on a pair of radial modes
(see also \refcite{kim:gate:transverse} for a M\o{}lmer-S\o{}rensen gate
performed on the radial modes).

The detunings from the COM and ROC (the equivalent to the STR mode in terms of
the axial motional modes) are chosen to have same absolute values
$\mode[\text{COM}]{\delta} = - \mode[\text{ROC}]{\delta} =
2 \pi \times \unit[65]{kHz}$ (see \figref{didi:ax:vs:rad}).
The geometric phases acquired on each motional mode are basically the same as in
\eqref{geo:phase:str:2}, where ``STR'' has to be replaced by ``ROC'',
and \eqref{geo:phase:com:2}.
(However, the signs change due to a change of the signs of the detunings.)
The contributions to the total differential geometric phase between
$\ket{\down\down}$/$\ket{\up\up}$ and $\ket{\down\up}$/$\ket{\up\down}$ due to
the COM and ROC mode are (approximately) equal now.
However, the dynamic phase (analogous to \eqref{dyn:phase:com:2}) arising from
the COM mode can no longer be neglected.

\figurehandler{\myfiguredidipulse}

The pulse scheme of the geometric phase gate is modified by adding a second
displacement pulse in the second gap of the spin-echo sequence (see
\figref{didi:pulsescheme} and \cmp{} \refcite{home:gate}).
The intensities of the beams are now adjusted for a differential geometric
phases due to each displacement pulse of
$\mode[\text{COM}]{\Phi}_{\down\down/\up\up} -
\mode[\text{ROC}]{\Phi}_{\down\up/\up\down} = \pi / 4$.
While the geometric phases of both displacement pulses add up to $\pi / 2$, the
dynamic phases cancel each other, as the $\pi$ pulse of the spin-echo sequence
interchanges $\ket{\down\down} \leftrightarrow \ket{\up\up}$ (and
$\ket{\down\up} \leftrightarrow \ket{\up\down}$).
Additionally, the more symmetric pulse scheme enhances the robustness of the
gate against uncompensated differential AC Stark shifts between $\ket{\down}$
and $\ket{\up}$.

\figurehandler{\myfiguredidievol}
\figurehandler{\myfiguredidiparity}

The total fluorescence from the two ions as a function of the total duration
of the displacements $2 T_D$ is shown in \figref{didi:evol}.
The gate duration due to the smaller detunings and the second displacement
pulse is more than a factor of $8$ longer than for the gate in
\refcite{schmitz:arch}.
Still, the fidelity exceeds $\unit[95]{\%}$ (see \figref{didi:parity}).

\subsection{Quantum Ising Hamiltonian}
\seclabel{spin:spin}

Above we have introduced $\pauli{z} \otimes \pauli{z}$ interactions that
are used in quantum gates.
In the following we will present a slightly different approach, in which
Ising spin--spin interactions are continuously induced by means of optical
forces.

The spin--spin interaction as proposed in \refcite{porras:eff-spin} and
experimentally realized in the simulation of a quantum Ising Hamiltonian in
\refcite{friedenauer:qmagnet} is identical to the interaction $\HI$
described in \secref{phase:space}.
(Note that a similar proposal involving
the same mathematics is given in \refcite{wunderlich:condspinres}.)
However, the quantum Ising Hamiltonian contains an additional (simulated)
magnetic field pointing in $x$-direction.
We will adapt our notation in this section and split the total interaction
Hamiltonian into the following terms:
$\HI[S]$ denotes the term that generates the spin--spin interaction and
$\HI[M]$ denotes the term leading to the simulated magnetic field.
The index ``$\text{I}$'' of the frequencies $\OmegaI$ and $\omegaI$ etc. is
changed to ``$\text{S}$'' or ``$\text{M}$'' accordingly in the respective
terms.
The complete interaction is described by the Hamiltonian $\HI = \HI[S]
+ \HI[M]$.
In the following, we will first derive the spin--spin interaction Hamiltonian
from $\HI[S]$ focusing on an Ising interaction ($\pauli{z} \otimes \pauli{z}$
only).
Afterwards, we will discuss the magnetic field term $\HI[M]$ and its
effect.

The derivation of the quantum Ising Hamiltonian 
\cite{porras:eff-spin,porras:2d:crystals:long} involves a slightly different
interaction picture compared to \secref{interaction:pic} by substituting $\H_0$
with $\H_\emptyset$:
\begin{equation}
  \begin{aligned}
    \H_0
      &= \H_\text{e} + \H_\text{m} \\
      &= \underbrace{\H_\text{e}
        + \sum_{m = 1}^{3 N} \hbar \omegaI[S] \mode{\ad} \mode{\a}}%
        _{\eqdef \H_\emptyset}
        \underbrace{- \sum_{m = 1}^{3 N} \hbar \mode{\delta} \mode{\ad}
        \mode{\a}}_{\eqdef \H_\delta}.
  \end{aligned}
\end{equation}
The term $\H_\delta$ is added to the interaction Hamiltonian.

To retrieve the representation of $\HI[S]$ in the newly defined
interaction picture,
\begin{equation}
  \HI[S]' \defeq \U_\emptyset^\dagger \HI[S] \U_\emptyset
  \ms \text{with} \ms \U_\emptyset \defeq \ee^{-\ii \H_\emptyset t / \hbar},
\end{equation}
we adapt the calculations from
\secsref{interaction:pic}{phase:space} accordingly:
The frequencies in the transformation \eqref{h:i:trans:a} are changed to
$\mode{\omega} \to \omegaI[S]$.
As a result, the substitution $\ee^{\pm \ii \mode{\omega} t}
\to \ee^{\pm \ii \omegaI[S] t}$ has to be applied to \eqref{h:i:i} (and
subsequent equations) and $\ee^{\pm \ii \mode{\delta} t} \to 1$ to
\eqref{rwa:ldr}.
Hence, $\HI[S]$ reads in the new interaction picture (including the
expansion to first order in the Lamb-Dicke parameters and the RWA):
\begin{multline}
  \rwa{\ldr{\HI[S]'}} \\
  = \sum_{i = 1}^N \sum_{m = 1}^{3 N}
  \ii \hbar \site{\OmegaI[S]} \site{\mode{\eta}}
  \ee^{\ii \site{\phiI[S]}} \mode{\ad}
  \left(\alpha_0 \site{\I} + \alpha_3 \site{\pauli{z}}\right) + \hc
\end{multline}
However, the full Hamiltonian in the interaction picture now also involves
\begin{equation}
  \H_\delta' = \U_\emptyset^\dagger \H_\delta \U_\emptyset = \H_\delta,
\end{equation}
where the transformation is the identity, because trivially
$\comm{\H_\emptyset}{\H_\delta} = 0$.

To gain the form of a spin--spin interaction, we apply a canonical
transformation (\cmp{} \refcite{porras:eff-spin}) to the Hamiltonian:
\begin{equation}
  \rwa{\ldr{\HI[S]''}} + \H_\delta'' \defeq
  \U_\text{c} \left(\rwa{\ldr{\HI[S]'}}
  + \H_\delta\right) \U_\text{c}^\dagger
\end{equation}
with
\begin{equation}
  \U_\text{c} \defeq \exp\left(- \sum_{i = 1}^N \sum_{m = 1}^{3 N}
  \frac{1}{\hbar \mode{\delta}} \left[
  \site{\mode{\op{\xi}}} \mode{\ad} - \site{\mode{\op{\xi}}}^\dagger \mode{\a}
  \right]\right)
\end{equation}
and
\begin{equation}
  \site{\mode{\op{\xi}}} \defeq \ii \hbar \site{\OmegaI[S]} \site{\mode{\eta}}
  \ee^{\ii \site{\phiI[S]}}
  \left(\alpha_0 \site{\I} + \alpha_3 \site{\pauli{z}}\right).
\end{equation}
Using the calculations from \secref{canonical:trans}, the transformed
Hamiltonian reads:
\begin{widetext}
  \begin{equation}
    \eqlabel{h:spin:spin}
    \begin{aligned}
      \rwa{\ldr{\HI[S]''}} + \H_\delta''
        &= \sum_{i = 1}^N \sum_{j = 1}^N \sum_{m = 1}^{3 N}
          \frac{1}{\hbar \mode{\delta}} \site{\mode{\op{\xi}}} \otimes
          \site[j]{\mode{\op{\xi}}}^\dagger 
          + \H_\delta \\
        &= \hbar \sum_{i = 1}^N \sum_{j = 1}^N \sum_{m = 1}^{3 N}
          \frac{\site{\OmegaI[S]} \site[j]{\OmegaI[S]} \site{\mode{\eta}}
          \site[j]{\mode{\eta}}}{\mode{\delta}}
          \ee^{\ii \left(\site{\phiI[S]} - \site[j]{\phiI[S]}\right)} 
          \left(\alpha_0 \site{\I} + \alpha_3 \site{\pauli{z}}\right) \otimes
          \left(\alpha_0 \site[j]{\I} + \alpha_3 \site[j]{\pauli{z}}\right)
           + \H_\delta
    \end{aligned}
  \end{equation}
\end{widetext}
The Hamiltonian can be expanded into a pure $\pauli{z} \otimes \pauli{z}$
interaction, a ``bias'' term with $\pauli{z}$ interaction, and a constant term
that can be neglected.
The ``bias'' term acts as a longitudinal magnetic field and leads to a
deviation from the quantum Ising model.
At first glance, this is not desired and it will be treated as an error in the
following discussion.
However, by including a ``bias'' term in a controlled way we could also explore
an extended phase diagram with the longitudinal field as an additional
parameter.

We want to stress the similarity between the spin--spin interaction according
to \eqref{h:spin:spin} and the $\pauli{z} \otimes \pauli{z}$ interaction
discussed in \secref{phase:space}.
The canonical transformation has the form of a displacement operator and
looks very similar to the first exponential function in \eqref{disp:geo:phase}
(except for the time dependence of the latter).
The similarity to the geometric phase term in \eqref{disp:geo:phase} can be
best seen comparing the time evolution operators.
As the Hamiltonian in \eqref{h:spin:spin} is time-independent, the time
evolution simply reads
\begin{widetext}
  \begin{equation}
    \eqlabel{u:spin:spin}
    \begin{aligned}
      \rwa{\ldr{\UI[S]''}}(t, t_0) \times \U_\delta(t, t_0)
        &= \exp\Biggl(- \ii \sum_{i = 1}^N \sum_{j = 1}^N \sum_{m = 1}^{3 N}
          \frac{\site{\OmegaI[S]} \site[j]{\OmegaI[S]} \site{\mode{\eta}}
          \site[j]{\mode{\eta}}}{\mode{\delta}^2} 
          \left(\alpha_0 \site{\I} + \alpha_3 \site{\pauli{z}}\right) \otimes
          \left(\alpha_0 \site[j]{\I} + \alpha_3 \site[j]{\pauli{z}}\right) \\
        &\quad \times \mode{\delta} (t - t_0)
          \ee^{\ii \left(\site{\phiI[S]} - \site[j]{\phiI[S]}\right)}\Biggr)
          \times \exp\left(\ii \sum_{m = 1}^{3 N} \mode{\delta} (t - t_0)
          \mode{\ad} \mode{\a} \right).
    \end{aligned}
  \end{equation}
\end{widetext}

Before we can apply the easier time evolution of \eqref{u:spin:spin}, in which
electronic states are decoupled from motional states, the state vector
$\ket{\psi}$ has to be transformed from the original picture to
$\ket{\psi}'' \defeq \U_\text{c} \ket{\psi}$.
As $\U_\text{c}$ depends on the electronic state, the transformation will in
general lead to an entangled state and the canonical transformation can be
interpreted as dressed-state picture (electronic states ``dressed'' with
motional states).
As the states for the simulation of the quantum Ising Hamiltonian
are prepared in the original (undressed) picture, but the Hamiltonian acts in
the dressed picture, an error is introduced into the simulation
(see, \eg{}, \refcite{deng:eff-spin:err}).
However, as long as the effect due to $\U_\text{c}$ is small
($|\site{\OmegaI[S]} \site{\mode{\eta}} \alpha_{l} / \mode{\delta}| \ll 1$), we
can use the approximation $\ket{\psi}'' \approx \ket{\psi}$.
In terms of the geometric phase gate this corresponds to
the case, when the circles in phase space are small and the entanglement
between electronic and motional states can be neglected at any time.

The same holds for the measurements of observables:
They are performed in the original (undressed) picture, in which electronic
states are entangled with the motional states, and in general a further error
is introduced in the simulation.
However, the measurement of the states is typically insensitive to the
motional states and involves a projection to one of the electronic states,
\eg{}, the $\ket{\down}$ state: $\site{\P} \defeq
\site{\ket{\down}}\site{\bra{\down}}$.
As $\comm{\U_\text{c}}{\site{\P}} = 0$, the projector does not change under the
canonical transformation and the readout of $\pauli{z}$ eigenstates (without
any rotations of the bases applied beforehand) does not introduce further
errors.

The form of the ``bias'' term proportional to $\pauli{z}$ can be simplified in
the case of a linear Paul trap with equal Rabi frequencies $\site{\OmegaI[S]}$
and equal phases $\site{\phiI[S]}$ for all ions:
The sum over $j$ extends over the Lamb-Dicke parameters $\site[j]{\mode{\eta}}$
only.
This sum is non-zero only for centre-of-mass modes, for which the
$\site[j]{\mode{\eta}}$ additionally are independent of the site $j$.
Hence, the three sums simplify to a sum over $\site{\pauli{z}}$ with constant
prefactor \cite{porras:eff-spin}:
\begin{equation}
  2 \OmegaI[S]^2 \hbar N \alpha_0 \alpha_3 \left(\sum_{m \in \set{\text{c.m.}}}
  \frac{\mode{\eta}^2} {\mode{\delta}}\right) \sum_{i = 1}^N \site{\pauli{z}}.
\end{equation}
However, this simplification does not necessarily hold for two-dimensional
arrays of individual traps for each ion.

In the following, we will discuss the magnetic field term, which originates
from a $\pauli{x}$ interaction described by \eqref{rwa:xy}
(with $\alpha_1 = 1$ and $\alpha_2 = 0$).
In principle, we have to apply the substitution
$\ee^{\pm \ii \mode{\omega} t} \to \ee^{\pm \ii \omegaI[S] t}$ due to the new
interaction picture here, too.
However, we consider a magnetic field term without motional dependence
in the following ($\site{\mode{\eta}} = 0$, \cmp{} \secref{rabi:sideband}) and
thus the terms containing the motional creation/annihilation operators vanish:
\begin{equation}
  \eqlabel{mag:field:i}
  \rwa{\HI[M]'} = \sum_{i = 1}^N \frac{\hbar}{2} \site{\OmegaI[M]}
  \ee^{\ii \left(-\left(\omegaI[M] - \omega_\updown\right) t
  + \site{\phiI[M]}\right)} \site{\pauli{+}} + \hc
\end{equation}

The canonical transformation can be rewritten as
\begin{equation}
  \U_\text{c} = \exp\left(\ii \sum_{i = 1}^N \site{\op{h}}
  \left(\alpha_0 \site{\I} + \alpha_3 \site{\pauli{z}}\right) \right)
\end{equation}
with the Hermitian operator
\begin{equation}
  \site{\op{h}} \defeq \sum_{m = 1}^{3 N}
  \left[\site{\mode{\zeta}} \mode{\ad} +
  \site{\mode{\zeta}}^\dagger \mode{\a} \right]
\end{equation}
and
\begin{equation}
  \site{\mode{\zeta}} \defeq - \frac{\hbar \site{\OmegaI[S]}
  \site{\mode{\eta}} \ee^{\ii \site{\phiI[S]}}}{\hbar \mode{\delta}}.
\end{equation}
Trivially, the commutator $\comm{\site{\op{h}}}{\site{\pauli{+}}} = 0$.
The canonical transformation of $\rwa{\HI[M]'}$
(see \eqref{mag:field:i}) is thus equivalent to a transformation of the
$\site{\pauli{\pm}}$ operator as in \eqsref{pauli:trans:p}{pauli:trans:m}:
\begin{equation}
  \begin{aligned}
    \rwa{\HI[M]''}
      &= \U_\text{c} \rwa{\HI[M]'} \U_\text{c}^\dagger \\
      &= \sum_{i = 1}^N \frac{\hbar}{2} \site{\OmegaI[M]}
        \ee^{\ii \left(- \left(\omegaI[M]
        - \omega_\updown\right) t + \site{\phiI[M]}\right)}
        \ee^{2 \ii \alpha_3 \site{\op{h}}} \site{\pauli{+}} \\
      &\quad + \hc
  \end{aligned}
\end{equation}
The expansion to first order in $\site{\mode{\zeta}}$ (and thus to
first order in $\site{\mode{\op{\eta}}}$) yields:
\begin{equation}
  \eqlabel{h:eff:b}
  \begin{aligned}
    \rwa{\HI[M]''}
      &\approx \sum_{i = 1}^N \frac{\hbar}{2} \site{\OmegaI[M]}
        \ee^{\ii \left(- \left(\omegaI[M]
        - \omega_\updown\right) t + \site{\phiI[M]}\right)} \\
      &\quad \times \left(1 + 2 \ii \alpha_3 \sum_{m = 1}^{3 N}
        \left[\site{\mode{\zeta}}
        \mode{\ad} + \site{\mode{\zeta}}^\dagger \mode{\a} \right]\right)
        \site{\pauli{+}} \\
      &\quad + \hc \\
      &\eqdef \rwa{\HI[M]'} + \HI[E]''.
  \end{aligned}
\end{equation}
The magnetic field term after the canonical transformation deviates to order
$\site{\OmegaI[S]} \site{\mode{\eta}} \alpha_3 / \mode{\delta}$ due to
$\HI[E]''$ from the pure
$\pauli{x}$ interaction $\rwa{\HI[M]'}$.
This introduces an additional error in the simulation.
If the condition $|\site{\OmegaI[S]} \site{\mode{\eta}} \alpha_{l}
/ \mode{\delta}| \ll 1$ is met, it can be small or even negligible and we
effectively will obtain the desired magnetic field term.

To summarize, the complete Hamiltonian is obtained by adding
\eqref{h:spin:spin} and \eqref{h:eff:b}.
It consists of a spin--spin interaction term and a simulated magnetic
field pointing in $x$-direction, which add up to the ideal quantum Ising
Hamiltonian.
Assuming a resonant interaction for the simulated magnetic field ($\omegaI[M] -
\omega_\updown = 0$) and neglecting the phases
($\site{\phiI[S]} = \site{\phiI[M]} = 0$), the quantum Ising part can be written
\begin{equation}
  \eqlabel{h:ising}
  \begin{aligned}
    \HI[QIsing]
      &\defeq \HI[B] + \HI[J] \\
      &= \sum_{i = 1}^N \site{B_x} \site{\pauli{x}}
        + \sum_{i = 1}^N \sum_{\substack{j = 1 \\ j \neq i}}^N
        \site[i, j]{J} \site{\pauli{z}} \otimes \site[j]{\pauli{z}},
  \end{aligned}
\end{equation}
where
\begin{equation}
  \eqlabel{b:j:ising}
  \site{B_x} \defeq \hbar \site{\OmegaI[M]} \ms \text{and} \ms
  \site[i, j]{J} \defeq \hbar \sum_{m = 1}^{3 N}
  \frac{\site{\OmegaI[S]} \site[j]{\OmegaI[S]}
  \site{\mode{\eta}} \site[j]{\mode{\eta}}}{\mode{\delta}} \alpha_3^2.
\end{equation}
(Note that the superscripts of $\site{B_x}$ and $\site[i, j]{J}$ indicating
the site will be omitted in the following sections, if the interaction
strengths for all ions are equal.)
In addition to the Ising part we obtain the following terms that lead to a
deviation from the ideal model (constant terms are omitted):
\begin{equation}
  \HI[Error] = 2 \sum_{i = 1}^N \sum_{j = 1}^N
  \frac{\alpha_0}{\alpha_3} \site[i, j]{J} \site{\pauli{z}}
  + \H_\delta + \HI[E]''.
\end{equation}
The first ``bias'' term is further discussed in the context of the
experimental realization, see \refcite{friedenauer:qmagnet} and
\secref{many:body:exp}.
The second term $\H_\delta$ can be interpreted as an energy offset, which
cancels by applying an appropriate redefinition of the energy scale.
As mentioned above, the last term leads to only a small or even negligible error
for $|\site{\OmegaI[S]} \site{\mode{\eta}} \alpha_{l} /
\mode{\delta}| \ll 1$.
For a more detailed discussion of the errors in the simulation of quantum spin
Hamiltonians we refer the reader to \refcite{deng:eff-spin:err}.

We want to emphasized that $\comm{\HI[B]}{\HI[J]} \neq 0$, such that the time
evolution of the total quantum Ising Hamiltonian $\HI[QIsing]$ cannot be
simply described by the time evolutions of $\HI[B]$ and $\HI[J]$ separately.

\ultrashortmacrosoff 


\ultrashortmacroson

\section{Operations Interpreted for Experimental QS}
\seclabel{interpretation}

To realize a QS for a quantum spin Hamiltonian, we have to (1)
simulate the spin, provide (2) its initialization and (3) the
interaction of this ``spin'' with a simulated magnetic field, (4)
realize an interaction between several spins (spin--spin interaction),
and (5) allow for efficient detection of the final spin state.
Additional diversity for QS arises by the capability of
precise initialization, control and readout of the motional states.

The mathematical derivation and description of the individual
operations have been described in \secref{theory}.
In this section, we explain in a simplified pictorial way the related
generic building blocks in terms of an adiabatic QS of a quantum spin
Hamiltonian within a linear chain of ions.
No specific ion species or trapping concept is required.
A well-suited system to illustrate the generic requirements and to
investigate the feasibility of QS in ion traps is given by the
quantum Ising Hamiltonian (see \eqref{h:ising}).
We want to note that the building blocks already suffice to implement a whole
family of quantum spin Hamiltonians.

\subsection{Simulating the Spin}

The mutual distance between the ions/spins in linear RF traps is typically
of the order of several micrometres (see \figref{coulomb:crystals}).
Therefore, the direct interaction between their electronic
states remains negligible, which is advantageous, because the related
interaction strength could hardly be tuned or even switched off.
Therefore, the spin-$1/2$ states are implemented like qubit states (see
\secref{tools:ions}).

\subsection{Simulating the Magnetic Field}

Implementing an artificial spin allows to shape artificial fields to
implement a precisely controllable interaction and related dynamics
between the ``spin'' and the ``field''.
To simulate an effective magnetic field, the two electronic
states $\ket{\down}$ and $\ket{\up}$ are coupled via electro-magnetic
radiation (see \secref{tools:operations}, operation (a)).
The related coherent oscillation of the state population between the two
levels can be described in terms of Rabi flopping.
In the Bloch sphere picture, the tip of the electronic
state vector rotates during one flop continuously from state $\ket{\down}$ to
$\ket{\up}$ and vice versa.
For continuous coupling this can be interpreted as the precession
of a spin exposed to a perpendicular magnetic field.

The rotation matrix in \eqref{rot:mat} exactly describes this interaction with
a single spin (see also \secsref{rabi:sideband}{phase:gate}).
For example, if we start with $\ket{\down}$ and apply a pulsed rotation
$\R(\pi /2, \pi / 2)$, we will obtain an eigenstate of $\pauli{x}$, which is
abbreviated by $\ket{\rgt} \defeq 1/\sqrt{2} (\ket{\down} + \ket{\up})$.
In the Bloch picture, this corresponds to a $\unit[90]{\degree}$ rotation
of the Bloch vector around the $y$-axis, such that it will point in direction
of the $x$-axis.
Continuing with a second identical rotation we just flip the spin to
$\ket{\up}$ as if we applied $\R(\pi, \pi / 2)$ or a $\unit[180]{\degree}$
rotation around the $y$-axis, respectively.
However, we can replace the second operation by $\R(\pi / 2, 0)$, which
corresponds to a rotation around the $x$-axis.
As the state $\ket{\rgt}$ is an eigenstate of $\sigma_x$ it will not be
affected.

Stroboscopic rotations have been introduced in \secref{phase:gate} to implement
single-qubit gates for a QC.
Continuous versions of these single-qubit operations can be interpreted in the
context of analogue QS as simulated magnetic field (first term of
\eqref{h:ising}).

\subsection{Simulating the Spin--Spin Interaction}

Let us first discuss the implementation of a basic spin--spin
interaction close to the original proposal in \refcite{porras:eff-spin}:
Two ions are confined in a linear RF trap and a standing wave provides
state-dependent dipole forces.
The ions are located at same phases ($\site{\phi} = 0$), such that ions in
different spin states are pulled/pushed in opposite directions.

If both ions are in the same spin state, they will be pulled in the same
direction.
Hence, their mutual distance and mutual Coulomb energy, respectively, remains
unchanged.
However, if the two spins are in different states, one ion will be pulled
and the other one pushed.
Their mutual distance and as a result their mutual Coulomb energy will change.
This is exactly the essence of a spin--spin interaction, where the energy
corresponding to a spin state depends on the states of its neighbours.
To interpret interactions as ferromagnetic or anti-ferromagnetic
it is advantageous to consider the mutual Coulomb energy in longer chains
of spins (see \figref{sw:ferro:antiferro}).

\figurehandler{\myfigureferroantiferro}

The technical realization in \refcite{friedenauer:qmagnet} avoids the
difficulties arising from standing waves and resonantly enhances the
interaction strengths by implementing the spin--spin interactions with
stimulated-Raman transitions as in the case of quantum gates
\cite{porras:2d:crystals:long} (see \secref{phase:gate}).
In a pictorial interpretation, the standing waves are replaced by ``walking''
waves and instead of static displacements we obtain driven oscillations of the
ions.
However, the mathematical description yields exactly the same spin model
in an appropriately chosen frame (see \secref{spin:spin}).
The sign of $J$ can additionally be changed by choosing a
different sign for the detunings $\mode{\delta}$ from the modes
(see \eqref{b:j:ising}).

\subsection{Geometric Phase Gate versus Adiabatic QS}

It might be helpful to emphasize the differences and
similarities of the interactions in QC and analogue QS:
To realize a phase gate operation on the radial modes of two qubits, as
described in \secref{phase:gate}, typically one or two isolated modes of
motion are selected.
The small detuning from the modes is chosen to obtain comparatively large
interaction strengths and thus motional excitations.
For ions being initialized in the motional ground state the displacements
in the respective phase space(s) lead to an average phonon number
$\mean{n} \sim 1$ and to a significant entanglement between electronic and
motional states at intermediate times.
However, this entanglement can be withdrawn until the end of the gate and
maximized the final entanglement between the qubit states only
(see \secref{phase:gate}).

In contrast, we consider an adiabatic evolution according to the quantum
Ising Hamiltonian in the case of analogue QS (see \secref{many:body:exp} for
the experimental protocol).
We have to make sure that the entanglement between electronic and
motional states remains small at any time during the simulation (see discussion
of errors in \secref{spin:spin}).
Additionally, running the simulation on many spins simultaneously will
result in contributions from many motional modes simultaneously.
As a result, a large detuning from all modes has to be chosen, such that the
difference of the radial frequencies can be neglected and a net effect from
all modes remains.
Choosing the right parameters allows to simulate spin--spin interactions of
different strength, different signs and even range of interaction
\cite{porras:eff-spin}.

Furthermore, a scan of the duration of the displacement pulses $T_D$ in
geometric phase gates leads to a periodic evolution from $\ket{\down\down}$ to
$\ket{\up\up}$ and vice versa (see \figref{didi:evol}).

In contrast, the distinct contributions ($\HI[B]$ and $\HI[J]$) of the quantum
Ising Hamiltonian are not stroboscopically alternated but applied
simultaneously.
As mentioned in \secref{spin:spin}, the time evolution according to the quantum
Ising Hamiltonian is not simply the time evolution according to $\HI[B]$ and
$\HI[J]$ separately.
As a consequence, applying the spin--spin interaction for a longer duration
and/or increasing its strength does not alter the state anymore.

\subsection{Note on Simulating (Virtual) Particles for QS}

Up to now we summarized how the tools developed for QC can be adapted,
used and interpreted as tools for analogue QS.
However, the toolbox for QS is substantially larger (see also
\secref{many:body:theo}).
(1) Phonons do not have to be restricted to mediate interactions in QC
and QS:
They where also proposed to simulate bosons, for example atoms in the
Bose--Hubbard model \cite{porras:bosehubbard} or charged particles
\cite{bermudez:gauge}.
(2) Topological defects in the zigzag structure of two-dimensional
Coulomb crystals (see \figref{kink}) are proposed to simulate solitons
\cite{landa:kink}.

\ultrashortmacrosoff


\ultrashortmacroson

\section{Towards Simulating Many-Body Physics}
\seclabel{many:body}

In the first part of this section we want to assemble the building
blocks described above to illustrate, how an analogue QS of a
quantum spin Hamiltonian can be implemented.
For this purpose, we will describe the realization of first
proof-of-principle experiments on the quantum Ising Hamiltonian (see
\eqref{h:ising}).
In the second part we aim to summarize, to the best of our
knowledge, the existing proposals addressing many-body physics with
the described and available toolbox.

\subsection{Proof-of-Principle Experiments on Quantum Spin Hamiltonians}
\seclabel{many:body:exp}

\figurehandler{\myfigureqm}

First, we will describe the basic implementation of the experimental protocol
on the axial modes for
the case of two spins \cite{friedenauer:qmagnet}, as illustrated in
\figref{qm}.
Subsequently, we will emphasize the differences and additional information
explored in \refscite{islam:qmagnet}{kim:frustration}.
For the details on the individual experimental parameters we refer to these
references.

For the case of two spins, the protocol has been realized following
five steps:
(1) The two ions are initialized by Doppler cooling, sideband cooling, and
optical pumping (see \secref{tools:readout}) in the state
$\ket{\down\down}\ket{\mode[\text{STR}]{n} = 0}$.
(2) Both spins are prepared by a common $\R(\pi/2, \pi/2)$ rotation
in the $\site{\pauli{x}}$ eigenstate $\ket{\rgt \rgt}
\ket{\mode[\text{STR}]{n} = 0}$.
(3) An effective magnetic field of amplitude $B_x$ is
applied equivalent to a continuous $\R(2 \OmegaI[M] t, 0)$ rotation (see
\eqsref{h:ising}{b:j:ising}).
At this step, the state $\ket{\rgt \rgt}\ket{\mode[\text{STR}]{n} = 0}$
represents the ground state of the first term of the quantum Ising Hamiltonian
in \eqref{h:ising} that can be ``easily'' prepared.
Note that the rotation is slightly off-resonant to mimic an additional
$\pauli{z}$ interaction counteracting the ``bias'' field (see also
\refcite{friedenauer:qmagnet}).
(4) The effective spin--spin interaction $J$ is ramped up adiabatically with
respect to the timescale $1 / \OmegaI[M]$ defined by the simulated magnetic
field, until $|J| \gg B_x$.
The system adiabatically evolves into its new ground state, which is
an equal superposition of the two energetically preferred states of the
ferromagnetic order:
$1 / \sqrt{2} (\ket{\down\down} + \ket{\up\up})$.
(5) Finally both interactions are switched off.
The readout of the final spin state is performed by state dependent
detection.
This projects the spin state to one out of the four eigenstates
of the measurement basis ($\ket{\down\down}, \ket{\down\up},
\ket{\up\down}, \ket{\up\up}$).
Steps (1) to (5) are repeated many times to obtain the populations related
to these states.

To investigate the degree of entanglement of the final spin state, an
additional parity measurement is performed as in the case of the geometric
phase gates (see \secref{phase:gate}).
The populations of $\ket{\down\down}$ and $\ket{\up\up}$ in dependence of
$|J| / B_x$ and the entanglement fidelity are summarized in \figref{qm}.

The experimentally observed entanglement of the final
states confirms that the transition from paramagnetic to ferromagnetic
order is not caused by thermal fluctuations that drive thermal phase
transitions, but by the so-called quantum fluctuations
\cite{sachdev:quantum:phase:trans:a,sachdev:criticality}
driving QPTs in the thermodynamic limit at zero temperature.
In this picture tunnelling processes induced by $B_x$ coherently
couple the degenerate states $\ket{\up}$ and $\ket{\down}$ with an amplitude
$\propto B_x / |J|$.
For $N$ spins the amplitude for the tunnelling process between
$\ket{\Psi_{N \up}} = \ket{\up\up\dotso\up}$ and $\ket{\Psi_{N \down}} =
\ket{\down\down\dotso\down}$ is proportional to $(B_x / |J|)^N$,
since all $N$ spins must be flipped.
In the thermodynamic limit ($N \rightarrow \infty$) the system
is predicted to undergo a QPT at $|J| = B_x$.
At values $|J| > B_x$ the tunnelling between $\ket{\Psi_{\infty \up}}$ and
$\ket{\Psi_{\infty \down}}$ is completely suppressed.
In our case of a finite system $\ket{\Psi_{2 \up}}$ and $\ket{\Psi_{2 \down}}$
remain coupled and the sharp QPT is smoothed into a gradual
change from paramagnetic to ferromagnetic order (see \figref{qm}).

It has to be noted that the performance of such a simulation on a
large number of spins in a one-dimensional chain requires several
technical improvements.
Recently, the group at the University of Maryland pioneered a substantial step
for scaling by investigating the emergence of magnetism in the quantum
Ising model using up to nine ions \cite{islam:qmagnet}.
To achieve these results they mediated the interactions via the radial
modes of motion \cite{porras:eff-spin,kim:gate:transverse} (see also
\secref{phase:gate}).
Furthermore, they implemented the effective spin--spin interactions in a
rotated frame using M\o{}lmer--S\o{}rensen interactions
\cite{soerensen:quantum:comp,lee:phase:control}
on robust hyperfine clock states.
Thereby, they do not depend on the phases $\site{\phi}$ of the
laser beams at the sites of the ions.
To perform an adiabatic transition, the simulated magnetic field has been
adiabatically turned off, while the effective spin--spin interaction remained
constant.

Their results allow already much more than simply increasing
the number of spins: they enter a new regime of intriguing questions.
The crossover of the quantum magnetization \cite{friedenauer:qmagnet} from
paramagnetic to ferromagnetic order is sharpening as the number of
ions is increased from two to nine, ``prefacing the expected
quantum phase transition in the thermodynamic limit'' \cite{islam:qmagnet}.
Even though the results can still be calculated on a classical computer,
they provide a possibility to critically benchmark QS aiming for only slightly
larger, but intractable systems.

Already increasing the number of ions to three and adapting the
individual spin--spin interactions including their signs allows to address
spin frustration
in the smallest possible magnetic network \cite{kim:frustration}.
Spin frustration of the ground state can be pictorially understood in a
two-dimensional triangular spin lattice featuring anti-ferromagnetic
spin--spin interactions.
Here, it becomes impossible for neighbouring ions to have pairwise opposite
states.
Classically, two ions will adopt different states, while the state of the third
one is undetermined.
During an adiabatic evolution of the quantum mechanical system
(starting from the paramagnetic order) nature
will choose a superposition of all degenerate states, leading to
massive entanglement in a spin-frustrated system.
In the realization of the experiment, the three ions are still trapped
in a one-dimensional chain.
However, almost complete control over the amplitudes and signs of
$\site[1, 2]{J}, \site[2, 3]{J}, \site[3, 1]{J}$ is gained by coupling to
particular collective modes of motion and choosing appropriate detunings
\cite{kim:frustration}.

It has to be mentioned that for an increased number of spins the
energetic gap between ground and excited states further shrinks and the
requirement on adiabaticity enforces longer simulation durations related to
a longer exposure to decohering disturbances.
Still, as mentioned in \secref{intro}, the influence of decoherence might
destroy the entanglement within the system, but this might not be relevant for
the observable of interest.
Here it will be crucial to investigate the role of the decoherence effects with
respect to the specific analogue QS.

With respect to digital (stroboscopic) QS it should be emphasized, that no
quantum error correction is required for proof-of-principle
experiments on a few ions.
Promising results of a ``trotterized'' version of the simulation of the
quantum Ising Hamiltonian with two spins have been performed recently
\cite{blatt:priv:comm}.

\subsection{Systems Featuring Many-Body Physics Proposed for Analogue QS}
\seclabel{many:body:theo}

Condensed matter met atomic, molecular, and optical physics not so
long ago, when trapping techniques for ultracold neutral atoms and
ions allowed experimentalists to generate lattices and crystals,
where models from solid state physics may be implemented.
Combining the fields has led to a very rich interdisciplinary research
activity, as well as to several misunderstandings between scientists
looking at the same system from different points of view.
In the particular case of trapped ion experiments, the outlook for
quantum simulation of many-body models is very exciting, but some
knowledge on the details of this physical system is required to
understand both the limitations and the amazing possibilities of
this setup.

In the following we review many-body models that have the potential
to be simulated with trapped ions.
There have been several contributions both from theory and experiments to this
research line.
Most of them share the common feature that they are inspired by known models
from condensed matter physics, but their implementation with trapped ions
turns out to lead to a rich variety of new physical phenomena, which may even
require new theoretical paradigms that go beyond the conventional ones in the
solid state.
The three main reasons for that are:
(1) Trapped ion experiments are naturally performed in a non-equilibrium
regime, whereas solid state physics typically deals with thermal equilibrium,
(2) trapped ion systems may in principle be controlled and measured at the
single particle level, (3) ion crystals are typically mesoscopic systems,
in the sense that they may reach a number of particles (spins,
phonons, \ldots) large enough to show emergent many-body physics, but
still finite size effects are important.
All those peculiarities have to be kept in mind, since they provide us with
unique features for analogue QS.

\subsubsection{Quantum Spin Models}

Following the experimental advances in QIP, the most natural degree of
freedom to be used for QS seems to be the electronic states for spins and the
phonons to mediate their mutual interactions.
However, one has to identify conditions where interesting phenomena arise,
such as, for example, quantum critical phases.
This has already lead to the promising proof-of-principle experiments
discussed above.

A unique feature that we can exploit with trapped ions is the fact
that the effective spin--spin interactions can be implemented showing
a dipolar decay, $\site[i, j]{J} \propto 1 / |i - j|^3$.
In the case of the Ising interaction, the cubic dependence does not change
the critical universality class of the model, as shown, for example, by the
numerical calculations in \refcite{deng:eff-spin:err}.
However, even in this case, long-range entanglement is induced by the
long-range interaction, which is absent in conventional nearest-neighbour
quantum Ising chains.
On the other hand, when considering other interacting schemes, like the
$xyz$ Hamiltonian, the dipolar interaction may lead to the formation of
quasi-crystalline phases of spin excitations \cite{hauke:devil}.

Several pieces have been added to the toolbox of quantum simulation,
which definitely allow us to explore physics beyond conventional
solid state paradigms.
For example, a theoretical proposal has been presented to implement models,
whose ground states show topological features \cite{milman:topologic:qubits}.
Also, methods to implement three-body spin--spin interactions have been
designed, see \refcite{bermudez:interactions}.
Finally, dissipation in trapped ion systems has been proved to be useful to
engineer quantum phases that arise as steady-state of dissipative
processes \cite{barreiro:open:system}.
The many-body physics of dissipative systems is a much more unexplored area
than equilibrium properties, even for theorists.
For that reason, adding dissipation to quantum magnetism opens an exciting
perspective for trapped ions.

\subsubsection{Interacting Boson Models}

A variety of exciting quantum many-body systems may also be simulated by
using the collective motional degrees of freedom (phonons) to realize models
of interacting bosons.
In particular, whenever the motional coupling between ions is small
compared to the trapping frequency, the phonon number is conserved and
becomes a good quantum number to characterize the quantum state of the system.
This principle was introduced and exploited in \refcite{porras:bosehubbard}
to show that the physics of radial modes in Coulomb chains is effectively
described by a Bose--Hubbard model.
Vibrational couplings between two ions, say $1$ and $2$, induced by the Coulomb
interaction, have a typical form $\propto \site[1]{\op{x}} \site[2]{\op{x}}$,
where $\site{\op{x}}$ is the ion displacement operator.
Under the approximation of phonon number conservation, those terms become
tunnelling couplings of the form $(\mode[1]{\ad} \mode[2]{\a} + \hc)$.
The same idea applies to quartic anharmonicities of the trap, which yield
Hubbard interactions, $(\mode{\ad} \mode{\a})^2$.
Anharmonicities may be induced and controlled with optical forces, as shown in
\refcite{porras:bosehubbard}.
This analogy between phonons and interacting bosons opens an exciting
avenue of research, where experiments might be relevant even with a
single ion, realizing a single anharmonic quantum oscillator.

The ground state of those phonon--Hubbard models in Coulomb chains
was extensively studied in \refcite{deng:quantum:phases}, where it was
shown that phonon Luttinger liquid phases may arise.
Very recent experiments show indeed the tunnelling of phonons between ions
trapped by different potentials, realizing thus an important step
towards the use of phonons for quantum simulation 
\cite{brown:oscillators,harlander:antennae}.
Exploiting phonon tight-binding models has been also shown to allow us to
implement models with disorder showing Anderson localization
\cite{bermudez:localization}, as well as synthetic gauge potentials by
using periodic driving of the trap frequencies, see \refcite{bermudez:gauge}.
Dipole forces acting on ions confined in a microtrap array (see
\secref{surface:traps}), motional couplings can be controlled such that
phonons experience synthetic gauge fields.
This idea would lead to the simulation of magnetism in quantum lattices with
trapped ions.

\subsubsection{Spin--Boson Models}

The natural convergence of the proposals presented above leads to the quantum
simulation of spin--boson models.
This is a paradigmatic model for quantum impurities in solids, which typically
describes a single spin coupled to a continuous bath of harmonic oscillators
with a power-law spectral density.
Surprisingly, the coupling of the electronic levels of a single ion to the
axial phonons of a Coulomb chain yields a spin--boson model with a quasi-ohmic
spectral density \cite{porras:spin:boson}.
The physics to be simulated here is equivalent to some celebrated models in
condensed matter physics, such as the Kondo effect.
The finite size effects that are intrinsic of trapped ion systems, turn out
to yield features beyond the conventional physics of these models, in particular
quantum revivals associated with the reflection of vibrational waves along the
chain.
Quite recently, it has been proposed to study a situation in which spins and
phonons are coupled, in such a way that a Jaynnes--Cummings--Hubbard model is
simulated \cite{ivanov:quantum:phase:trans}.
In this model phonons follow a tight-binding Hamiltonian and, in addition,
they are locally coupled to spins.
The system has been shown to undergo a superfluid--Mott insulator QPT.

\subsubsection{Inhomogeneous Many-Body Models: Impurities and Topological
Defects}

\figurehandler{\myfigurekink}

The tools for QS in ion traps are not restricted to electronic and motional
degrees of freedom only.
It has been proposed to exploit impurities in the Coulomb crystal.
On the one hand, for example, by embedding ion(s) of a different species
(different mass) into the crystal and taking advantage of the altered spectrum
of the modes and scattering of phonons \cite{ivanov:doped:crystals} and the
option to include larger simulated spins ($S > 1/2$) \cite{ivanov:qmagnet}.
On the other hand, by creating localized topological defects within the more
dimensional structure of the Coulomb crystal (see \figref{kink}).
In \refcite{landa:kink} it was suggested to induce a structural phase
transition from a linear chain of ions (see \figref{coulomb:crystals}b) to a
zigzag structure (see \figref{coulomb:crystals}c), for example, by lowering
the radial confinement.
Changing the parameters in a non-adiabatic way (fast compared to the phonons
mediating information within the crystal) should cause independent domains of
``zigzag'' and ``zagzig'' structure, respectively.
At their clash, topological defects were predicted and have recently been
observed (see \figref{kink}).
The number of the created defects should scale according to the
Kibble--Zurek prediction
\cite{kibble:topology,zurek:cosmology:helium,campo:kinks}.
The defects themselves can be interpreted as solitons \cite{landa:kink}.
Solitons are defined as localized solutions of nonlinear systems,
which depend essentially on the nonlinearity.
Such solitons have a unique spectrum of frequencies with modes which are
localized to the soliton and whose frequency is separated by a gap from the
other phonons.
A quantum mechanical time evolution of these modes was calculated numerically
and it is expected to remain coherent for hundreds of oscillations
\cite{marcovitch:frenkel:kontorova}.
QS could allow to explore their potential applications for QIP
\cite{landa:kink} as well as the quantum behaviour of these ``objects''
themselves.
Solitons appear in all branches of the natural sciences and have been
extensively investigated in solid state systems \cite{kartashov:solitons}.
Among others, classical solitons were observed in waveguide arrays
\cite{eisenberg:solitons,fleischer:solitons} and Bose--Einstein condensates
\cite{trombettoni:solitons}, where they are mean field solutions.
Discrete solitons were investigated in the Frenkel--Kontorova (FK)
model \cite{frenkel:kontorova,kivshar:frenkel:kontorova}, which describes
chains of coupled particles interacting with a local nonlinear potential.
In a different realization, a variant of the FK model can also
be realized in the ion trap by adding an optical lattice to a linear chain
\cite{garcia:mata:frenkel:kontorova,benassi:nanofriction,%
pruttivarasin:optical:lattice}.

The important requirement to address any of these intriguing models will be to
increase the number of ions and the dimensionality of the system.
Trapping ions in two-dimensional arrays would allow to study hard-core boson
phases, showing the effect of frustration, quantum spin liquid phases, and
quantum states with chiral ordering \cite{schmied:ion:optical:lattice}.
Two approaches for scaling will be described in more detail in the following
sections.

\ultrashortmacrosoff


\section{Scaling Analogue QS in Arrays of RF Surface Electrode Traps}
\seclabel{surface:traps}

One possible way to overcome the limitations on scalability of trapped ions
in a common potential well (see \secref{tools:traps}) is to store them in an
array of individual RF traps.

\figurehandler{\myfiguretwodproposal}

\subsection{One-Dimensional RF Surface Electrode Traps}

Conventional RF traps with their three-dimensional geometry of electrodes (see
\figref{rf:trap}) individually fabricated with conventional machining were
unique ``masterpieces'' with unique characteristics.

In 2005 and 2006, the group at NIST pioneered the miniaturization of
RF traps by projecting the electrodes onto a surface
\cite{chiaverini:surface-trap,seidelin:surface-trap} (see
\figref{two:d:proposal}a.), very similar to chip traps for neutral atoms
\cite{folman:chip:traps}.
Introducing photo-lithographic techniques for the trap fabrication opened up
exceptional precision and the production of small series of identical traps,
see \eg{} \refscite{stick:surface:electrode:trap:arxiv}{moehring:y:junction}.
Within these linear RF surface electrode traps, motional ground state cooling
was achieved at a height of the ion over the electrode surface of
$h = \unit[40]{\micro m}$ and with a comparatively small motional heating rate
of the order of $\unit[1]{quantum/ms}$ \cite{seidelin:surface-trap}.

Motional heating rates scale with $\sim h^{-4}$ \cite{turchette:heating}.
The exact heating mechanisms are not yet fully understood and the groups at
NIST, in Berkeley, at MIT and others currently put a lot of effort in
further investigations.
However, the groups at MIT \cite{labaziewicz:heating}, NIST, and the University
of Maryland demonstrated a significant reduction of the heating rates in
cryogenic (surface electrode) traps for QC purposes (see also
\refcite{wineland:bible}).
The inverse of these heating rates is long compared to typical operational
durations of a QC of tens of microseconds (see also \secref{phase:gate}).

For scaling towards a universal QC it might suffice to interconnect
linear ion traps via junctions on a two-dimensional surface to a
network of one-dimensional traps \cite{wineland:review:qip}, realizing the
``multiplex ion trap architecture'' \cite{kielpinski:architecture}.
That is, ions are proposed to be shuttled between processor and memory
traps only interacting in the processor traps.
This would allow to subdivide the large total number of ions into small
groups in many individual traps and to reduce the local requirements to a
technically manageable effort.
One-dimensional RF surface electrode traps with more than 150 individual DC
electrodes and several junctions have been realized \cite{wineland:review:qip},
allowing to shuttle ions at moderate heating rates.

In addition the opportunity arose to deliver identical traps to
different groups.
One example is the linear RF surface electrode trap (denoted by ``Sandia Linear
Trap'' in the following) \cite{stick:surface:electrode:trap:arxiv}, which was
designed by the groups of Oxford, Innsbruck and Sandia National Laboratories.
The latter fabricated a small series of identical replicas.
The traps have been tested in several laboratories and the individually
measured trapping parameters are in good agreement with the
design values.
Publications are in preparation by the groups at Oxford and Sandia (see also
\refcite{stick:surface:electrode:trap:arxiv}).

It has to be emphasized that pursuing the multiplex approach for
scaling universal QC is not applicable to the proposed analogue QS, where
the ensemble of spins is supposed to evolve uniformly as a whole.

\subsection{Optimized Two-Dimensional Arrays of RF Surface Electrode Traps}

Shortly after the invention of RF surface electrode traps it was proposed to
concatenate linear traps sufficiently close, such that the
ions experience mutual Coulomb interaction in two dimensions
\cite{schaetz:qsim} (see \figref{two:d:proposal}b).
However, for a real two-dimensional lattice at sufficiently small and
uniform ion distances of $d \leq \unit[40]{\micro m}$ in two dimensions,
this proposal requires the ions to approach the disturbing surface
to $h \leq d / 2 = \unit[20]{\micro m}$ \cite{chiaverini:surface-trap}.

R.~Schmied \etal{} implemented a method to calculate
the global optimum of the electrode shapes for arbitrary trap locations and
curvatures (originally only for periodic boundary conditions)
\cite{schmied:optimal}.
The gaps between neighbouring electrodes were neglected.
The authors exemplary optimized a trap array with comparatively stiff
horizontal confinement.

The idea of optimizing electrode structures can also be used for designing
traps for analogue QSs with partially converse requirements.
In a collaboration of R.~Schmied, NIST, Sandia National Laboratories, and us,
such a surface electrode trap has been designed and is currently in fabrication.
The trap will provide three trapping zones arranged in a triangle (similar
to \figref{triangle:height}) and is intended as a first step towards larger
arrays of ions.
For this purpose, the optimization method was extended to finite-sized traps.

It has to be emphasized that there are currently several proposals and
approaches for arrays of surface electrode traps mainly for QC.
Groups in Berkeley and Innsbruck aim at trap arrays with individually
controlled RF electrodes.
They have the advantage of selectively lowerable trap frequencies for individual
traps and thus increasable interaction strengths between ions in different
traps, while especially the height of the ions above the surface can be larger
and the trap depth of other traps can be sustained
\cite{kumph:rf:array} (see also discussion in the following subsections).
This approach can in principal be extended to quasi micromotion-free shuttling
of ions in
arrays of RF traps \cite{karin:switable:rf} at the expense of a precise control
of the RF voltage for each RF electrode.
Another proposal suggests individual coils to be included for each trap to
allow for laser-less interactions mainly for QS \cite{chiaverini:laserless}.
Different trap geometries specifically for QSs are designed in the group in
Sussex \cite{hennsinger:priv:comm}.
Arrays of Penning traps with surface electrodes are advanced by the groups at
Imperial College \cite{castrejon:penning} and the University of Mainz
\cite{hellwig:penning}.

In the following subsections we discuss the optimization goals for a surface
electrode trap for an analogue QS, their implications and the perspectives
for scalability of this approach.

\subsubsection{Maximization of Interaction Strengths}

The crucial prerequisite for QSs is to maximize the interaction strength
(see \secref{theory}), however, opposed to QC between individually trapped
ions, while still outrunning decoherence rates.
The increased mutual ion distances in arrays of individual traps
substantially reduces the strength of the effective spin--spin interaction.
It has be taken into account that the conditional forces have a limited
strength, \eg{}, because the laser power is limited or the
assumptions in the theoretical model impose constraints as for the quantum
Ising Hamiltonian (see \secref{spin:spin}).
However, a reduced stiffness of the individual potentials compared to the
example \cite{schmied:optimal} (trap frequencies on the order of
$2 \pi \times \unit[20]{MHz}$) results in larger displacements of the ions
by the same forces.
This is related to an increased mutual Coulomb energy and thus larger
interaction strengths.
Still, a lower bound for the trap frequencies (on the order of
$2 \pi \times \unit[1]{MHz}$ for $\atom{Mg}[+]$) is imposed by the constraints
for efficient ground state cooling.

\subsubsection{Minimization of Decoherence}

The ions will inevitably approach the disturbing electrode surfaces, if
the distance between the individual traps is reduced.
We now reinvest the reduced requirements on the stiffness of
the horizontal confinement to increase the height of the ions above the surface
$h$ keeping the mutual ion distances $d$ constant.
Some results for the scenario of a basic triangular lattice are
depicted in \figref{triangle:height}, which demonstrates the adapted shape of
the electrodes due to different optimization goals.
Note that the influence from electrodes of neighbouring traps increases for
an increased height $h$.
The optimization allows for an increase of the height by more then a factor of
two, still maintaining realistic trapping parameters (see below).
Hence, the related motional heating rates (in units of energy per time) are
expected to be reduced by more than an order of magnitude.
In addition, the increased $h$ should help to protect the electrodes from the
high intensity of the laser beams parallel to the electrode surfaces.

\figurehandler{\myfiguretriangleheight}

We additionally include required isolating gaps between electrodes
into subsequent simulations to deduce deviations in the resulting trapping
potential \cite{schmied:gap} (see \figref{triangle:height}).
The influence of the gaps turned out to be negligible for the example shown in
\figref{triangle:height}, however, for further miniaturized traps these
influences will gain of importance due to technically limited gap sizes.

\subsubsection{Maximization of the Lifetime of Trapped Ions}

The reduced frequencies and increased height above the surface come
at the price of a reduced trap depth.
First, sufficiently deep potentials have to be provided to
assure adequate loading rates out of thermal atomic beams,
preferably via efficient photo-ionization
\cite{hurst:ionization,kjaergaard:ionization}.
Second, sufficient lifetimes for many ions within the
potentials of scaled traps have to be achieved.
Currently, the average lifetime in a room temperature surface electrode
trap exceeds one hour (for the Sandia Linear Trap operated in our laboratory).

Deeper trapping potentials for surface traps were already achieved
by a conductive mesh with controlled voltage ($\unit[85]{\%}$ transmittance)
few millimetres above the electrode surface \cite{pearson:planartrap}.
It has also been successfully tested for the Sandia Linear Trap.
The mesh shields the ions from charges on the window and provides a
wavelength-independent alternative to a conductive coating
(see, \eg{} \refcite{allcock:surface:electrode:trap}).

\figurehandler{\myfiguretriangletilt}

\subsubsection{Control of the Symmetry of Interaction}

We additionally gain control over the individual orientations of trap axes or
the relative orientations of axes of different traps, respectively (see
\figref{triangle:tilt}).
This allows to shape the interaction for a given direction
of motional excitation between ions in different traps (see \secref{theory}).
It also allows for cooling of all spatial degrees of
freedom with laser beams, which have to propagate parallel to the trap surface
to minimize scattering off the surface.
We can rotate the individual trap axes from pointing towards the centre of the
structure (see \figref{triangle:tilt}a) into a parallel alignment and
additionally include the required tilt of the vertical ($Z$) axes, which
will result in a different symmetry of the electrodes (see
\figref{triangle:tilt}b).

\subsubsection{Control of the Potential in Individual Traps}

First, splitting DC electrodes into several separately controllable
segments allows for the individual compensation of displacements of the ions
from the minima of the pseudopotential due to stray fields and space charge
effects (\cmp{} \secref{tools:traps} and see
\refscite{berkeland:micromotion}{allcock:surface:electrode:trap} for schemes
of micromotion compensation).
Second, for further scaling, these electrodes can be used
to compensate boundary effects.
Due to larger number of inner ions, outer ones would be shifted to larger
mutual distances.
The further increased density of electrodes on the surface requires
their connections in a multilayer structure with vertical wiring (vias)
\cite{stick:surface:electrode:trap:arxiv,moehring:y:junction}.

\subsubsection{Estimation of Parameters}

We estimate the strength of simulated spin--spin interactions for
the case of $\atom{Mg}[+]$ ions in such devices with currently available laser
equipment.
We assume an typical laser power of $\unit[400]{mW}$ (max.
$\unit[600]{mW}$ are available) at $\unit[280]{nm}$ from an all solid state
laser source \cite{friedenauer:shg}.
We further assume the beam to have a cylindrical profile with waists of
$\unit[10]{\micro m} \times \unit[100]{\micro m}$ and an electrode structure
as depicted in \figref{triangle:tilt}b ($d = h = \unit[40]{\micro m}$).
For a trap depth of $\unit[100]{meV}$ and a minimal oscillation
frequency of the ions of $2 \pi \times \unit[2]{MHz}$, the interaction
strengths exceed by far $2 \pi \hbar \times \unit[1]{kHz}$.

In a different approach, we could think of using the motional degrees of
freedom for QSs.
This scheme would have the advantage that bare motional couplings are
already in the $\unit[5]{kHz}$ regime.
In that sense, they are stronger than effective spin--spin
interactions, since the latter are slowed down with respect to the original
motional couplings by the requirement of adiabaticity.
A recent theoretical proposal by some of us has shown that by using periodic
modulations of the trapping frequencies, some phenomena from solid state
physics may be simulated, such as photon assisted tunnelling
\cite{bermudez:gauge} (see \secref{many:body:theo}).

\figurehandler{\myfiguretrianglescaled}

\subsection{Perspectives of Our Approach}

As depicted in \figref{triangle:tilt}b, in a first step three ions will reside
on the vertices of a triangle and the interaction between the spins can be
simulated as in \refscite{friedenauer:qmagnet}{kim:frustration} (\cmp{}
\secref{spin:spin}) or \refcite{bermudez:gauge} (\cmp{}
\secref{many:body:theo}).
The above parameter estimates should already suffice for proof-of-principle
experiments and mesoscopically scaled QS.
Motional modes in two-dimensional trap arrays will behave similarly to radial
modes in linear RF traps for all three dimensions
\cite{porras:eff-spin,porras:2d:crystals:long} and the effective spin--spin
interaction will prefer anti-ferromagnetic order for far, red detuning from all
modes.
Thus, the systems should give us the possibility to study spin frustrations in
a spatial, triangular configuration (see also \secref{many:body:exp}).

Based on the results of these investigations further scaling of the
surface trap architecture to large-scale (triangular) lattices of tens or even
hundreds of spins might be pursued (see \figref{triangle:scaled}).
Besides of the optimization of the trapping parameters, further technical
difficulties have to be considered:

Decoherence due to motional heating as a result of the vicinity to the
electrode surfaces could be mitigated within a cryogenic setup
\cite{wineland:bible,labaziewicz:heating}.
The reduced vacuum pressure could additionally help to increase the
lifetime of $\atom{Mg}[+]$, which is currently limited by photo-chemical
reactions with hydrogen (mostly $\atom{H}_2 + \atom{Mg}[+*]
\to \atom{MgH}[+] + \atom{H}[*]$) and collisions with heavy components of the
rest gas.
The reaction can also be inverted by pulsed laser beams \cite{bertelsen:mgh}.
However, scaling the system to tens or hundreds of ions will still require
frequent and efficient reloading.
Increasing the loading efficiency and preserving the vacuum conditions could
be achieved by photo-ionizing cold atoms from a magneto-optical trap (MOT)
\cite{cetina:coldionsource}.

Currently, the available laser power should not impose any restrictions on
the realization of systems of few tens of ions (see \refcite{friedenauer:shg}
providing currently up to $\unit[600]{mW}$).
Higher laser powers for magnesium are in reach
\cite{feng:150w:laser,feng:25w:laser,toptica:priv:comm} and could
allow for even larger arrays of simultaneously coupled ions.
Besides of that, efforts in optics, \eg{} arrays of lenses
\cite{dumke:trap:array,streed:optics}, fibres integrated into the trap
\cite{vandevender:fiber:trap}, or integrated mirrors
\cite{shu:mirror:trap,noek:optics:trap,herskind:optics:trap}, could
provide individual addressing and high light intensities at the position of the
ions.
To further mitigate the problem of scattered light from surfaces,
one could think of realizing surface traps on partially transparent
substrates \cite{seidelin:surface-trap}.
Alternatively, laser-less coupling could be used as mentioned
in \secref{tools:operations} \cite{wunderlich:long-wave,johanning:addressing,%
chiaverini:laserless,ospelkaus:magnetic:gates,brown:gate,ospelkaus:gate}.

Last but not least, it still has to be identified how to measure observables
that permit the verification of frustration effects without the need for full
(exponentially complex) state tomography.


\section{Scaling QS Based on Ions in Optical Lattices}
\seclabel{lattices}

Some groups aim to merge the two fields of QS based on
ions in RF traps and atoms confined in optical lattices.
It had already been proposed to combine Coulomb crystals in a
harmonic confinement of a common RF trap of three-dimensional geometry with
(commensurate) optical lattices to shape anharmonic trapping
potentials providing new possibilities to simulate interactions
\cite{schmied:ion:optical:lattice}.
Another proposal deals with the simulation of the Frenkel--Kontorova model
using a standing wave aligned with the trap axis
\cite{pruttivarasin:optical:lattice}.
C.~Kollath et al. suggested to exploit a trapped ion to
coherently couple (like a scanning microscope) to the atoms confined
in an optical lattice \cite{kollath:microscope}.

\figurehandler{\myfigureperspectives}

Optical ion trapping was realized with a single $\atom{Mg}[+]$ ion
trapped in a dipole trap \cite{schneider:dipole:trap}.
We can now dream of spanning an array of ions (even simultaneously with
neutral atoms) within an optical lattice.
It has to be emphasized that the smaller trap depth of optical traps
(see \figref{optical:rf:trapping}) renders it highly unlikely that optically
trapping charged atoms will allow to outperform the achievable trapping
parameters or coherence times of both, ions in RF traps and of
optically trapped neutral atoms.
However, in our opinion, this is not required.
The advantage of equally and closely spaced traps might be combined with
individual addressability and, most important, the long-range interaction
provided by Coulomb forces between the ions.

In the following, we will first describe how trapping of an ion in a dipole
trap was achieved.
Still facing a huge variety of challenges, the new possibilities
will be discussed afterwards.

\subsection{Trapping of an Ion in a Dipole Trap}

The procedure used in \refcite{schneider:dipole:trap} to load a magnesium ion
($\atom[24]{Mg}[+]$) into a dipole trap consists of the following steps:
An atom is photo-ionized out of a thermal beam and trapped and Doppler cooled
in a conventional RF trap.
Next, stray electric fields are minimized at the site of the ion using the ion
as a sensor.
Then a Gaussian laser beam providing the dipole trap is focused onto the ion
and the RF drive of the RF trap is switched off.
From that time on, the ion is confined in the dipole trap in the directions
perpendicular to the beam propagation.
The depth of the dipole trap potential amounts to
$U_0 \approx 2 \pi \hbar \times \unit[800]{MHz}$ or $U_0 \approx \kB \times
\unit[38]{mK}$, respectively, and the detuning of the dipole trap beam
from the relevant transition ($\level{S}[1/2] \leftrightarrow \level{P}[3/2]$)
to $\Delta \approx - 6600 \Gamma$, where $1 / \Gamma$ determines the
lifetime of the $\ket{\level{P}[3/2]}$ state.
Static electric fields provide the confinement in the direction of beam
propagation.
After a few milliseconds the RF drive is switched on again and the presence
of the ion can be verified via its fluorescence during Doppler cooling.

For the given parameters a half-life of approximately $\unit[2.5]{ms}$ is
achieved.
This value is in very good agreement with the theoretical
predictions, assuming exclusively the heating process related to
off-resonant scattering of the trapping light by the ion, the
so-called recoil heating.
It can be concluded that the heating and subsequent loss of ions from
the optical potential is not dominated by heating effects related to the
charge of the ion, \eg{}, due to the vicinity of electrodes or fluctuating stray
electric fields.
Thus, state of the art techniques for neutral atoms should allow to
effectively enhance the lifetime and coherence times \cite{grimm:dipole:trap}.

\subsection{Lifetime and Coherence Times of Optically Trapped Ions}

We aim to increase the lifetime by cooling the ion in the dipole trap.
Due to the large AC Stark shift and its large
position dependency, simple Doppler cooling within the existing
setup is challenging.
Possibilities of cooling the ions directly towards the ground state of motion
within the dipole potential are currently investigated theoretically and
experimentally.

An alternative approach suggests to use cold atoms or even a BEC to
sympathetically cool ions \cite{zipkes:ion:bec}.
On longer timescales the approach of cavity assisted cooling of
ions in conventional RF traps reported in \refcite{leibrandt:cavity:cooling}
might also provide long lifetimes without affecting the electronic
state of the ion.

Currently, the coherence time of the electronic state of the ion is limited
to few microseconds due to the high spontaneous emission rate.
If longer coherence times are required (which is not necessarily the case
for every scenario), they can be achieved in two ways:
(1) As for two-photon stimulated-Raman transitions the spontaneous emission
rate can be reduced by increasing the detuning.
A larger beam intensity could sustain the potential depth.
(2) Another option would be to work with blue detuned light, where the
potential depths can remain identical, however, the ions seek for low intensity
and exhibit less spontaneous emission.

\subsection{Towards Ions and Atoms in a Common Optical Lattice}

It has still to be demonstrated that one or several ions can be confined
within one- or more-dimensional optical lattices.
Regarding the currently available parameters, we expect a mutual ion distance
of the order of a few micrometres only, which corresponds to one ion at
approximately every 50th lattice site.
Therefore the mutual ion distance could remain comparable to the distance
within one (tight) common potential of a RF trap and smaller than the
currently envisioned distances between neighbouring traps in the RF surface
electrode trap approach (see \secref{surface:traps}).

Since the photo-ionization scheme applied so far ionizes out of a
thermal beam of magnesium atoms, the average kinetic energy of the atoms is
much larger than the depth of the optical potential and, in addition, the
local vacuum is severely affected.
The loading efficiency for RF traps could be largely enhanced by ionizing
$\atom{Mg}$ atoms from a magneto-optical trap (MOT) \cite{rehbein:quenching},
which would also allow to directly load atoms into an optical trap.
In addition, after loading neutral atoms into the lattice, some of them
could be photo-ionized on site.

Ions and atoms confined in a common optical lattice could offer an approach
to exploit the physics of charge transfer reactions.
This might allow for a complete new class of QS, for example, of solid-state
systems, where atoms in a completely occupied lattice (at an initially small
density of ions) share electrons by tunnelling causing highly entangled
states of the compound system with most interesting quantum dynamics governed
by the Bose--Hubbard Hamiltonian \cite{cirac:priv:comm,zoller:priv:comm}.


\section{Conclusions}
\seclabel{outlook}

In the last few years the basic building blocks for a scalable
architecture of a quantum information processor (QC) with trapped
ions have been demonstrated for a few qubits.
Additionally, a large variety of new techniques is already tested
that might considerably extend the available toolbox.
For example, interactions based on magnetic field gradients and RF fields,
fibre-coupled optical support on chip or economically and
technologically facilitated cryogenic environments.
Despite the fact that it will be a non-trivial challenge to scale the system to
approximately $10^5$ qubits, no fundamental limitations can be
identified so far.

On a shorter timescale, intriguing problems might be studied by
realizing analogue quantum simulators (QS), by far exceeding the
capabilities of classical computers.
They can be based on similar techniques as a potential QC, but with
less severe constraints on the fidelity of operations and the number
of required ions.

Currently, available operational fidelities are predicted to allow
for studying many-body physics, for example in systems described by
quantum spin Hamiltonians, the Bose--Hubbard and the spin--boson
models.
First proof-of-principle experiments simulating Ising type interactions
with few ions were already successfully demonstrated.

The required increase of the the number of ions and the
accessible dimensions is proposed within two-dimensional
arrays of RF surface electrode traps.
However, the approach is still at the level of proof-of-principle
experiments and further challenges might arise during its development.
Alternative approaches include Penning traps or optical lattices.

Even though the enthusiasm within this quickly growing field seems
to be justified, it has to be emphasized that efficient analogue
quantum simulators still require more than simply scaling.
Examples of other important challenges are:
(1) To investigate carefully the influence of different sources of
decoherence on the fidelity of the simulation.
Thus, it must be distinguished for the dedicated application, which decoherence
the simulation will be robust against, which decoherence can be
considered in the simulation and which decoherence is even essential
to be included.
(2) To identify possibilities to cross-check the validity of the
output or to benchmark it against other QS approaches, as soon as
the achieved output is not accessible with a classical
computer anymore.

In the future it might be beneficial to combine advantages of
several systems for hybrid QS.
On longer timescales, the experiences gained by developing an analogue
QS based on trapped ions might culminate in approaches incorporating
solid-state devices that might allow for ``easier'' scaling.
With the realization of a universal QC, universal QS will also become
accessible.


\iftwocolumn{}{\clearpage}


\section*{Acknowledgement}

C.~S. and T.~S. acknowledge support by the Max-Planck-Institut f\"{u}r
Quantenoptik (MPQ),
Max-Planck-Gesellschaft (MPG), Deutsche Forschungsgemeinschaft (DFG)
(SCHA 973/1-6), the European Commission (The Physics of Ion Coulomb Crystals:
FP7 2007--2013. grant no. 249958) and the DFG Cluster of Excellence
``Munich Centre for Advanced Photonics''.
D.~P. acknowledges support from \mbox{C.A.M.} Project QUITEMAD, RyC Contract
Y200200074, and MICINN FIS2009-10061.
Martin Enderlein, Thomas Huber, and Hector Schmitz have participated in the
measurements of the geometric phase gate utilizing the radial modes of motion
presented in \secref{phase:gate}.
G\"{u}nther Leschhorn and Steffen Kahra have observed and taken the picture of
the structural defect in a Coulomb crystal of $\atom{Mg}[+]$ ions
(see \figref{kink}b).
Benni Reznik and Haggai Landa provided the corresponding simulation
(see \figref{kink}a).
Roman Schmied kindly provided several simulation results and illustrations of
optimized electrode structures of surface electrode traps
(see \secref{surface:traps}).
We want to thank Geza Giedke and Alex Retzker for discussions and
Martin Enderlein, Thomas Huber, and Dietrich Leibfried for comments on the
manuscript.
We also thank Ignacio Cirac and Gerhard Rempe for their intellectual and
financial support.


\iftwocolumn{}{\clearpage}

\appendix


\ultrashortmacroson 

\section{Normal Modes and Frequencies}
\seclabel{modes:freqs}

This section describes a more general derivation of the normal modes and
frequencies compared to the one-dimensional treatment for the linear Paul trap
as in \refcite{james:iontraps}.
The equations are extended to three dimensions and an arbitrary trap potential,
as long as the potential at the equilibrium position $\site{\vec x_0}$
of each ion can be well approximated by a harmonic potential.

The position of the $i$-th ion is expressed in the Cartesian coordinates of
the laboratory frame
\begin{equation}
  \site{\vec r} = r_i \vec e_X + r_{i + N} \vec e_Y + r_{i + 2 N} \vec e_Z.
\end{equation}
The decomposition into the equilibrium position $\site{\vec x_0}$ and
displacements $\site{\vec x}$ yields
\begin{align}
  \site{\vec r}
    &= \site{\vec x_0} + \site{\vec x} \\
    &= \left(x_{0, i} + x_i\right) \vec e_X
      + \left(x_{0, i + N} + x_{i + N}\right) \vec e_Y \nonumber\\
    &\quad + \left(x_{0, i + 2 N} + x_{i + 2 N}\right) \vec e_Z.
\end{align}

The Lagrangian for $N$ ions takes the form
\begin{equation}
  \eqlabel{lagrangian}
  \L = \frac{1}{2} M \Biggl[\sum_{k = 1}^{3 N} \dot x_k^2 - \sum_{k = 1}^{3 N}
    \sum_{l = 1}^{3 N} \underbrace{\frac{1}{M}
    \left(\frac{\partial^2 V}{\partial r_k \partial r_l}\right)%
    _{x_k = x_l = 0}}_{\eqdef a_{kl}}
    x_k x_l \Biggr],
\end{equation}
where $M$ denotes the mass of an ion, the index of the partial derivatives
signifies its evaluation at the equilibrium positions and $V$ denotes the
potential consisting of the trap potential $V_0$ and the Coulomb potentials of
the ions:
\begin{equation}
  V = V_0 + \frac{Q^2}{8 \pi \epsilon_0} \sum_{i = 1}^N
  \sum_{\substack{j = 1 \\ j \neq i}}^N \frac{1}{\left|\site{\vec r}
  - \site[j]{\vec r} \right|}.
\end{equation}
Here, $Q$ denotes the charge and $\epsilon_0$ the electric constant.

For practical purposes the trap potential can be expressed by the harmonic
terms corresponding to each ion:
\begin{equation}
  V_0 = \frac{1}{2} M \sum_{i = 1}^N \sum_{j = 1}^3 \site{\omega_j}^2
  \left(\site{\vec r} - \site{\vec p}\right) \site{\vec d_j} \otimes
  \site{\vec d_j} \left(\site{\vec r} - \site{\vec p}\right).
\end{equation}
Here, $\site{\omega_j}$ denotes the $j$-th frequency of the harmonically
approximated potential of the $i$-th ion, $\site{\vec d_j}$ the unity vector of
the principle axis corresponding to $\site{\omega_j}$, and $\site{\vec p}$ the
position of the local minimum of the potential for the $i$-th ion.
Note that the frequencies, the vectors of the principle axes, and the minima of
the potentials become equal for all ions in the special case of a linear Paul
trap.

The eigenvalues of the Hessian $A \defeq \left(a_{kl}\right)$
(see \eqref{lagrangian}) yield the squares of the frequencies $\mode{\omega}$
of the normal modes and its eigenvectors $\vec b_m$ determine the ions' motion
of the $m$-th mode:
\begin{equation}
  q_m = \vec b_m \cdot \vec x \ms \text{with} \ms
  \vec x \defeq \left(x_1, \dotsc, x_{3 N} \right).
\end{equation}
With the abbreviations $\vec q \defeq \left(q_1, \dotsc, q_{3N}\right)$ and
$B \defeq \transpose{\left(\vec b_1, \dotsc, \vec b_{3N}\right)}$, where the
$\vec b_m$ shall be understood as rows of $B$, we can express the relation in a
more compact way:
\begin{equation}
  \eqlabel{modes:trans}
  \vec q = B \vec x \ms \Leftrightarrow \ms \vec x = \transpose B \vec q.
\end{equation}
Typically, the eigenvalues and eigenvectors of $A$ have to be determined
numerically.

\section{Transformations of Pauli Operators}
\seclabel{pauli:trans}

The definitions of the Pauli operators is repeated here to avoid confusions
concerning their normalization.

The Pauli operators are defined by:
\begin{equation}
  \eqlabel{pauli}
  \pauli{x} \defeq \begin{pmatrix}
      0 & 1 \\
      1 & 0
    \end{pmatrix}, \ms
  \pauli{y} \defeq \begin{pmatrix}
      0 & -\ii \\
      \ii & 0
    \end{pmatrix}, \ms
  \pauli{z} \defeq \begin{pmatrix}
      1 & 0 \\
      0 & -1
    \end{pmatrix}.
\end{equation}
The Pauli operators obey the relations
\begin{align}
  \comm{\pauli{i}}{\pauli{j}} &= 2 \ii \epsilon_{ijk} \pauli{k} \\
  \acomm{\pauli{i}}{\pauli{j}} &= 2 \delta_{ij} \\
  \pauli{i}^2 &= \I
\end{align}

A more convenient notation in some contexts is
\begin{align}
  \eqlabel{pauli:p}
  \pauli{+} &\defeq \pauli{x} + \ii \pauli{y} = \begin{pmatrix}
      0 & 2 \\
      0 & 0
    \end{pmatrix}, \\
  \eqlabel{pauli:m}
  \pauli{-} &= \pauli{x} - \ii \pauli{y} = \begin{pmatrix}
      0 & 0 \\
      2 & 0
    \end{pmatrix}
\end{align}
with the normalization as in \refcite{wineland:bible}.
They fulfil the following relations
\begin{align}
  \comm{\pauli{\pm}}{\pauli{\mp}} &= \pm 4 \pauli{z} \\
  \comm{\pauli{z}}{\pauli{\pm}} &= \pm 2 \pauli{\pm}
\end{align}

The transformations of the Pauli operators into the interaction picture involve
terms of the form:
\begin{equation}
  \pauli{i}' \defeq
  \ee^{\ii \kappa \pauli{z}} \pauli{i} \ee^{-\ii \kappa \pauli{z}}.
\end{equation}
The transformation leaves $\I$ and $\pauli{z}$ unchanged.
The non-trivial cases $i = x$ and $i = y$ can be calculated using the
Baker-Campbell-Hausdorff formula
\begin{equation}
  \begin{aligned}
    \ee^{-\op{B}} \op{A} \ee^{\op{B}}
      &= \sum_n \frac{1}{n!} \comm{\op{A}}{\op{B}}^{\{n\}} \\
      &= \op{A} + \comm{\op{A}}{\op{B}} + \frac{1}{2}
        \comm{\comm{\op{A}}{\op{B}}}{\op{B}} + \dotsb
  \end{aligned}
\end{equation}
with $\op{B} = - \ii \kappa \pauli{z}$, $\op{A} = \pauli{x/y} =
\frac{\pauli{+} \pm \pauli{-}}{m_\pm}$, $m_+ \defeq 2$, and $m_- \defeq 2 \ii$.

The commutators are given by
\begin{align}
  \comm{\pauli{+} \pm \pauli{-}}{- \ii \kappa \pauli{z}}
    &= \ii 2 \kappa \left(\pauli{+} \mp \pauli{-}\right) \\
  \comm{\pauli{+} \pm \pauli{-}}{- \ii \kappa \pauli{z}}^{\{2\}}
    &= \ii 2 \kappa \comm{\left(\pauli{+} \mp \pauli{-}\right)}%
      {- \ii \kappa \pauli{z}} \nonumber\\
    &= \left(\ii 2 \kappa\right)^2 \left(\pauli{+} \pm \pauli{-}\right) \\
  \comm{\pauli{+} \pm \pauli{-}}{- \ii \kappa \pauli{z}}^{\{2n - 1\}}
    &= \left(\ii 2 \kappa\right)^{2n - 1}
      \left(\pauli{+} \mp \pauli{-}\right) \\
  \comm{\pauli{+} \pm \pauli{-}}{- \ii \kappa \pauli{z}}^{\{2n\}}
    &= \left(\ii 2 \kappa\right)^{2n}
      \left(\pauli{+} \pm \pauli{-}\right).
\end{align}
Hence, the Pauli operators in the interaction picture read
\begin{equation}
  \eqlabel{pauli:trans}
  \begin{aligned}
    \ee^{\ii \kappa \pauli{z}} \pauli{x/y} \ee^{-\ii \kappa \pauli{z}}
      &= \ee^{\ii \kappa \pauli{z}} \frac{\pauli{+} \pm \pauli{-}}{m_\pm}
        \ee^{-\ii \kappa \pauli{z}} \\
      &= \frac{1}{m_\pm} \sum_n \frac{\left(\ii 2 \kappa\right)^{2n + 1}}%
        {(2 n + 1)!} \left(\pauli{+} \mp \pauli{-}\right) \\
      &\quad + \frac{1}{m_\pm} \sum_n \frac{\left(\ii 2 \kappa\right)^{2 n}}%
        {(2 n)!} \left(\pauli{+} \pm \pauli{-}\right) \\
      &= \frac{1}{m_\pm} \sum_n \frac{\left(\ii 2 \kappa\right)^{n}}{n!}
        \pauli{+} \\
      &\quad \pm \frac{1}{m_\pm} \sum_n \frac{\left(- \ii 2 \kappa\right)^{n}}%
        {n!} \pauli{-} \\
      &= \frac{1}{m_\pm} \left(\ee^{\ii 2 \kappa} \pauli{+} \pm
        \ee^{- \ii 2 \kappa} \pauli{-}\right).
  \end{aligned}
\end{equation}

The operators $\pauli{+}$ and $\pauli{-}$ transform:
\begin{align}
  \eqlabel{pauli:trans:p}
  \ee^{\ii \kappa \pauli{z}} \pauli{+} \ee^{-\ii \kappa \pauli{z}}
    &= \ee^{\ii 2 \kappa} \pauli{+} \\
  \eqlabel{pauli:trans:m}
  \ee^{\ii \kappa \pauli{z}} \pauli{-} \ee^{-\ii \kappa \pauli{z}}
    &= \ee^{-\ii 2 \kappa} \pauli{-}.
\end{align}

\section{Transformations of Motional Operators}
\seclabel{motional:trans}

The creation operator $\a$ and the annihilation operator $\ad$ fulfil the
relations:
\begin{align}
  \a \ket{n} = \sqrt{n} \ket{n - 1} \\
  \ad \ket{n} = \sqrt{n + 1} \ket{n + 1} \\
  \comm{\a}{\ad} &= \I
\end{align}

The transformation of the Hamiltonians into the interaction picture requires
the knowledge of the transformation of $\ee^{\ii \xi \left(\a + \ad\right)}$.

It can be performed using the special case of the Baker-Campbell-Hausdorff
formula from \secref{pauli:trans} again.
The commutators appearing in the formula are
\begin{align}
  \comm{\a + \ad}{- \ii \lambda \ad \a}
    &= - \ii \lambda \left(\a - \ad\right) \\
  \comm{\a + \ad}{- \ii \lambda \ad \a}^{\{2\}}
    &= - \ii \lambda \comm{\a - \ad}{- \ii \lambda \ad \a} \nonumber\\
    &= (- \ii \lambda)^2 \left(\a + \ad\right) \\
  \comm{\a + \ad}{- \ii \lambda \ad \a}^{\{n\}}
    &= (- \ii \lambda)^n \left(\a + (-1)^n \ad\right) \\
\end{align}
Hence, the full transformation reads
\begin{equation}
  \begin{aligned}
    \ee^{\ii \lambda \ad \a} \left(\a + \ad\right) \ee^{- \ii \lambda \ad \a}
      &= \sum_n \frac{(- \ii \lambda)^n}{n!} \left(\a + (-1)^n \ad\right) \\
      &= \a \ee^{- \ii \lambda} + \ad \ee^{\ii \lambda}.
  \end{aligned}
\end{equation}

From this relation we can immediately derive
\begin{equation}
  \begin{aligned}
    \ee^{\ii \lambda \ad \a} \left(\a + \ad\right)^n \ee^{- \ii \lambda \ad \a}
      &= \left[\ee^{\ii \lambda \ad \a} \left(\a + \ad\right)
        \ee^{- \ii \lambda \ad \a} \right]^n \\
      &= \left[\a \ee^{- \ii \lambda} + \ad \ee^{\ii \lambda} \right]^n
  \end{aligned}
\end{equation}
by making use of the unitarity of $\ee^{\ii \lambda \ad \a}$.
We obtain for the transformation:
\begin{equation}
  \eqlabel{exp:aad:trans}
  \begin{aligned}
    \ee^{\ii \lambda \ad \a} \ee^{\ii \xi \left(\a + \ad\right)}
        \ee^{- \ii \lambda \ad \a}
      &= \ee^{\ii \lambda \ad \a} \sum_n
        \frac{(\ii \xi)^n}{n!} \left(\a + \ad\right)^n
        \ee^{- \ii \lambda \ad \a} \\
      &= \sum_n \frac{(\ii \xi)^n}{n!}
        \left[\a \ee^{- \ii \lambda} + \ad \ee^{\ii \lambda} \right]^n \\
      &= \exp\left(\ii \xi \left[\a \ee^{- \ii \lambda}
        + \ad \ee^{\ii \lambda}\right]\right).
  \end{aligned}
\end{equation}

\section{Matrix Elements of Displacement Operator}
\seclabel{matrix:disp}

The following derivation is based on \refcite{wineland:cooling}.
A difference is that we do not restrict the displacement to purely
imaginary $\lambda$ in the following.
A similar derivation can be found in appendix B of \refcite{cahill:expansions}.

The simple Baker-Campbell-Hausdorff formula
\begin{equation}
  \ee^{A + B} = \ee^A \ee^B \ee^{-\comm{A}{B} / 2}
\end{equation}
holds for $\comm{A}{\comm{A}{B}} = \comm{B}{\comm{B}{A}} = 0$.

We obtain for the annihilation operator:
\begin{equation}
  \a^m \ket{n} = \begin{cases}
    \sqrt{\frac{n!}{(n - m)!}} \ket{n - m}
      &\text{for } m \le n \\
    0 &\text{else}
  \end{cases}.
\end{equation}
Using the above form of the Baker-Campbell-Hausdorff formula, we can rewrite
the displacement operator as
\begin{equation}
  \D(\lambda) = \ee^{\lambda \ad - \lambda^* \a}
  = \ee^{-|\lambda|^2 / 2} \ee^{\lambda \ad} \ee^{- \lambda^* \a}.
\end{equation}
With
\begin{equation}
  \begin{aligned}
    \ee^{- \lambda^* \a} \ket{n}
      &= \sum_m \frac{\left(- \lambda^*\right)^m}{m !} \a^m \ket{n} \\
      &= \sum_m \frac{\left(- \lambda^*\right)^m}{m !}
        \sqrt{\frac{n!}{(n - m)!}} \ket{n - m}
  \end{aligned},
\end{equation}
this yields for $n' \ge n$:
\begin{equation}
  \eqlabel{mat:disp:a}
  \begin{aligned}
    \braket{n' | \D(\lambda) | n}
      &= \ee^{-|\lambda|^2 / 2}
        \braket{n' | \ee^{\lambda \ad} \ee^{- \lambda^* \a} | n} \\
      &= \ee^{-|\lambda|^2 / 2} \sum_{m'} \sum_m \braket{n' - m' | n - m}
        \frac{\lambda^{m'}}{m' !} \\
      &\quad \times \frac{\left(- \lambda^*\right)^m}{m !}
        \sqrt{\frac{n'!}{(n' - m')!}} \sqrt{\frac{n!}{(n - m)!}} \\
      &= \ee^{-|\lambda|^2 / 2} \lambda^{n' - n} \sum_{m = 0}^{n}
        \frac{(-1)^m |\lambda|^{2 m}}{m ! (n' - n + m)!} \\
      &\quad \times \frac{\sqrt{n'! n!}}{(n - m)!} \\
      &= \ee^{-|\lambda|^2 / 2} \lambda^{n' - n} \sqrt{\frac{n !}{n' !}}
        L_n^{(n' - n)}\left(|\lambda|^2\right),
  \end{aligned}
\end{equation}
where $L_n^{(\alpha)}(x)$ denotes the generalized Laguerre polynomials
\cite{abramowitz:stegun}.
Analogously, we obtain for $n' \le n$:
\begin{equation}
  \eqlabel{mat:disp:b}
  \braket{n' | \D(\lambda) | n}
  = \ee^{-|\lambda|^2 / 2} \left(-\lambda^*\right)^{n - n'}
  \sqrt{\frac{n' !}{n !}} L_{n'}^{(n - n')}\left(|\lambda|^2\right).
\end{equation}

For values $\lambda = \ii \eta \ee^{\ii \omega t}$ with
$\eta \in \nset{R}$, we can write \eqsref{mat:disp:a}{mat:disp:b} as
\begin{equation}
  \begin{aligned}
    \braket{n' | \D\left(\ii \eta \ee^{\ii \omega t}\right) | n}
      &= \ee^{-\eta^2 / 2} \left(\ii \eta\right)^{|n' - n|}
        \ee^{\ii \omega (n' - n) t} \\
      &\quad \times \sqrt{\frac{n_< !}{n_> !}}
        L_{n_<}^{|n' - n|}\left(\eta^2\right),
  \end{aligned}
\end{equation}
where $n_< \defeq \min(n', n)$ and $n_> \defeq \max(n', n)$.

\section{System of Differential Equations of Rabi Problem}
\seclabel{rabi:dgl}

The Rabi problem consists of the following system of differential equations:
\begin{equation}
  \eqlabel{rabi:dgl}
  \left|\begin{aligned}
    \dot c_2 &= \lambda \ee^{-\ii \omega t} c_1 \\
    \dot c_1 &= - \lambda^* \ee^{\ii \omega t} c_2
  \end{aligned}\right|.
\end{equation}
It can be solved by differentiating with respect to $t$
\begin{equation}
  \left|\begin{aligned}
    \ddot c_2 &= \lambda \ee^{-\ii \omega t} \dot c_1
      - \ii \omega \lambda \ee^{-\ii \omega t} c_1 \\
    \ddot c_1 &= - \lambda^* \ee^{\ii \omega t} \dot c_2
      - \ii \omega \lambda^* \ee^{\ii \omega t} c_2
  \end{aligned}\right|
\end{equation}
and inserting \eqref{rabi:dgl}:
\begin{align}
  \eqlabel{rabi:dgl:2:a}
  \ddot c_2 &= - \ii \omega \dot c_2 - |\lambda|^2 c_2 \\
  \eqlabel{rabi:dgl:2:b}
  \ddot c_1 &= \ii \omega \dot c_1 - |\lambda|^2 c_1.
\end{align}

Using the ansatz $c_i = a_i \ee^{\ii \kappa_i t}$ we obtain the characteristic
equations
\begin{align}
  - \kappa_2^2 &= \omega \kappa_2 - |\lambda|^2 \\
  - \kappa_1^2 &= - \omega \kappa_1 - |\lambda|^2,
\end{align}
which have the solutions
\begin{align}
  \kappa_{2, \pm}
    &= - \frac{\omega}{2} \pm \sqrt{\frac{\omega^2}{4} + |\lambda|^2}
      \defeq - \frac{\omega}{2} \pm \kappa' \\
  \kappa_{1, \pm}
    &= \frac{\omega}{2} \pm \sqrt{\frac{\omega^2}{4} + |\lambda|^2}
      \defeq \frac{\omega}{2} \pm \kappa' = - \kappa_{2, \mp}.
\end{align}
Here, we introduced the abbreviation $\kappa' \defeq \sqrt{\frac{\omega^2}{4}
+ |\lambda|^2}$.
The solutions of \eqsref{rabi:dgl:2:a}{rabi:dgl:2:b} read
\begin{align}
  &\begin{aligned}
    c_2
      &= a_{2, +} \ee^{\ii \kappa_{2, +} t} + a_{2, -}
        \ee^{\ii \kappa_{2, -} t} \\
      &= \left(a_{2, +} \ee^{\ii \kappa' t} + a_{2, -}
        \ee^{- \ii \kappa' t}\right) \ee^{- \ii \omega t / 2}
  \end{aligned}\\
  &\begin{aligned}
    c_1
      &= a_{1, +} \ee^{\ii \kappa_{1, +} t} + a_{1, -}
        \ee^{\ii \kappa_{1, -} t} \\
      &= \left(a_{1, +} \ee^{\ii \kappa' t} + a_{1, -}
        \ee^{- \ii \kappa' t}\right) \ee^{\ii \omega t / 2}.
  \end{aligned}
\end{align}

Inserting them into the original system of differential equations
\eqref{rabi:dgl}, we obtain the following relations for the constants
$a_{i, \pm}$:
\begin{align}
  \eqlabel{rabi:const:rel}
  \ii \kappa_{2, \pm} a_{2, \pm} &= \lambda a_{1, \pm} \\
  \ii \kappa_{1, \pm} a_{1, \pm} &= - \lambda^* a_{2, \pm}
\end{align}
We replace $a_{1, \pm}$ by $a_{2, \pm}$ using \eqref{rabi:const:rel} and
obtain
\begin{align}
  \eqlabel{rabi:sol:pre:a}
  c_2 &= \left(a_{2, +} \ee^{\ii \kappa' t} + a_{2, -}
    \ee^{- \ii \kappa' t}\right) \ee^{- \ii \omega t / 2} \\
  \eqlabel{rabi:sol:pre:b}
  c_1 &= \left(\mu_+ a_{2, +} \ee^{\ii \kappa' t} + \mu_- a_{2, -}
    \ee^{- \ii \kappa' t}\right) \ee^{\ii \omega t / 2},
\end{align}
where we introduced the (temporary) abbreviation $\mu_\pm \defeq
\ii \kappa_{2, \pm} / \lambda$.

The constants $a_{2, \pm}$ can now be expressed in terms of the initial
values $c_{20} \defeq c_2(t = 0)$ and $c_{10} \defeq c_1(t = 0)$.
Setting $t = 0$ in \eqsref{rabi:sol:pre:a}{rabi:sol:pre:b},
we obtain a system of linear equations with the solutions
\begin{align}
  a_{2, +}
    &= \frac{\mu_- c_{20} - c_{10}}{\mu_- - \mu_+} \\
  a_{2, -}
    &= - \frac{\mu_+ c_{20} - c_{10}}{\mu_- - \mu_+}.
\end{align}
Hence,
\begin{align}
  c_2
    &= \frac{(\mu_- c_{20} - c_{10}) \ee^{\ii \kappa' t}
      - (\mu_+ c_{20} - c_{10}) \ee^{- \ii \kappa' t}}{\mu_- - \mu_+}
      \ee^{- \ii \omega t / 2} \\
  c_1 &= \frac{\mu_+ (\mu_- c_{20} - c_{10}) \ee^{\ii \kappa' t}
    - \mu_- (\mu_+ c_{20} - c_{10}) \ee^{- \ii \kappa' t}}{\mu_- - \mu_+}
    \ee^{\ii \omega t / 2},
\end{align}
and by expressing $\mu_\pm$ in terms of $\omega$, $\kappa'$, and $\lambda$:
\begin{align}
  &\begin{aligned}
    c_2(t)
      &= \left(\cos(\kappa' t) + \frac{\omega}{2} \frac{\ii}{\kappa'}
        \sin(\kappa' t)\right) \ee^{- \ii \omega t / 2} c_2(0) \\
      &\quad + \frac{\lambda}{\kappa'} \sin(\kappa' t) \ee^{- \ii \omega t / 2}
        c_1(0)
  \end{aligned}\\
  &\begin{aligned}
    c_1(t)
      &= - \frac{\lambda^*}{\kappa'} \sin(\kappa' t) \ee^{\ii \omega t / 2}
        c_2(0) \\
      &\quad + \left(\cos(\kappa' t) - \frac{\omega}{2} \frac{\ii}{\kappa'}
        \sin(\kappa' t)\right) \ee^{\ii \omega t / 2} c_1(0).
  \end{aligned}
\end{align}

\section{Time Evolution Operator}
\seclabel{time:evol}

The calculation of the time evolution operator involves terms of the form
\begin{equation}
  \mode{\site{\H}}(t) = \ii \site{\mode{\xi}} \ee^{\ii\left(-\mode{\delta} t
  + \site{\phi}\right)} \mode{\ad} + \hc,
\end{equation}
where $\site{\mode{\xi}} \in \nset{R}$ and the total Hamiltonian reads
$\H(t) \eqdef \sum_{i = 1}^N \sum_{m = 1}^{3 N} \mode{\site{\H}}(t)$.
(More generally, the constants $\site{\mode{\xi}}$ represent Hermitian
operators $\site{\mode{\op{\xi}}}$ with $\comm{\site{\mode{\op{\xi}}}}%
{\site[j]{\mode[n]{\op{\xi}}}} = 0 \Forall i, j, m, n$.)

The commutator of two of these terms will trivially vanish for all $i, j$ and
all times $t', t''$, if both terms belong to different modes $m \neq n$:
\begin{equation}
  \comm{\mode{\site{\H}}(t')}{\mode[n]{\site[j]{\H}}(t'')} = 0
  \ms \text{for} \ms m \neq n.
\end{equation}
However, for $m = n$, the commutators do not vanish.
Using the relation
\begin{multline}
  \comm{\ee^{\ii \lambda} \ad - \ee^{-\ii \lambda} \a}%
    {\ee^{\ii \lambda'} \ad - \ee^{-\ii \lambda'} \a} \\
  = \ee^{\ii\left(\lambda - \lambda'\right)}
    \left(\a \ad - \ad \a\right)
    - \ee^{-\ii\left(\lambda - \lambda'\right)}
    \left(\a \ad - \ad \a\right) \\
  = 2 \ii \sin\left(\lambda - \lambda'\right) \I
\end{multline}
yields
\begin{multline}
  \comm{\mode{\site{\H}}(t')}{\mode{\site[j]{\H}}(t'')} \\
  = 2 \ii \site{\mode{\xi}} \site[j]{\mode{\xi}}
    \sin\left(\mode{\delta} (t' - t'')
    - \left(\site{\phi} - \site[j]{\phi}\right)\right).
\end{multline}

The time evolution operator can be calculated using a Magnus expansion
\cite{magnus:expansion,blanes:magnus:expansion}.
As commutators with higher ``nesting level'' trivially vanish, the expansion
simplifies to:
\begin{equation}
  \begin{aligned}
    \U(t, t_0)
      &= \exp\Biggl(- \frac{\ii}{\hbar} \int_{t_0}^t \di t' \H(t') \\
      &\quad- \frac{1}{2 \hbar^2} \int_{t_0}^t \di t' \int_{t_0}^{t'} \di t''
        \comm{\H(t')}{\H(t'')}\Biggr).
  \end{aligned}
\end{equation}
The single integrals yield
\begin{equation}
  \begin{aligned}
    \int_{t_0}^t \di t' \mode{\site{\H}}(t')
      &= -\frac{\site{\mode{\xi}}}{\mode{\delta}}
        \left(\ee^{-\ii \mode{\delta} (t - t_0)} - 1\right)
        \ee^{-\ii \mode{\delta} t_0} \ee^{\ii \site{\phi}} \mode{\ad} \\
      &\quad + \hc
  \end{aligned}
\end{equation}
and the double integrals of the commutators yield
\begin{equation}
  \begin{aligned}
    &\int_{t_0}^t \di t' \int_{t_0}^{t'} \di t''
      \comm{\mode{\site{\H}}(t')}{\mode{\site[j]{\H}}(t'')} \\
    &= 2 \ii \site{\mode{\xi}} \site[j]{\mode{\xi}} \\
    &\quad \times \int_{t_0}^t \di t' \int_{t_0}^{t'} \di t''
      \sin\left(\mode{\delta} (t' - t'')
      - \left(\site{\phi} - \site[j]{\phi}\right)\right) \\
    &= \frac{2 \ii \site{\mode{\xi}} \site[j]{\mode{\xi}}}{\mode{\delta}}
      \int_{t_0}^t \di t' \left[\cos\left(\site{\phi} - \site[j]{\phi}\right)
      \right.\\
    &\quad \left.- \cos\left(\mode{\delta} (t' - t_0)
      - \left(\site{\phi} - \site[j]{\phi}\right)\right)\right] \\
    &= \frac{2 \ii \site{\mode{\xi}} \site[j]{\mode{\xi}}}{\mode{\delta}^2}
      \left[\mode{\delta} (t - t_0) \cos\left(\site{\phi}
      - \site[j]{\phi}\right) \right. \\
    &\quad \left.
      - \sin\left(\mode{\delta} (t - t_0) - \left(\site{\phi}
      - \site[j]{\phi}\right)\right)
      - \sin\left(\site{\phi} - \site[j]{\phi}\right)\right].
  \end{aligned}
\end{equation}
Note that in the time evolution operator corresponding to $\H(t)$ the terms
$\sin\left(\site{\phi} - \site[j]{\phi}\right) =
- \sin\left(\site[j]{\phi} - \site{\phi}\right)$ cancel each other.

\section{Canonical Transformation}
\seclabel{canonical:trans}

The unitary operator of the canonical transformation has the form
\begin{equation}
  \U_\text{c} \defeq \ee^{- \left(\lambda \ad - \lambda^* \a\right)}
  \ms \text{with} \ms \lambda \defeq \frac{\xi}{\kappa}
\end{equation}
and is applied to
\begin{equation}
   \H \defeq \underbrace{\left(\xi \ad + \xi^* \a\right)}_{\eqdef \H_1}
   + \underbrace{\left(-\kappa \ad \a\right)}_{\eqdef \H_2}
   \ms \to \ms \H' \defeq \U_\text{c} \H \U_\text{c}^\dagger.
\end{equation}
Here, the constants $\xi \in \nset{C}$ and $\kappa \in \nset{R}$.

We use the Baker-Campbell-Hausdorff formula from \secref{pauli:trans} to do
the transformation.
The commutators for $\H_1$ yield
\begin{equation}
  \eqlabel{h:1:comm}
  \begin{aligned}
    \comm{\xi \ad + \xi^* \a}{\lambda \ad - \lambda^* \a}
      &= \xi \lambda^* \left(- \ad \a + \a \ad\right) \\
      &\quad - \xi^* \lambda \left(\ad \a - \a \ad\right) \\
      &= \left(\xi \lambda^* + \xi^* \lambda\right) \I \\
      &= \frac{2 \xi \xi^*}{\kappa} \I.
  \end{aligned}
\end{equation}
Commutators with higher ``nesting levels'' trivially vanish.
Hence, the complete transformation of $\H_1$ reads
\begin{equation}
  \H_1' \defeq \U_\text{c} \H_1 \U_\text{c}^\dagger =
  \left(\xi \ad + \xi^* \a\right) + \frac{2 \xi \xi^*}{\kappa} \I
\end{equation}

The commutators for $\H_2$ read
\begin{equation}
  \begin{aligned}
    \comm{- \kappa \ad \a}{\lambda \ad - \lambda^* \a}
      &= - \kappa \left(\ad \comm{\a}{\lambda \ad - \lambda^* \a}\right. \\
      &\quad \left.+ \comm{\ad}{\lambda \ad - \lambda^* \a} \a\right) \\
      &= - \kappa \left(\lambda \ad + \lambda^* \a\right) \\
      &= - \left(\xi \ad + \xi^* \a\right)
  \end{aligned}
\end{equation}
and
\begin{equation}
  \begin{aligned}
    \comm{- \kappa \ad \a}{\lambda \ad - \lambda^* \a}^{\{2\}}
      &= \comm{- \left(\xi \ad + \xi^* \a\right)}%
        {\lambda \ad - \lambda^* \a} \\
      &= - \frac{2 \xi \xi^*}{\kappa} \I,
  \end{aligned}
\end{equation}
where we used \eqref{h:1:comm}.
Higher order terms in the expansion trivially vanish again.
The complete transformation of $\H_2$ reads
\begin{equation}
  \H_2' \defeq \U_\text{c} \H_2 \U_\text{c}^\dagger =
  - \kappa \ad \a - \left(\xi \ad + \xi^* \a\right)
  - \frac{\xi \xi^*}{\kappa} \I
\end{equation}

Hence, the transformation of the full Hamiltonian reads.
\begin{equation}
  \begin{aligned}
    \H' = \H_1' + \H_2' = \frac{\xi \xi^*}{\kappa} \I - \kappa \ad \a.
  \end{aligned}
\end{equation}
We want to stress that the canonical transformation is exact in this case.

\ultrashortmacrosoff 


\iftwocolumn{}{\clearpage}

\bibliography{quantum_magnet,numerics,priv_comm}
\bibliographystyle{\mybibstyle}


\iftwocolumn{}{%
  \ultrashortmacroson 
  \clearpage
  \section*{Figures}
  \thefigurepages
  \ultrashortmacrosoff 
}


\iftwocolumn{}{\clearpage}

\therevisionchanges

\end{document}